\documentclass[10pt,a4paper]{article}

\usepackage{amsmath,amsgen,latexsym}
\usepackage{amstext,amssymb,amsfonts,latexsym}
\usepackage{theorem}
\usepackage{pifont}
\usepackage{graphicx}

 \newcommand{\bs}{\bigskip}
 \newcommand{\ms}{\medskip}
 \newcommand{\n}{\noindent}
 \newcommand{\s}{\smallskip}
 \newcommand{\hs}[1]{\hspace*{ #1 mm}}
 \newcommand{\vs}[1]{\vspace*{ #1 mm}}

 \newcommand{\setempty}{\mathrm{\O}}
 \newcommand{\real}{\mathbb{R}}

 \newcommand{\nat}{\mathbb{N}}
 
 \newcommand{\integer}{\mathbb{Z}}

 \newcommand{\prob}{{\mathrm{Prob}}}

 \newcommand{\co}{\mathrm{co}\mbox{-}}

 \newcommand{\ie}{\textrm{i.e.},\hspace*{2mm}}
 \newcommand{\eg}{\textrm{e.g.},\hspace*{2mm}}

 \newcommand{\CC}{{\cal C}}
 \newcommand{\FF}{{\cal F}}
 \newcommand{\DD}{{\cal D}}

 \newcommand{\PP}{{\cal P}}

 \newcommand{\dl}{\mathrm{L}}

 \newcommand{\np}{\mathrm{NP}}

 \newcommand{\bpp}{\mathrm{ BPP}}

 \newcommand{\oneflin}{1\mbox{-}\mathrm{FLIN}}

 \newcommand{\reg}{\mathrm{REG}}
 \newcommand{\cfl}{\mathrm{CFL}}

 \newcommand{\cflsvt}{\mathrm{CFLSV_t}}
 \newcommand{\cflsv}{\mathrm{CFLSV}}
 
 \newcommand{\cflmv}{\mathrm{CFLMV}}
 \newcommand{\cflmvtwo}{\mathrm{CFLMV(2)}}
 \newcommand{\cflsvtwo}{\mathrm{CFLSV(2)}}

 \newcommand{\IFF}{\Longleftrightarrow}
\theoremstyle{plain}
\theoremheaderfont{\bfseries}
 \newtheorem{theorem}{Theorem}[section]
 \newtheorem{lemma}[theorem]{Lemma}
 \newtheorem{proposition}[theorem]{Proposition}
 \newtheorem{corollary}[theorem]{Corollary}

 {\theorembodyfont{\rmfamily}
  \newtheorem{definition}[theorem]{Definition}}
 {\theorembodyfont{\rmfamily} }
 {\theorembodyfont{\rmfamily} }

 \newtheorem{claim}{Claim}

 \newenvironment{proof}{\par \noindent
            {\bf Proof. \hs{2}}}{\hfill$\Box$ \vspace*{3mm}}

 \newenvironment{proofsketch}{\par \noindent
            {\bf Proof Sketch. \hs{2}}}{\hfill$\Box$ \vspace*{3mm}}

 \newenvironment{proofof}[1]{\vspace*{5mm} \par \noindent
         {\bf Proof of #1.\hs{2}}}{\hfill$\Box$ \vspace*{3mm}}

 \newcommand{\ceilings}[1]{\lceil #1 \rceil}
 \newcommand{\floors}[1]{\lfloor #1 \rfloor}
 \newcommand{\pair}[1]{\langle #1 \rangle}

\newif\ifnotesw\noteswtrue% T to show box & marginal notes; F suppresses.
   {\ifnotesw\marginpar[\hfill\(\top\)]{\(\top\)}\fi}%
      {\ifnotesw\marginpar[\hfill\(\bot\)]{\(\bot\)}\fi}
      
\newcommand{\mnote}[1]%
   {\ifnotesw\marginpar%
	  [{\scriptsize\begin{minipage}[t]{\marginparwidth}
	  \raggedleft#1%
		  \end{minipage}}]%
	  {\scriptsize\begin{minipage}[t]{\marginparwidth}
	  \raggedright#1%
		  \end{minipage}}%
    \fi}
\newcommand{\ignore}[1]{}

\newcommand{\track}[2]{[\:\begin{subarray}{c} #1 \\%
      #2 \end{subarray} ]}

\newcommand{\cent}{|\!\! \mathrm{c}}
\newcommand{\dollar}{\$}

\begin{document}
%%%%%%%%%%%%%%%%%%
%%%%%%%%%%%%%%%%%%
%%%%%%%%%%%%%%%%%%

\pagestyle{plain}
\setcounter{page}{1}

\begin{center}
{\Large {\bf Pseudorandom Generators Against Advised \s\\
Context-Free Languages}} \bs\ms\\
{\sc Tomoyuki Yamakami}\footnote{Department of Information Science, University of Fukui, 3-9-1 Bunkyo, Fukui 910-8507, Japan} \ms\\
\end{center}

%%%%%%%%%%%%%%%%%%%%

\begin{quote}
{\bf Abstract.}\hs{1}
Pseudorandomness has played a central role in modern cryptography, finding theoretical and practical applications to various fields of computer science. A function that generates pseudorandom strings from shorter but truly random seeds is known as a pseudorandom generator. Our generators are designed to fool languages (or equivalently, Boolean-valued functions).
In particular, our generator fools advised context-free languages, namely, context-free languages assisted by external information known as  advice, and moreover our generator is made almost one-to-one, stretching $n$-bit seeds to $n+1$ bits.
We explicitly construct such a pseudorandom generator, which is computed by a deterministic Turing machine using logarithmic space and also belongs to CFLMV(2)/$n$---a functional extension of the 2-conjunctive closure of CFL with the help of appropriate deterministic advice.
In contrast, we show that there is no almost one-to-one pseudorandom generator against context-free languages if we demand  that it should be computed by a nondeterministic pushdown automaton equipped with a write-only output tape. Our generator naturally extends known pseudorandom generators against advised regular languages. Our proof of the CFL/$n$-pseudorandomness of the generator is quite elementary, and in particular, one part of the proof  utilizes a special feature of the behaviors of nondeterministic pushdown automata, called a swapping property, which is interesting in its own right, generalizing the swapping lemma for context-free languages.

\ms

{\bf Keywords:} context-free language, advice,  pseudorandom generator, pushdown automaton, pseudorandom language, swapping property

{ACM Subject Classification:} F.4.3, F.1.1, F.1.3
\end{quote}

%%%%%%%%%%%%%%%%%%%%%%%%%%%%%%%%%%%%%%%%
\section{Our Challenges and Contributions}

\sloppy

Regular and context-free languages are unarguably considered as the most fundamental notions in formal language and automata theory. Those special languages  have been extensively studied since the 1950s and a large volume of work has been devoted to unearthing quite intriguing features of their behaviors and powers. Underlying finite(-state) automata that recognize those languages can be further assisted by external information, called {\em (deterministic) advice}, which is given besides input instances in order to enhance the computational power of the automata. One-way deterministic finite  automata (or dfa's, in short) and their associated regular languages that are appropriately supplemented by advice strings of size $n$ in parallel to input instances of length $n$ naturally form an advised language family, which is dubbed as $\reg/n$ first in  \cite{TYL10} and further studied in \cite{Yam08,Yam10,Yam11,Yam12}.
In a similar fashion, one-way nondeterministic pushdown automata (or npda's) and their corresponding context-free languages with appropriate advice naturally induce another  advised language family $\cfl/n$ \cite{Yam08,Yam10}.
The notion of advice endows the underlying machines with a  non-uniform nature of computation; for instance, advised regular languages
are recognized by non-uniform series of length-dependent dfa's and also characterized in \cite{Yam10} in terms of length-dependent {\em non-regularity}. Beyond the above-mentioned advice, recent studies further dealt with its important variants: {\em randomized advice} \cite{Yam10,Yam12} and {\em quantum advice} \cite{Yam12}.

In an analysis of the behaviors of languages, their corresponding {\em functions} defined on finite strings over certain alphabets have sometimes played a supporting role. Types of those functions vary considerably from an early example of functions computed by Mealy machines \cite{Mea55} and Moore machines \cite{Moo56} to more recent examples of acceptance probability functions (\eg \cite{Mac93}) and counting functions \cite{TYL10} and to an example of functions computed by npda's equipped with write-only output tapes \cite{Yam11,Yam14a,Yam14b}.
Nonetheless, a field of such functions has been largely unexplored in formal language and automata theory, and our goal to the full understandings of   structural properties of those functions still awaits to be fulfilled.
Our particular interest in this paper rests in one of those structural properties, known as {\em pseudorandomness} against advised language families \cite{Yam11}, and its theoretical application to {\em pseudorandom generators}.

The notion of pseudorandom generator dates back to early 1980s and it has since then become a key ingredient in modern cryptography and also it has made a significant impact on the development of computational complexity theory. An early generator that Blum and Micali \cite{BM84} proposed is designed to produce
a sequence in which any reasonably powerful adversary hardly predicts the sequence's next bit.  Yao's \cite{Yao82} generator, on the contrary, produces a sequence that no  adversary distinguishes from a uniformly random sequence with a small margin of error. Those two formulations---unpredictability and indistinguishability---are essentially equivalent and the generators that are formulated accordingly are now known as {\em pseudorandom generators}. Since their introduction, the pseudorandom generators have played key roles in constructing various secure  protocols as an important cryptographic primitive.  However, the existence of a (polynomial-time computable) pseudorandom generator is still unknown unless we impose certain unproven complexity-theoretical assumptions, such as $\np\nsubseteq\bpp$ or the existence of polynomial-time one-way functions (see, \eg \cite{Gol01}).

Within a framework of formal language and automata theory, a recent study \cite{Yam11} was focused on a specific type of pseudorandom generator, whose adversaries are represented in a form of languages  (or equivalently, $\{0,1\}$-valued functions), compared to standard ``probabilistic algorithms.'' Such a generator also appears when the generator's adversaries are  ``Boolean circuits'' that produce one-bit outputs.
Intuitively, given an arbitrary alphabet $\Sigma$, a (single-valued total) function $G:\Sigma^*\to\Sigma^*$, which stretches $n$-symbol seeds to $s(n)$-symbol strings, is said to {\em fool} language $A$ over $\Sigma$ if the characteristic function\footnote{The characteristic function $\chi_{A}$ of a language $A$ is defined as $\chi_{A}(x)=1$ if $x\in A$ and $\chi_{A}(x)=0$ otherwise, for every input string $x$.}
$\chi_{A}$ of $A$ cannot distinguish between the output distribution of  $\{G(x)\}_{x\in\Sigma^n}$ and a truly random distribution  of $\{y\}_{y\in\Sigma^{s(n)}}$ with {\em non-negligible} success probability.
We call $G$
a {\em pseudorandom generator} against language family $\CC$ if $G$ fools every language $A$ over $\Sigma$ in $\CC$.
As our limited adversaries, we intend to take regular languages and context-free languages assisted further by advice. An immediate advantage of dealing with such weak adversaries is that we can actually construct corresponding pseudorandom generators {\em without any unproven assumption}.

A fundamental question that naturally arises from the above definition is whether there exists an {\em efficiently computable} pseudorandom generator against a ``low-complexity'' family of languages. In an early study \cite{Yam11},  a single-valued total function
computed by an appropriate npda equipped with a write-only output tape (where the set of those functions is briefly denoted $\cflsvt$, an automaton-analogue of $\mathrm{NPSV_t}$ \cite{BLS84}) was proven to be a pseudorandom generator against $\reg/n$.
This pseudorandom generator actually stretches truly random seeds of $n$ bits to strings of $n+1$ bits and, moreover, it is made one-to-one for all but a negligible fraction of their domain instances (called {\em almost one-to-one}, or almost 1-1). The existence of such a  restricted pseudorandom generator is closely linked to the {\em $\reg/n$-pseudorandomness} of languages in $\cfl$ (context-free language family) \cite{Yam11}. Regarding the computational complexity of the generator, one may wonder if such a generator can be computed much more efficiently.  Unfortunately, as shown in \cite{Yam11}, no pseudorandom generator against $\reg$ (regular language family) can be computed by single-tape linear-time Turing machines as long as the generator is almost 1-1 and stretches $n$-bit seeds to $(n+1)$-bit strings. Notice that  almost one-to-oneness and a small stretch factor are a key to establish those results, because any generator satisfying those properties become pseudorandom if and only if its range (viewed as a language) is pseudorandom \cite{Yam11} (see also Lemma \ref{generator-pseudorandom}).

A critical question left unsolved in \cite{Yam11} is whether an efficient  pseudorandom generator of small stretch factor actually exists  against $\cfl/n$.
A simple and natural way to construct such a specific generator is to apply  a so-called {\em diagonalization technique}: first enumerate all advised languages in $\cfl/n$ and then diagonalize them one by one to determine an outcome of the generator. Such a technique  gives a generator that can be   computed deterministically in exponential time. For each language in $\cfl/n$, since it can be expressed as a family of polynomial-size Boolean circuits, a design-theoretic method of
Nisan and Wigderson \cite{NW94} can be used to construct a pseudorandom generator against those polynomial-size circuits, however, at a cost of super-polynomial running time.
With a much harder effort in this paper, we intend to give an explicit construction of a pseudorandom generator  against $\cfl/n$ whose computational complexity is simultaneously in $\mathrm{FL}$ (logarithmic-space function class) and in  $\cflmvtwo/n$---a functional analogue of $\cfl(2)/n$ (which coincides with  the {\em 2-conjunctive closure} of $\cfl/n$ by Claim \ref{CFL(2)-equivalent-def}) as well as a natural extension of $\cflmv$ (multiple-valued partial CFL-function class) given in \cite{Yam11}.

\ms
\n{\bf [First Main Theorem]} A pseudorandom generator $G$ against all advised context-free languages exists in $\mathrm{FL}\cap\cflmvtwo/n$. More strongly, $G$ can be made almost 1-1 with stretch factor $n+1$. (Theorem \ref{generator-CFL}.)
\ms

With no use of diagonalization techniques, our construction of the desired generator described in this first main theorem is  rather elementary and our proof of its pseudorandomness demands no complex arguments customarily found in a polynomial-time setting.
In particular,  the proof will require only two previously known results: a discrepancy upper bound of the inner-product-modulo-two function and a behavioral property of npda's. In particular, from the latter property, we can derive a so-called {\em swapping property} of npda's (Lemma \ref{swapping-property}), which is also interesting in its own right in connection to the {\em swapping lemma for context-free languages}  \cite{Yam08} (re-stated as Corollary \ref{swapping-lemma-CFL}).
Our pseudorandom generator $G$ against $\cfl/n$ is actually based on a special language $IP_{3}$, which embodies the {\em (binary) inner-product-modulo-two function}.  Based upon the aforementioned close tie between pseudorandom generators and pseudorandom languages, our major task of this paper becomes proving that $IP_{3}$ is a
$\cfl/n$-pseudorandom language. The most portion of this paper will be devoted to carrying out this task. Since $IP_{3}$ is in
$\mathrm{L}\cap\cfl(2)/n$ (Proposition \ref{IP-dtime}), an immediate consequence of the $\cfl/n$-pseudorandomness of $IP_{3}$ is a new class separation of $\cfl(2)\nsubseteq \cfl/n$ (Corollary \ref{CFL(2)-vs-CFL/n}), which is in fact incompatible with an earlier separation of $\co\cfl\nsubseteq\cfl/n$, proven in \cite{Yam08}.

To guide the reader through the proof of the first main theorem, here we shall give a proof outline.

\ms

\n{\bf Outline of the Proof of the First Main Theorem.}
Our desired generator $G$ that stretches $n$-bit seeds to $(n+1)$-bit strings will be formulated, in Section \ref{sec:random-generator}, based on a special language $IP_{3}$, which is defined by the (binary) inner product operation. For technical reason, we shall actually use its variant, called $IP^{+}$. We shall show in Claim \ref{property-of-G} that $G$ is almost 1-1 and its range $rang(G)$ coincides with $IP^{+}$.  In Proposition \ref{G-complexity}, $G$ will be proven to fall into $\mathrm{FL}\cap\cflmvtwo/n$.
To show that $G$ is indeed a pseudorandom generator against $\cfl/n$,
it suffices by Lemma \ref{generator-pseudorandom} to prove that $rang(G)$  is a $\cfl/n$-pseudorandom language.  Since $rang(G)$ equals $IP^{+}$, which is essentially $IP_{3}$ (Lemma \ref{IP_{3}-implies-IP_3}), we shall aim only at verifying that $IP_{3}$ is $\cfl/n$-pseudorandom (Proposition \ref{IP-pseudorandom}).
To achieve this goal, we shall pick an arbitrary advised context-free language $S$. By taking a close look at
its behavior, we shall demonstrate, in Section \ref{sec:swapping}, its  useful structural property, named as the {\em swapping property lemma} (Lemma \ref{swapping-property}), that each subset of $S$ restricted to input instances of length $4n$ can be expressed as a union of a small number of product sets $\{A_e\times B_e\mid e\in\Delta_{j_0,k_0,4n}\}$ with an appropriate index set $\Delta_{j_0,k_0,4n}$ (after suitable rearrangement of input bits). Those product sets help decompose this subset of $S$ into a finite series $\{S_e\}_{e\in \Delta_{j_0,k_0,4n}}$.
The swapping property lemma is derived from a crucial assertion of \cite{Yam08} (Lemma \ref{height-interval}), which was used for proving the {\em swapping lemma for context-free languages} \cite{Yam08} (Corollary \ref{swapping-lemma-CFL}).
In Section \ref{sec:pseudo-IP*}, we shall introduce a basic notion of {\em discrepancy}.
The $\cfl/n$-pseudorandomness of $IP_{3}$ is in fact proven by exhibiting a ``small'' discrepancy between $S_e\cap IP_{3}$ and $S_e\cap\overline{IP_{3}}$. Unfortunately, we are unable to apply a well-known discrepancy bound (see, \eg \cite{AB09}) directly to $S_e$'s. To overcome this difficulty, we shall introduce their substitutions  $\{T_e\}_{e\in \Delta_{j_0,k_0,4n}}$, whose close correspondence to $S_e$'s will be shown in Claim \ref{T_e-S}. For this set $T_e$, we shall claim a key lemma (Lemma \ref{disc_M(T)-bound}), which gives a good discrepancy upper-bound of $T_e$. This  bound will finally lead to the desired small discrepancy between $S_e\cap IP_{3}$ and $S_e\cap\overline{IP_{3}}$. The remaining proof of Lemma \ref{disc_M(T)-bound} will be given independently in Section \ref{sec:discrepancy}, completing the proof of Proposition \ref{IP-pseudorandom} and therefore the proof of the first main theorem.

\ms

To complement our first main theorem further in a ``uniform'' setting, we shall prove that any almost 1-1 pseudorandom generator  against  $\cfl$ cannot be efficiently computed by npda's equipped with write-only output tapes.
This result marks a complexity limitation of the efficiency of pseudorandom generators  against $\cfl$.

\ms
\n{\bf [Second Main Theorem]} There is no pseudorandom generator against $\cfl$ in $\cflmv$, if the generator is demanded to be almost 1-1 with stretch factor $n+1$. (Theorem \ref{no-generator}.)
\ms

We strongly expect that this paper will open a door to a full range of extensive research on {\em structural properties of functions} in formal language and automata theory and on their applications to other areas of computer science.

%%%%%%%%%%%%%%%%%%%%%%%%%%%%%%%%%%%%%%%%
%%%%%%%%%%%%%%%%%%%%%%%%%%%%%%%%%%%%%%%%
\section{Fundamental Notions and Notations}\label{sec:notions}

Let $\nat$ denote the set of all {\em nonnegative integers} (called {\em natural numbers}) and set $\nat^{+}$ for $\nat-\{0\}$.
Given two integers $m$ and $n$ with $m\leq n$, the {\em integer interval} $[m,n]_{\integer}$ is a set $\{m,m+1,m+2,\ldots,n\}$. For example, $[2,5]_{\integer} = \{2,3,4,5\}$. As a special case, we set $[n]$ to be  $[1,n]_{\integer}$ for any $n\in\nat^{+}$. We write $\real$ and $\real^{\geq0}$ respectively for the sets of all {\em real numbers} and of all {\em nonnegative real numbers}.
A (single-valued total) function $\mu$ from $\nat$ to $\real^{\geq0}$ is {\em negligible} if, for every positive(-valued) polynomial $p$, there exists a positive number $n_0$ for which $\mu(n) \leq 1/p(n)$ holds for any integer  $n\geq n_0$, where polynomials are always assumed to take integer coefficients.
Given two sets $A$ and $B$, their {\em symmetric difference}  $A\triangle B$ is the set $(A-B)\cup(B-A)$.

Let $\Sigma$ be an {\em alphabet} (\ie a finite nonempty set). A {\em string}  $x$ is a finite sequence of symbols taken from $\Sigma$ and the {\em empty string} is always denoted by $\lambda$. The {\em length} of a string $x$, denoted by $|x|$, is the number of (not necessarily distinct) symbols in $x$. Let $\Sigma^*$ be the set of all strings over $\Sigma$ and let  $\Sigma^n$  be the set of all strings of length exactly $n$ for each number $n\in\nat$. Furthermore, the notation $\Sigma^{\leq n}$ (resp., $\Sigma^{\geq n}$) expresses the union $\bigcup_{k\in[0,n]_{\integer}} \Sigma^k$ (resp., $\bigcup_{k\in\nat\wedge k\geq n} \Sigma^k$).
Given any string $x = x_1x_2\cdots x_{n-1}x_n$ with $x_1,\ldots,x_n\in\Sigma$, the notation $x^{R}$ denotes the {\em reverse} of $x$; that is, $x^R = x_nx_{n-1}\cdots x_2x_1$.
A {\em language} over $\Sigma$ is a subset of $\Sigma^*$.
Given a language $S$ over $\Sigma$ and a number $n\in\nat$, the notation $dense(S)(n)$ expresses the cardinality of the set $S\cap\Sigma^n$; that is, $dense(S)(n) = |S\cap\Sigma^n|$.
The notation $\chi_{S}$ denotes the {\em characteristic function} of $S$; namely, $\chi_{S}(x)=1$ if $x\in S$ and $\chi_{S}(x)=0$ otherwise.
For any pair of symbols $\sigma\in\Sigma_1$ and $\tau\in\Sigma_2$ over alphabets $\Sigma_1$ and $\Sigma_2$, the {\em track notation}  $\track{\sigma}{\tau}$ denotes a new symbol made from $\sigma$ and $\tau$. Given two strings $x=x_1x_2\cdots x_n$ and $y=y_1y_2\cdots y_n$ of the same length $n$, the notation $\track{x}{y}$ is shorthand for the concatenation  $\track{x_1}{y_1}\track{x_2}{y_2} \cdots\track{x_n}{y_n}$. See \cite{TYL10} for further details.

Given two languages $A$ and $B$ over $\Sigma$ and a string $a\in\Sigma^*$, the notation $aB$ (resp., $Ba$) expresses the set $\{ax\mid x\in B\}$ (resp., $\{xa\mid x\in B\}$) and the {\em concatenation} $AB$ of $A$ and $B$ is the set $\{xy\mid x\in A, y\in B\}$.  Given two binary strings $x$ and $y$ of the same length $n$, $x\oplus y$ denotes the {\em bitwise exclusive-or} of
$x$ and $y$.
For any string $x$ of length $n$, let $pref_{i}(x)$ denote the string consisting of the first $i$ symbols of $x$ and similarly let $suf_{j}(x)$ be the string made up from the last $j$ symbols of $x$. Moreover, we denote by $midd_{i,j}(x)$ the string obtained from $x$ by deleting the first $i$ symbols as well as the last $n-j$ symbols. Note that$x$ equals $pref_{i}(x) midd_{i,j}(x) suf_{j}(x)$ for any $i,j\in[0,n]_{\integer}$
with $i\leq j$.

Let $\reg$ and $\cfl$ denote respectively the family of {\em  regular languages} and the family of {\em context-free languages}.
It is well known that regular languages and context-free languages are characterized by {\em one-way one-head deterministic finite automata} (or dfa's, in short)  and {\em one-way one-head nondeterministic pushdown automata} (or npda's), respectively.
In a machine model with one-way head moves, for simplicity, we demand that each input string provided on an input tape is initially surrounded by two endmarkers, $\cent$ (left-endmarker) and $\dollar$ (right-endmarker), a tape head is initially located at the left-endmarker, and a machine halts just after the tape head scans the right-endmarker. Moreover, we allow the machine's tape head to {\em stay stationary}; however, we demand that all computation (both accepting and rejecting) paths of the machine on every input should terminate in $O(n)$ steps, where $n$ is the input length
(refer to Section \ref{sec:features} for reasoning).  A {\em finite conjunctive closure of $\cfl$} is a natural extension of $\cfl$. Languages, each of which is expressed as the intersection of two context-free languages, form a language family $\cfl(2)$.  It is well known that $\cfl\subsetneq \cfl(2)$ since $\cfl(2)$ contains non-regular languages, such as $L_{3eq} = \{a^nb^nc^n\mid n\in\nat\}$ (see, \eg \cite{HMU01}). The language family $\mathrm{L}$ is composed of any language that is recognized by an appropriate two-way deterministic off-line Turing machine equipped with a read-only input tape and a read/write work tape using only logarithmic space on the work tape.

Here, we wish to give a machine-independent definition of advised language families.
An {\em advice function} is a map $h$ from $\nat$ to $\Gamma^*$, where $\Gamma$ is an appropriate alphabet (called an {\em advice alphabet}).
Generally speaking, based on a given language family $\CC$, an advised class $\CC/n$ expresses a collection of all languages $L$, each of which over alphabet $\Sigma$ requires the existence of another alphabet $\Gamma$, an advice function $h$ from $\nat$ to $\Gamma^*$, and a language $S\in\CC$ over the {\em induced alphabet}  $\Sigma_{\Gamma} = \{\track{\sigma}{\tau}\mid \sigma\in\Sigma,\tau\in\Gamma\}$ satisfying that, for every length $n\in\nat$,  (1) $|h(n)|=n$ and (2)  for every string $x\in\Sigma^n$, $x\in L$ iff $\track{x}{h(n)}\in S$.
For our convenience, an advice function $h$ is called {\em length-preserving} if $|h(n)|=n$ holds for all numbers $n\in\nat$.
By setting $\CC=\reg$ and $\CC=\cfl$, two important advised language families $\reg/n$ \cite{TYL10,Yam08} and $\cfl/n$ \cite{Yam08,Yam10} are obtained.  Likewise, by choosing $\CC=\cfl(2)$, we obtain another important advised language family $\cfl(2)/n$ \cite[Section 7]{Yam10}, which is also  characterized by Claim \ref{CFL(2)-equivalent-def} in a slightly different way.

Since the main theme of this paper is pivoted around $\cfl/n$, we assume that the reader is familiar with fundamental definitions and properties of  npda's (refer to, \eg \cite{HMU01}). Later in Section \ref{sec:swapping}, we shall place more restrictions on the behaviors of npda's to make our argument simpler. Besides finite automata, we shall use a restricted model of {\em one-tape one-head two-way off-line deterministic Turing machine}, which is used to accept/reject an input string or to produce an output string on this single tape whenever the machine halts with an accepting state.
Let $\oneflin$ (whose prefix ``$1\mbox{-}$'' emphasizes a ``one-tape'' model) denote the set of all single-valued total functions computable by those one-tape Turing machines running in time $O(n)$ \cite{TYL10}.

In the case of a one-way machine having an unique output tape, we say that such an output tape is {\em write only} if  (1) initially, all the tape cells are blank,  (2) its tape head can write symbols (from a fixed output alphabet), (3) the tape head can stay on a blank cell until it starts writing a non-blank symbol, and (4) whenever the tape head writes down a non-blank symbol, it should step forward to the  next cell.  In other words, the tape head is  allowed neither to go back nor to read any already-written non-blank symbol on the output tape.

Analogously to the nondeterministic polynomial-time function classes $\mathrm{NPMV}$, $\mathrm{NPSV}$, and $\mathrm{NPSV_t}$  \cite{BLS84,Sel94} studied for decades in computational complexity theory,
three function classes $\cflmv$, $\cflsv$, and $\cflsvt$ were introduced in \cite{Yam11}, where ``MV,'' ``SV,'' and $\mathrm{SV_t}$'' respectively stand for ``multi-valued,'' ``single-valued,'' and ``single-valued and total.''
To define those classes, we need to consider a special npda\footnote{An automaton that can produce outputs is sometimes called a transducer.}  $M$ that is equipped with a single write-only output tape, running in linear time (refer to \cite{Yam14a,Yam14b} for more details).  Such an npda $M$ generally produces numerous output strings along different computation paths.
For convenience, we say that an output string $x$ written on the output tape in a particular computation path is {\em valid} if the path is an accepting computation path; otherwise, $x$ is {\em invalid}.
The notation $\cflmv$ denotes the set of all {\em multi-valued partial functions} $f$, each of which satisfies the following condition: there are alphabets $\Sigma$ and $\Gamma$ for which $f$ maps from $\Sigma^*$ to $\Gamma^*$ and there exists an npda $M$ equipped with a write-only output tape
such that, for every input $x\in\Sigma^*$, $f(x)$ is a  set of all valid output strings produced by $M$. Whenever $f(x)$ is empty, we always treat $f(x)$ as being {\em undefined}, and thus $f$ becomes a ``partial'' function. Next, $\cflsv$ is the set composed of all functions  $f$ in $\cflmv$ such that  $f$ is a {\em single-valued} function (\ie $f(x)$ is always a singleton). Finally, $\cflsvt$ is composed of all {\em total} functions $f$ (\ie $f(x)$ is defined for all inputs $x$). In the case that   $f(x)$ is a singleton, say, $\{y\}$, we conventionally write $f(x)=y$ instead of $f(x)=\{y\}$.

It follows from the above definitions that $\cflsvt\subseteq \cflsv\subseteq \cflmv$. Concerning those function classes, as the next lemma suggests, they can be viewed as a functional extension of $\cfl\cap\co\cfl$, rather than $\cfl$.\footnote{Lemma \ref{CFL-cap-coCFL-CFLSVt} was first stated in \cite[Section 2]{Yam11} without any proof, but the statement therein erroneously cited ``$\cfl\cap\co\cfl$'' in this lemma as ``$\cfl$.''}

\begin{lemma}\label{CFL-cap-coCFL-CFLSVt}
Let $A$ be an arbitrary language. It holds that $A\in\cfl\cap\co\cfl$ if and only if $\chi_{A}\in\cflsv$. Moreover, $\cflsv$ can be replaced by $\cflsvt$ and $\cflmv$.
\end{lemma}

\begin{proof}
Let $\Sigma$ be any alphabet and let $A$ be any language over $\Sigma$. Since $\chi_{A}$ is a single-valued total function, the second part of the lemma immediately follows from the first part.

\s

(Only If--part) Assume that $A$ is in $\cfl\cap\co\cfl$ and take two npda's  $M_0$ and $M_1$ that respectively recognize $\overline{A}$ and $A$.
We define a new npda $M$, equipped with a write-only output tape, as follows. On input $x$, $M$ first guesses (\ie nondeterministically chooses) a bit $b$, writes down $b$ on its output tape, and then simulates $M_b$ on $x$.
If $M_b$ halts in an accepting state, then $M$ enters its own accepting state; otherwise, $M$ enters a rejecting state. It is easy to verify that
$M(x)$ always produces a single valid output bit, which matches  the value $\chi_{A}(x)$. Hence, $\chi_{A}$ is in $\cflsvt\subseteq \cflsv$.

\s

(If--part) Assume that $\chi_{A}\in\cflsv$. There exists an npda $M$, equipped with a write-only output tape, computing the single-valued total function $\chi_{A}$. Since $M$ eventually produces a single valid bit on the output tape, we can modify this $M$ so that, instead of writing down the  output bit $1$ in a certain accepting state on each input instance, it ``accepts'' the input, and it ``rejects'' the input in any other case.  The npda, say, $M_1$ obtained by this modification requires no output tape and it obviously recognizes $A$ since $\chi_{A}$ is single-valued and total.
Thus, $A$ belongs to $\cfl$. Likewise, we can define another npda $M_2$ from $M$ by flipping the role of $b$ in the above definition of $M_1$.
This new npda $M_2$ obviously recognizes $\overline{A}$, and thus $A$ is in $\co\cfl$. Therefore, $A$ belongs to $\cfl\cap\co\cfl$.
\end{proof}

To compute a given multi-valued partial function $f$, we may provide its  underlying npda  $M$ with a piece $h(n)$ of {\em (deterministic) advice} together with any length-$n$ input instance $x$ in the form of $\track{x}{h(n)}$; that is, for any string $y$, $y$ is in $f(x)$ if and only if  $M(\track{x}{h(|x|)})$ outputs $y$ along a certain accepting computation path. All functions $f$ computed by npda's $M$ with the help of such advice functions $h$ form a function class, dubbed as $\cflmv/n$. In a more general fashion, given any function class $\FF$, a multi-valued partial function $f$ is in $\FF/n$ if and only if there exist a multi-valued partial function $g\in\FF$ and a length-preserving advice function $h$ satisfying $f(x) = g(\track{x}{h(|x|)})$ for all $x$.
Two other advised classes  $\cflsv/n$ and $\cflsvt/n$ are naturally introduced by setting $\FF=\cflsv$ and $\FF=\cflsvt$, respectively.

In comparison with Lemma \ref{CFL-cap-coCFL-CFLSVt}, the following lemma exemplifies a clear difference between $\cflsv$ and $\cflmv$ in the presence of advice. Note that, since $\co(\cfl/n)$ coincides with $(\co\cfl)/n$, we simply express this language family as $\co\cfl/n$.

\begin{lemma}\label{CFL-cap-coCFL-CFLMV/n}
For any language $A$, it holds that $A\in\cfl/n\cap\co\cfl/n$ if and only if $\chi_{A}\in\cflmv/n$.
\end{lemma}

\begin{proof}
Let $A$ be any language over alphabet $\Sigma$.

\s

(Only If--part) Assume that $A\in\cfl/n\cap\co\cfl/n$. Since $A$ is in $\cfl/n$, there are an npda $M_1$ and a length-preserving advice function $h_1$ for which $A=\{x\in\Sigma^* \mid M_1\text{ accepts }\track{x}{h_1(|x|)}\}$.
Similarly, we can take $M_0$ and $h_0$ for $\overline{A}$ because of  $\overline{A}\in\cfl/n$. A new advice function $g$ is set to satisfy  $g(n)=\track{h_0(n)}{h_1(n)}$ for every length $n\in\nat$. Furthermore, we shall prepare a new npda $N$ with a write-only output tape that behaves as follows. On input $\track{x}{u}$ with $u=\track{z_0}{z_1}$ and $|x|=|z_0|=|z_1|$, $N$ guesses
a bit $b$, writes $b$ on the output tape, and then simulates $M_b$ on the input $\track{x}{z_b}$. Whenever $M_b$ enters either an accepting state or a rejecting state,  $N$ also enters the same type of inner state. It is obvious that $M(\track{x}{g(|x|)})$ produces $\chi_{A}(x)$ on its output tape. Unfortunately, this npda $M$ may have no valid output or have multiple valid outputs when $u$ is  different from $g(|x|)$. As a consequence, $\chi_{A}$ must belong to $\cflsv/n\subseteq \cflmv/n$.

\s

(If--part) Assuming that $\chi_{A}\in\cflmv/n$, we take an npda $M$ with a write-only output tape and a length-preserving advice function $h$ such that, for every string $x\in\Sigma^*$ and every bit $y\in\{0,1\}$, $\chi_{A}(x) =y$ if and only if $M(\track{x}{h(|x|)})$ produces $y$ on its output tape in an accepting state.  Let us define another npda $N$ with no output tape as follows. On input $\track{x}{u}$ with $|x|=|u|$, $N$ simulates $M(\track{x}{u})$ using  its ``imaginary'' output tape.
Note that, when $u=h(|w|)$,  $M$ writes down only a single symbol (either $0$ or $1$) along accepting computation paths by the time $M$ halts. In this case, $N$ can remember this output in the form of inner state.
To handle any other string $u$, we additionally demand that $N$ should reject immediately whenever $M$ starts writing more than one bit on the imaginary output tape. When $M$ enters an accepting state with a valid outcome of $1$,  $N$ enters an appropriate accepting state and halts. In any other case, $N$ rejects the input. Since $A=\{x\mid \chi_{A}(x)=1\}$, $N$ accepts $\track{x}{h(|x|)}$ if and only if $x\in A$. This implies that $A\in\cfl/n$. In a similar way, we can show that $\overline{A}\in\cfl/n$ by exchanging the roles of accepting states and of rejecting states of $N$. Overall, we conclude that $A$ belongs to $\cfl/n\cap\co\cfl/n$.
\end{proof}

Finally, the notation $\mathrm{FL}$ denotes the collection of all single-valued total functions, each of which can be computed by a certain three-tape deterministic Turing machine $M$, which is equipped with a read-only input tape, a read/write work tape, and a write-only output tape, where two tape heads on the input and work tapes can move in two directions,  using only logarithmic space on the work tape and polynomial space on the output tape, where the last space bound is needed to prevent the function from producing exceptionally long strings.

%%%%%%%%%%%%%%%%%%%%%%%%%%%%%%%
%%%%%%%%%%%%%%%%%%%%%%%%%%%%%%%
\section{Pseudorandom Generators and Pseudorandom Languages}

To state our first main theorem explicitly as Theorem \ref{generator-CFL},
we shall formally introduce the notion of {\em pseudorandom generator} whose adversaries
are particularly languages  (which are essentially equivalent to $\{0,1\}$-valued functions). Of those languages, we are particularly interested in {\em advised context-free languages} (\ie context-free languages supplemented with advice).  Pseudorandom generators that are limited to be {\em almost 1-1} have a close relationship to pseudorandom languages \cite{Yam11}.  This fact will be used to prove the pseudorandomness of a specially designed generator, later called $G$.

%%%%%
%%%%%
\subsection{Pseudorandom Generators}\label{sec:random-generator}

A {\em generator} is, in general, a single-valued total function mapping from $\Sigma^*$ to $\Sigma^*$ for an alphabet $\Sigma$. Given a (single-valued total) function $s:\nat\rightarrow\nat$, a generator $G$ from  $\Sigma^*$ to $\Sigma^*$ is said to have {\em stretch factor} $s(n)$ if $|G(x)|=s(|x|)$ holds for any string $x\in\Sigma^*$. Informally, we also say that $G$ {\em stretches $n$-symbol strings (or seeds) to $s(n)$-symbol strings}, where $n$  refers to its input size. We use the notation $\prob_{x\in\Sigma^n}[\PP(x)]$ to denote the probability, over a random variable $x$ distributed uniformly over $\Sigma^n$, that the property $\PP(x)$ holds. When the probability space $\Sigma^n$ is clear from the context, we omit the script ``$\Sigma^n$'' altogether throughout the sections.

\begin{definition}
Let $\Sigma$ be our arbitrary alphabet. A generator  $G:\Sigma^*\rightarrow\Sigma^*$ with stretch factor $s(n)$ is said to {\em fool} language $A$ over $\Sigma$ if the function
\[
\ell(n) =_{def} \left| \prob_{x}[\chi_{A}(G(x))=1] - \prob_{y}[\chi_{A}(y)=1]  \right|
\]
is negligible, where $x$ and $y$ are random variables over $\Sigma^n$ and $\Sigma^{s(n)}$, respectively.
A function $G$ is
called a {\em pseudorandom generator} against language family $\CC$ if $G$ fools every language $A$ over the alphabet $\Sigma$ in $\CC$.
\end{definition}

In this paper, we shall be particularly focused on generators whose stretch factor is $s(n)=n+1$.  The existence of almost 1-1 pseudorandom generators against $\reg/n$ was extensively discussed in \cite{Yam11}, where a generator $G$ is
called {\em almost one-to-one} (or almost 1-1) if there exists a negligible function $\varepsilon:\nat\rightarrow\real^{\geq0}$  satisfying the equality  $|\{G(x)\mid x\in\Sigma^n\}|=|\Sigma^{n}|(1-\varepsilon(n))$ for
all numbers $n\in\nat^{+}$.
Notably, it is known that certain almost 1-1 pseudorandom generator against $\reg/n$ with stretch factor $n+1$  are found even in the function class $\cflsvt$; however, no function in $\oneflin$ can become a similar kind of pseudorandom generator against $\reg$  \cite{Yam11}.
The existence of an  efficient  pseudorandom generator against $\cfl/n$ has been listed in \cite[Section 7]{Yam11} as an open problem. Our first main theorem naturally extends the above-mentioned results of \cite{Yam11} and answers this particular open problem affirmatively.

To describe our answer, we need to introduce a new function class, called $\cflmvtwo$, which naturally extends $\cflmv$.
A multi-valued partial function $f$ is in $\cflmvtwo$ if there are two multi-valued partial functions $g_1,g_2\in\cflmv$ for which $f$ satisfies the equality $f(x) = g_1(x)\cap g_2(x)$ for every input $x$.
An advised version of $\cflmvtwo$, denoted by $\cflmvtwo/n$, is composed of all multi-valued partial functions $f$, each of which meets the following criterion: there exist a function $g\in\cflmvtwo$ and a length-preserving advice function $h$ satisfying $f(x) = g(\track{x}{h(|x|)})$ for all  inputs $x$. Obviously,  it holds that $\cflmv\subseteq \cflmvtwo$ and $\cflmv/n\subseteq \cflmvtwo/n$.

Here, let us assert that an almost 1-1 pseudorandom generator against $\cfl/n$ actually exists in the intersection of both $\mathrm{FL}$ and $\cflmvtwo/n$.

\begin{theorem}\label{generator-CFL}
There exists an almost 1-1 pseudorandom generator in $\mathrm{FL}\cap \cflmvtwo/n$ against $\cfl/n$ with stretch
factor $n+1$.
\end{theorem}

Hereafter, we shall prove Theorem \ref{generator-CFL} by constructing the desired  pseudorandom generator, say, $G$ against $\cfl/n$. We fix $\Sigma=\{0,1\}$. Our construction of $G:\Sigma^*\to\Sigma^*$ is essentially based on a special language called $IP_{3}$\footnote{In \cite{Yam11}, a language called $IP_{*}$ was introduced and proven to be a pseudorandom language against $\reg/n$. To distinguish our language from it, we intentionally use the current notation $IP_3$. The subscript ``$3$'' in ``$IP_3$'' emphasizes the fact that each element in $IP_3$ is made of essentially three segments $x$, $y$, and $z$.} over $\Sigma$, which possesses a certain type of pseudorandomness. Let us begin with a formal description of $IP_{3}$, in which we intend to calculate the (binary) inner product. Here, the {\em (binary) inner product} $x\odot y$ between two binary strings $x=x_1x_2\cdots x_n$ and $y=y_1y_2\cdots y_n$ of length $n$ is defined as $x\odot y =\sum_{i=1}^{n}x_iy_i$. With this conventional notation, $IP_{3}$ is formally described as
\[
IP_{3} = \{axyz\mid a\in\Sigma^{\leq3},x,y,z\in\Sigma^+,|x|=|z|,|y|=2|x|,(xz)\odot y^R = 1\;(\mathrm{mod}\;2)\}.
\]
Note that, in the above definition of $IP_{3}$, we use the term
``$(xz)\odot y^R$'' instead of a much simpler form ``$(xz)\odot y$'' because, otherwise, it cannot be computed in $\cflmvtwo/n$ (cf. Proposition \ref{IP-dtime}) because of a limitation of stack operations.

%%%%%%%%%%

In what follows, we shall construct our pseudorandom generator $G$ with stretch factor $n+1$.
A well-known method (cf. \cite{Gol01}) to obtain such a generator $G$ is to define it as $G(w)=w\cdot \chi_{IP_3}(w)$ (concatenation) for all $w\in\Sigma^*$. Obviously, $G$ is a one-to-one function. Furthermore, in a similar fashion to Lemma \ref{IP_{3}-implies-IP_3}, it is possible to prove that its range $rang(G) = \{G(w)\mid w\in\Sigma^*\}$ is $\cfl/n$-pseudorandom if so is $IP_3$. By Lemma \ref{generator-pseudorandom} and Proposition \ref{IP-pseudorandom}, we thus conclude that $G$ is indeed a pseudorandom generator against $\cfl/n$. Although $G$ can be computed deterministically in logarithmic space, unfortunately, we are unable to prove that $G$ belongs to $\cflmvtwo/n$.

For our purpose of proving Theorem \ref{generator-CFL}, we need to seek a different type of generator $G$.
{}From $IP_3$, we first consider another useful language $IP^{+} =  \Sigma^{\leq 8}1 \cup (IP_{3}\cap\Sigma^{\geq 8})\Sigma^2$; in particular, $IP^{+}\cap \Sigma^{\geq10} = \bigcup_{e\in\Sigma^2} (IP_{3}\cap\Sigma^{\geq 8})e$. An intimate relationship between $IP_{3}$ and $IP^{+}$ in terms of pseudorandomness will be given later in Lemma \ref{IP_{3}-implies-IP_3}.
Our generator $G$ will be defined so that  $rang(G)$ coincides with $IP^{+}$.  Intuitively, we want to make four bits of each output string of $G$ quite difficult for npda's to calculate. Let $w$ be any input instance to $G$. When $|w|\leq 8$, we simply set $G(w)=w1$. Hereafter, we assume that $|w|\geq9$.
The input $w$ can be seen as a string of the form $w=axbyze$ altogether with $a\in\Sigma^{\leq 3}$, $b\in\Sigma$, $e\in\Sigma^2$, $|x|=|z|+1$, and $|by|=2|x|$. Note that $|xz|=|y|=2|x|-1$ holds.
For ease of the following description of $G$, let $y=y_1y_2$ with $|y_1|=|x|-1$ and $|y_2|=|x|$ and let $\hat{e} = e\oplus (dd')$, where $d=x\odot y_2^R\;(\mathrm{mod}\;2)$ and $d' = z\odot y_1^R\;(\mathrm{mod}\;2)$.  First, let us consider the simplest case where $a=\lambda$ and $|x|=n\geq2$.
The notation $\tilde{x}_{[i]}$ (resp., $\tilde{z}_{[i]}$) expresses a string obtained from $x$ (resp., $z$) by flipping its $i$th bit; namely, $\tilde{x}_{[i]} = x_1x_2\cdots x_{i-1}\overline{x_{i}}x_{i+1}\cdots x_n$ if $x=x_1x_2\cdots x_{i-1}x_{i}x_{i+1}\cdots x_n$. (resp., $\tilde{z}_{[i]} =  z_1z_2\cdots z_{i-1}\overline{z_{i}}z_{i+1}\cdots z_{n-1}$ if $z=z_1z_2\cdots x_{i-1} z_{i}x_{i+1}\cdots z_{n-1}$), where $\overline{0}=1$ and $\overline{1}=0$. The output string $G(w)$ is defined in the following way.

\begin{itemize}\vs{-1}
  \setlength{\topsep}{-2mm}%
  \setlength{\itemsep}{0mm}% original = 1mm
  \setlength{\parskip}{0cm}%

\item[1.] If $w=xbyze$ and $(xz)\odot y^R = 1\;(\mathrm{mod}\;2)$, then let $G(w)= xbyz\overline{b}\hat{e}$.

\item[2.] If $w=x1yze$ and $(xz)\odot y^R = 0\;(\mathrm{mod}\;2)$, then let $G(w)= x1yz1\hat{e}$.

\item[3.] If $w=x0yze$ and $(xz)\odot y^R = 0\;(\mathrm{mod}\;2)$, then let $i$ be the minimal index satisfying $y_i=1$ (if any), where $y=y_1y_2\cdots y_{2n-1}$.

\begin{itemize}\vs{-1}
  \setlength{\topsep}{-2mm}%
  \setlength{\itemsep}{0mm}% original = 1mm
  \setlength{\parskip}{0cm}%

\item[3a.] If such an $i$ exists in $[1,n-1]_{\integer}$, then let $G(w) = x0y\tilde{z}0\hat{e}$, where $\tilde{z} = \tilde{z}_{[n-i]}$.

\item[3b.] If such an $i$ exists in $[n,2n-1]_{\integer}$, then let $G(w) = \tilde{x}0yz0\hat{e}$, where $\tilde{x} = \tilde{x}_{[2n-i]}$.

\item[3c.] If no such $i$ exists, then let $G(w) = x1yz1\hat{e}$.
\end{itemize}
\end{itemize}\vs{-1}
Notice that $|G(w)|=4n+2$ holds since $G$ always outputs strings of length $|w|+1$. Moreover, when $a\neq\lambda$ and $|x|=n\geq2$, we additionally  define $G(axbyze) = a G(xbyze)$. Clearly, $G$ has stretch factor $n+1$.

Next, let us prove the following two fundamental properties.  For any nonempty string $x$ and any number $i$ with $1\leq i\leq |x|$, the notation $x^{(j)}$ denotes the string obtained from $x$ by removing its $j$th bit.

%%%%
%%%%

\begin{claim}\label{property-of-G}
\begin{enumerate}
\item $G$ is an almost 1-1 function.
\vs{-2}
\item $rang(G)=IP^{+}$.
\end{enumerate}
\end{claim}

\begin{proof}
Note that $rang(G)\cap\Sigma^{\leq9} = \{G(w)\mid w\in \Sigma^{\leq 8}\} = \Sigma^{\leq 8}1 = IP^{+}\cap \Sigma^{\leq 9}$. These equalities allow us to concentrate on proving the following two assertions: (1) $G$ is almost 1-1 on the domain $\Sigma^{\geq 10}$ and (2) $rang(G)\cap \Sigma^{\geq10} = IP^{+}\cap \Sigma^{\geq10}$.
For readability, we shall prove them only for the basic case of $a=\lambda$ because the other case $a\neq\lambda$ follows immediately from this basic case.

Fix a number $n\geq 2$ arbitrarily. Let $x\in\Sigma^n$, $b\in\Sigma$, $y\in\Sigma^{2n-1},$ $z\in\Sigma^{n-1}$, and $e\in\Sigma^2$ be arbitrary strings and set $w=xbyze$. Notice that $|w|=4n+1$. Moreover, partition  $y$ into $\hat{y}_1\hat{y}_2$ satisfying both $|\hat{y}_1|=n-1$ and $|\hat{y}_2|=n$ and define $d= x\odot \hat{y}_2^R\;(\mathrm{mod}\;2)$ and $d'=z\odot \hat{y}_1^R\;(\mathrm{mod}\;2)$.

\s

(1) By inspecting the aforementioned definition of $G$, in all cases except for Case 3c of the definition, we can show that $G$ is one-to-one on its domain. Given each pair  $(x,z)\in\Sigma^{n}\times\Sigma^{n-1}$, $G$ maps $\{x0^{2n}ze,x10^{2n-1}ze\}$ to $x10^{2n-1}z1\hat{e}$, making itself two-to-one on this particular domain.
Since there are exactly $2^{2n-1}$ such pairs $(x,y)$, we conclude that $|\{G(w')\mid w'\in\Sigma^{4n+1}\}|\geq 2^{4n+2}-2^{2n-1} = 2^{4n+2}(1-\varepsilon(n))$, where $\varepsilon(n)=1/2^{2n+3}$. Since $\varepsilon(n)$ is a negligible function, $G$ is indeed almost 1-1.

\s

(2) Henceforth, we want to show two inclusions, $rang(G)\subseteq IP^{+}$ and $IP^{+} \subseteq rang(G)$, separately.

($rang(G)\subseteq IP^{+}$) Let $u\in rang(G)\cap\Sigma^{4n+2}$ and assume that $G(w)=u$ for a certain string $w\in\Sigma^{4n+1}$.
When Case 2 of the definition of $G$ occurs, it must follow that $w=x1yze$,   $(xz)\odot y^R = 0\;(\mathrm{mod}\;2)$, and $u=x1yz1e'$ for a certain string $e'$ satisfying $e=e'\oplus (dd')$.  For convenience, we set $z'=z1$ and $y'=1y$; thus, $u=xy'z'e'$ holds. Since
\[
(xz')\odot (y')^R =  (xz1) \odot (y^R1) = (xz)\odot y^R  + 1\odot 1
= 0 + 1 =  1\;(\mathrm{mod}\;2),
\]
the string $xy'z'e'$ must be in $IP^{+}$, in other words,
$u\in IP^{+}$.

Next, consider Case 3a. Assume that $w = x0yze$, $(xz)\odot y^R =  0\;(\mathrm{mod}\;2)$, and $u=x0y\tilde{z}0e'$ for an appropriate $e'$ with   $e = e'\oplus(dd')$. Let $i$ be the minimal index in $[n-1]$ such that  $y_i$ (\ie the $i$th bit of $y$) equals $1$.
For this index $i$, we obtain  $\tilde{z}^{(n-i)} = z^{(n-i)}$ by the definition of $\tilde{z}$. If we set $y' = 0y$ and $z'=\tilde{z}0$, it follows that
\begin{eqnarray*}
(xz')\odot (y')^R &=&  (xz^{(n-i)})\odot (y^{(i)})^R  + \overline{z_{n-i}}\odot y_i + 0\odot 0
\;\;=\;\;  (xz^{(n-i)})\odot (y^{(i)})^R + z_{n-i}\odot y_i + 1 \\
&=&   (xz)\odot y^R +1 \;\;=\;\;  0+1 \;\;=\;\; 1 \;(\mathrm{mod}\;2)
\end{eqnarray*}
since $\overline{z_{2n-i}}\odot y_i = z_{2n-i}\odot y_i +1\;(\mathrm{mod}\;2)$.  Clearly, those equalities imply $u\in IP^{+}$.

Moreover, let us focus on Case 3b. In this case, it holds that
$w=x0yze$, $(xz)\odot y^R = 0\;(\mathrm{mod}\;2)$, and $u= \tilde{x}0yz0e'$, where $e=e'\oplus(dd')$, $\tilde{x} = \tilde{x}_{[2n-i]}$, and $y_i=1$ for the minimal index $i$. Notice that $i$ must exist in $[n,2n-1]_{\integer}$.
Letting  $y'=0y$ and $z'=z0$, we obtain
\begin{eqnarray*}
(\tilde{x}z')\odot (y')^R &=&  (x^{(2n-i)}z)\odot (y^{(i)})^R
+ \overline{x_{2n-i}}\odot y_i  + 0\odot 0
\;\;=\;\;  (x^{(2n-i)}z)\odot (y^{(i)})^R + x_{2n-i}\odot y_i + 1 \\
&=& (xz)\odot y^R + 1 \;\;=\;\; 0 + 1 \;\;=\;\;
1\;(\mathrm{mod}\;2).
\end{eqnarray*}
Thus, $u$ should belong to $IP^{+}$.

The other cases are similarly proven. Therefore, the desired inclusion  $rang(G)\subseteq IP^{+}$ follows.

($IP^{+} \subseteq rang(G)$) Take an arbitrary string $u$ in $IP^{+}\cap\Sigma^{4n+2}$ and assume that $u=xy'z'e'$ and $(xz')\odot (y')^R = 1\;(\mathrm{mod}\;2)$, where $x,z'\in\Sigma^n$, $y'\in\Sigma^{2n}$, and $e'\in\Sigma^2$.
If a certain bit $b$ satisfies both $y'=by$ and $z'=z\overline{b}$, then it  must hold that $(xz)\odot y^R = (xz')\odot (y')^R = 1\;(\mathrm{mod}\;2)$. Using a partition $y= \hat{y}_1\hat{y}_2$, we set $d=x\odot \hat{y}_2^R\;(\mathrm{mod}\;2)$ and $d'=z\odot \hat{y}_1^R\;(\mathrm{mod}\;2)$ and we further define $e = e'\oplus (dd')$, which is equivalent to $e'= e\oplus (dd')$. Since this case corresponds to Case 1 of the definition of $G$, by setting $w=xbyze$, we immediately obtain  $G(w)=u$, indicating that $u\in rang(G)$.

Next, assume that $y'=0y$ and $z'=\tilde{z}0$. Since $y\in\Sigma^{2n-1}$, let  $y = y_1y_2\cdots y_{2n-1}$.  Here, let us consider the case where the minimal index $i$ satisfying $y_i=1$ actually exists in $[2n-1]$. If $i< n$, then let $w=x0yze$, where  $z$ is obtained from $\tilde{z}$ by flipping its $(n-i)$th bit, and $e = e'\oplus (dd')$.  We express $z$ as  $z_1z_2\cdots z_{n-1}$.  Clearly, it holds that
\begin{eqnarray*}
(xz)\odot y^R +1 &=& (xz^{(n-i)})\odot (y^{(i)})^R + z_{n-i}\odot y_i +1
\;\;=\;\;  (xz^{(n-i)})\odot (y^{(i)})^R + \overline{z_{n-i}}\odot y_i \\
&=&  (x\tilde{z})\odot y^R  \;\;=\;\; (xz')\odot (y')^R \;\;=\;\;  1\;(\mathrm{mod}\;2).
\end{eqnarray*}
{}From these equalities, we conclude that $(xz)\odot y^R = 0\;(\mathrm{mod}\;2)$. Since this is exactly Case 3a, it should follow that $G(w)=u$, and thus we obtain the desired membership $u\in rang(G)$.

Since the other cases are similar, we therefore conclude that $IP^{+} \subseteq rang(G)$, as requested.
\end{proof}

%%%%%%
\subsection{Pseudorandom Languages}\label{sec:random-language}

A key idea developed in \cite{Yam11} for a technical construction of pseudorandom generator against $\reg/n$ is the pseudorandomness of  particular languages in $\cfl$. Those languages are generally called {\em pseudorandom languages} \cite{Yam11} and it is shown to have an intimate connection to the existence of pseudorandom generator. We wish to exploit this connection to prove the pseudorandomness of the generator $G$, defined in Section \ref{sec:random-generator}.

To describe the notion of pseudorandom language, we consider an arbitrary  language family $\CC$ containing languages over a certain alphabet $\Sigma$ of cardinality at least $2$.

\begin{definition}{\rm \cite{Yam11}}
Let $\CC$ be any language family and let $\Sigma$ be any alphabet with $|\Sigma|\geq2$. A language $L$ over $\Sigma$ is said to be {\em $\CC$-pseudorandom}  if the function $\ell(n) = \left| \frac{dense(L\triangle A)(n)}{|\Sigma^n|} - \frac{1}{2} \right|$ is negligible for every language $A$ over $\Sigma$ in $\CC$. A language family $\DD$ is called {\em  $\CC$-pseudorandom} if it contains a certain  $\CC$-pseudorandom language.
\end{definition}

The notion of pseudorandomness satisfies a {\em self-exclusion property}, in which a language family $\CC$ cannot be $\CC$-pseudorandom.
For instance, the language family  $\cfl$ is known to be  $\reg/n$-pseudorandom \cite{Yam11} but $\reg$ cannot be $\reg/n$-pseudorandom. There is another logically-equivalent formulation of the $\CC$-pseudorandomness, given in \cite{Yam11} under a term ``pseudorandom version'' of \cite[Lemma 5.1]{Yam11}. For a purpose of later referencing, we shall state this formulation as a lemma.

\begin{lemma}\label{dense-dense-lemma}{\rm \cite{Yam11}}\hs{1}
Let $\Sigma$ be any alphabet with $|\Sigma|\geq2$ and let $\CC$ be any language family. Assume that $\CC$ contains the language $\Sigma^*$. A language $L$ over  $\Sigma$ is $\CC$-pseudorandom if and only if, for every language $A$ over $\Sigma$ in $\CC$, the function $\ell''(n) = \frac{\left| dense(L\cap A)(n) - dense(\overline{L}\cap A)(n) \right|}{\left|\Sigma^n\right|}$
is negligible.
\end{lemma}

Two properties of ``almost one-to-oneness'' and ``restricted stretch factor'' make it possible to connect pseudorandom generators to pseudorandom languages.
In fact, an equivalence between the  $\CC$-pseudorandomness and the existence of a pseudorandom generator against $\CC$ with those two  properties was shown in \cite[Lemma 6.2]{Yam11}.  Since this equivalence is an important ingredient of proving our first main theorem, it is re-stated as a lemma.

\begin{lemma}\label{generator-pseudorandom}{\rm \cite{Yam11}}\hs{1}
Let $\Sigma=\{0,1\}$. Let $\CC$ be any language family containing $\Sigma^*$. Let $G$ be any function from $\Sigma^*$ to $\Sigma^*$
with stretch factor $n+1$. Assume that $G$ is almost 1-1.
The function $G$ is a pseudorandom generator against $\CC$ if and only if the set $rang(G)$ is  $\CC$-pseudorandom.
\end{lemma}

The above lemma clearly says that, as far as a generator $G$ is almost 1-1 stretching $n$-bit seeds to $(n+1)$-bit strings, the pseudorandomness of $G$ can be proven indirectly by establishing the pseudorandomness of the  range of $G$. Now, let us recall the generator $G$ defined in Section \ref{sec:random-generator}.
To prove that this $G$ is actually a pseudorandom generator against $\cfl/n$, it thus suffices for us to show that its range---$IP^{+}$---is $\cfl/n$-pseudorandom. Moreover, as the following lemma shows, we need only the $\cfl/n$-pseudorandomness of the language $IP_{3}$, which is an essential part of $IP^{+}$.

\begin{lemma}\label{IP_{3}-implies-IP_3}
If $IP_{3}$ is $\cfl/n$-pseudorandom, then $IP^{+}$ is also $\cfl/n$-pseudorandom.
\end{lemma}

\begin{proof}
We shall prove the contrapositive of the lemma. Our starting point is the assumption that $IP^{+}$ is not $\cfl/n$-pseudorandom. With this assumption, Lemma \ref{generator-pseudorandom} ensures the existence of  a language $A\in\cfl/n$ over the alphabet $\Sigma=\{0,1\}$, a positive polynomial $p$, and an infinite set $N\subseteq\nat^{+}$ such that, in particular,
\[
\ell''(n+2) =_{def} \frac{\left| dense(IP^{+}\cap A)(n+2) - dense(\overline{IP^{+}}\cap A)(n+2) \right|}{\left|\Sigma^{n+2}\right|}\geq \frac{1}{p(n+2)}
\]
holds for all numbers $n\in N$. Since $N$ is infinite, we can assume without loss of generality that the smallest element in $N$ is more than $7$.
Moreover, choose a constant $c>0$ satisfying $p(n+2)\leq c\cdot p(n)$ for all numbers $n\in N$. We then define another polynomial $q$ as $q(n) = \ceilings{c}p(n)$ for all $n\in\nat$.

In the following argument, we fix a number $n\in N$ arbitrarily. Note that $IP^{+}\cap \Sigma^{n+2} = \bigcup_{s\in\Sigma^2}(IP_{3}\cap\Sigma^n)s$
since $IP^{+}\cap\Sigma^{\geq10} = IP_{3}\Sigma^2\cap\Sigma^{\geq10}$. For each string $s\in\Sigma^2$, we abbreviate as $B_s$ the set $\{x\mid xs\in A\}$.
It then follows that $A\cap\Sigma^{n+2} = \bigcup_{s\in\Sigma^2}(B_s\cap\Sigma^n)s$, and thus $IP^{+}\cap A\cap\Sigma^{n+2} = \bigcup_{s\in\Sigma^2}[(IP_{3}\cap B_s)\cap\Sigma^n]s$. As a consequence, we obtain the equality  $dense(IP^{+}\cap A)(n+2) = \sum_{s\in\Sigma^2}dense(IP_{3}\cap B_s)(n)$. Similarly, it follows that  $dense(\overline{IP^{+}}\cap A)(n+2) = \sum_{s\in\Sigma^2}dense(\overline{IP_{3}}\cap B_s)(n)$. {}From those two equalities together with $\ell''(n+2)\geq 1/p(n+2)$, we conclude
\begin{eqnarray*}
\frac{2^{n+2}}{p(n+2)} &\leq& | dense(IP^{+}\cap A)(n+2) - dense(\overline{IP^{+}}\cap A)(n+2)| \\
&\leq& \sum_{s\in\Sigma^2} | dense(IP_{3}\cap B_s)(n) - dense(\overline{IP_{3}}\cap B_s)(n)| \\
&\leq& 4 \cdot \max_{s\in\Sigma^2}| dense(IP_{3}\cap B_s)(n) - dense(\overline{IP_{3}}\cap B_s)(n)|.
\end{eqnarray*}
The inequality $p(n+2)\leq q(n)$ leads to a lower bound:
\[
\max_{s\in\Sigma^2}| dense(IP_{3}\cap B_s)(n) - dense(\overline{IP_{3}}\cap B_s)(n)| \geq \frac{2^{n}}{p(n+2)} \geq \frac{2^n}{q(n)}.
\]

To eliminate the ``max'' operator in the above inequality, we choose a string $s_n$ ($\in\Sigma^2$) for each length $n\in N$ so that it  satisfies
\[
| dense(IP_{3}\cap B_{s_n})(n) - dense(\overline{IP_{3}}\cap B_{s_n})(n)|   = \max_{s\in\Sigma^2}| dense(IP_{3}\cap B_s)(n) - dense(\overline{IP_{3}}\cap B_s)(n)|.
\]
For any other length $n$ not in $N$, we automatically set $s_n$ to be $0^2$. Using the newly obtained series $\{s_n\}_{n\in \nat}$, we define a new language $B = \{x\mid xs_{|x|}\in A\}$. By the choice of $s_n$ for every length $n\in N$, it follows that
\begin{eqnarray*}
\lefteqn{| dense(IP_{3}\cap B)(n) - dense(\overline{IP_{3}}\cap B)(n)|}\hs{10} \\
&=&
\max_{s\in\Sigma^2}| dense(IP_{3}\cap B_s)(n) - dense(\overline{IP_{3}}\cap B_s)(n)| \;\;\geq\;\; \frac{2^n}{q(n)}.
\end{eqnarray*}
Therefore, the following inequality holds:
\[
\frac{| dense(IP_{3}\cap B)(n) - dense(\overline{IP_{3}}\cap B)(n)| }{|\Sigma^n|} \geq \frac{1}{q(n)}.
\]

Finally, we want to prove that $B$ belongs to $\cfl/n$.  Since $A\in\cfl/n$, $A$ can be expressed as a set $\{x\mid \track{x}{h(|x|)}\in C\}$ for a certain language $C\in\cfl$ and a certain length-preserving advice function $h:\nat\rightarrow\Gamma^*$, where $\Gamma$ is an appropriate advice alphabet.
For every length $n\in \nat$ and every string $x\in\Sigma^n$, it holds that
$x\in B$ $\IFF$ $xs_n\in A$ $\IFF$ $\track{xs_n}{h(n+2)}\in C$.
Next, let us define a new advice function $g$. Here, we prepare a new symbol $\pair{abc}$ to express each length-3 string $abc$. Given each index $n\in\nat$, let $g(n) = h_1h_2\cdots h_{n-1}\track{\pair{s_n\#}}{\pair{h_{n}h_{n+1}h_{n+2}}}$ if
$h(n+2) = h_1h_2\cdots h_{n+2}$, where each $h_i$ is a symbol in $\Gamma$.
Notice that $|g(n)|=n$ holds for all $n\in\nat$.
Furthermore, we introduce another language $D$ as
\[
D = \{ \track{x}{y}\mid \exists v_1,v_2,v_3\in\Gamma\, \exists s\in\Sigma^2\,  \exists z,u \text{ s.t. } |xs|=|z|\, \wedge\, z = uv_1v_2v_3 \, \wedge\, y = u\track{\pair{s\#}}{\pair{v_1v_2v_3}} \, \wedge\,  \track{xs}{z}\in C \}.
\]
Since $\track{xs_{n}}{h(|xs_n|)}\in C$ if and only if $\track{x}{g(|x|)}\in D$,  $B$ is expressed as a set $\{x\mid \track{x}{g(|x|)}\in D\}$.   It is not difficult to show that $D\in\cfl$, and thus $B$ belongs to $\cfl/n$.

In conclusion, $IP_{3}$ cannot be $\cfl/n$-pseudorandom.
\end{proof}

To prove the $\cfl/n$-pseudorandomness of $rang(G)$, Lemma \ref{IP_{3}-implies-IP_3} (with Claim \ref{property-of-G}(2)) helps us set our goal to prove the following proposition regarding $IP_{3}$. However, since the proof of the proposition  is  lengthy, we shall  postpone it until Sections \ref{sec:swapping}--\ref{sec:proof-prop}.

\begin{proposition}\label{IP-pseudorandom}
The language $IP_{3}$ is  $\cfl/n$-pseudorandom.
\end{proposition}

With the help of Proposition \ref{IP-pseudorandom} and Lemmas \ref{generator-pseudorandom} and \ref{IP_{3}-implies-IP_3} as well as Claim \ref{property-of-G}(2), the proof of Theorem \ref{generator-CFL} is now immediate.

\begin{proofof}{Theorem \ref{generator-CFL}}
Recall the generator $G$ introduced in Section \ref{sec:random-generator}. We wish to verify that this generator $G$ is indeed the desired pseudorandom generator of the theorem.
By Claim \ref{property-of-G}(1), $G$ is an almost 1-1 function. To show that $G$ fools every language in $\cfl/n$, we need to prove by Lemma \ref{generator-pseudorandom} that $rang(G)$ is  $\cfl/n$-pseudorandom.
Since $rang(G)=IP^{+}$ by Claim \ref{property-of-G}(2), Lemma \ref{IP_{3}-implies-IP_3} indicates that it is enough to show the $\cfl/n$-pseudorandom property of $IP_3$. This property comes from Proposition \ref{IP-pseudorandom}. Moreover, the efficient computability of $G$ will be given in Proposition \ref{G-complexity}. We therefore conclude that Theorem \ref{generator-CFL} truly holds.
\end{proofof}

%%%
%%%

Since Proposition \ref{IP-pseudorandom} is a key to our first main theorem, it is worth discussing the computational complexity of $IP_{3}$.
In what follows, we shall demonstrate that $IP_{3}$ belongs to both $\mathrm{L}$ and $\cfl(2)/n$. Notice that $\cfl(2)/n$ has been introduced in Section \ref{sec:notions} as the collection of languages $L$ for which there are a language $S\in\cfl(2)$ and a length-preserving advice function $h$ satisfying  $L=\{x\mid \track{x}{h(|x|)}\in S\}$.

\begin{proposition}\label{IP-dtime}
The language $IP_{3}$ belongs to $\mathrm{L}\cap \cfl(2)/n$.
\end{proposition}

\begin{proof}
Firstly, we shall show that $IP_{3}$ belongs to $\mathrm{L}$. To compute  $IP_{3}$, let us consider the following deterministic Turing machine $M$ equipped with a read-only input tape and a read/write work tape. Let $w=axyz$ be any input string, provided that $a\in\Sigma^{\leq3}$,
$|x|=|z|$, $|y|=2|x|$, and $y$ is of the form $y=y_1y_2$ with $|y_1|=|y_2|$.
 When $|w|\leq3$, since $w=a$, we simply force $M$ to accept it. Henceforth, we should consider only the case where $|x|=n\geq 1$.
In the first phase, $M$ calculates the size $|a|$ by scanning the entire input $w$ without using the work tape.  Note that it is possible for $M$ to locate all boundaries among $a$, $x$, $y$, and $z$ of the string $axyz$ using space $O(\log{n})$.
In the second phase, $M$ computes two values $x\odot y_2^R \;(\mathrm{mod}\;2)$ and $z\odot y_1^R \;(\mathrm{mod}\;2)$ by moving the tape head back and forth using $O(\log{n})$ memory bits.
In the last phase, $M$ accepts the input $w$ exactly when the sum of those two values modulo two equals $1$. This case is equivalent to the membership
$axyz\in IP_{3}$.
When we run $M$, it requires only $O(\log{n})$ work space, and therefore $IP_{3}$ is indeed in  $\mathrm{L}$.

Secondly, we wish to prove that $IP_{3}$ belongs to $\cfl(2)/n$. For each index $b\in\{0,1\}$, we introduce two auxiliary sets $A_b$ and $B_b$ as follows.
\begin{itemize}
\item $A_b = \{axyz \mid a\in\Sigma^{\leq3}, y_1,y_2,z\in\Sigma^{|x|},  y=y_1y_2, x\odot y_2^R = b\;(\mathrm{mod}\;2)\}$.
\vs{-2}
\item $B_b = \{axyz \mid a\in\Sigma^{\leq3}, y_1,y_2,z\in\Sigma^{|x|},  y=y_1y_2, z\odot y_1^R = b\;(\mathrm{mod}\;2)\}$.
\end{itemize}
To see that $A_b\in\cfl/n$, we take an advice alphabet $\Gamma=\{0,1,2\}$ and an advice function $h_{A}$ defined as $h_{A}(4n+i) =2^i1^n0^n1^n0^n$, where $i\in[0,3]_{\integer}$.
It is easy to construct an npda that, on any input of the form $\track{axyz}{h_A(4n+|a|)}$ with $|a|\leq 3$, $|x|=|z|=n$, $y=y_1y_2$, and $|y_1|=|y_2|=n$, first locates  two segments $\track{x}{1^n}$ and $\track{y_2}{1^n}$, computes the value $v=x\odot y_2^R \;(\mathrm{mod}\;2)$ using the npda's stack properly, and accepts the input exactly when $v=b$. This implies that $A_b$ falls into $\cfl/n$. Similarly, using another advice function $h_B(4n+i) = 2^i0^n1^n0^n1^n$, we can show that $B_b$ is also in $\cfl/n$.
Next, we define $C$ to be $(A_0\cap B_1)\cup(A_1\cap B_0)$. Since  $(xz)
\odot y^R = x\odot y_2^R + z\odot y_1^R$ holds for any three strings  $x,z\in\Sigma^n$ and $y\in\Sigma^{2n}$ with $y=y_1y_2$ and $|y_1|=|y_2|$, the equality $C=IP_{3}$ follows immediately.

Toward the desired goal, we need to argue that $C$ actually belongs to  $\cfl(2)/n$. For this purpose, let us recall the definition of $\cfl(2)/n$: a language $L$ is in $\cfl(2)/n$ if and only if $L=\{x\mid \track{x}{h(|x|)}\in S\}$ for a certain language $S\in\cfl(2)$ and a certain length-preserving advice function $h$.
Instead of using this original definition, we consider another simple way of defining the same language family $\cfl(2)/n$ in terms of $\cfl/n$.

\begin{claim}\label{CFL(2)-equivalent-def}
For any language $L$, $L$ is in $\cfl(2)/n$ if and only if there are two languages $L_1,L_2\in\cfl/n$ satisfying $L = L_1\cap L_2$.
\end{claim}

\begin{proof}
Let $L$ be any language over alphabet $\Sigma$.

\s

(Only if--part) Assuming that $L\in\cfl(2)/n$, take a language $S$ in $\cfl(2)$ and a length-preserving advice function $h$ satisfying $L=\{x\mid \track{x}{h(|x|)}\in S\}$. Since $S\in\cfl(2)$, there are two context-free languages $S_1$ and $S_2$ for which $S = S_1\cap S_2$.  Now, let us define $L_i = \{x\mid \track{x}{h(|x|)}\in S_i\}$ for each index $i\in\{1,2\}$. The equality  $L = L_1\cap L_2$ thus follows instantly. Obviously,  both $L_1$ and $L_2$ belong to $\cfl/n$, as requested.

\s

(If--part) Assume that $L=L_1\cap L_2$ for two languages $L_1,L_2\in\cfl/n$.
For each index $i\in\{1,2\}$, there exists a language $S_i\in\cfl$ and a length-preserving advice function $h_i$ for which  $L_i$ coincides with the set $\{x\mid \track{x}{h_i(|x|)}\in S_i\}$. Since $L = L_1\cap L_2$, it holds that $L = \{x\mid \track{x}{h_1(|x|)}\in S_1 \,\wedge\, \track{x}{h_2(|x|)}\in S_2\}$. To simplify the description of $L$, we set $h(n) = \track{h_1(n)}{h_2(n)}$ for every length $n\in\nat$ and we define $S'_i$ for each index $i\in\{1,2\}$  as $S'_i = \{\track{x}{y}\mid y = \track{z_1}{z_2} \,\wedge\, \track{x}{z_i}\in S_i\}$, which is clearly context-free since so is $S_i$.   If we set $S'$ to be $S'_1\cap S'_2$, the following equivalence holds: for any string $x$, $x\in L$ if and only if $\track{x}{h(|x|)}\in S'$. Since $S' = S'_1\cap S'_2 \in\cfl(2)$, $L$ must belong to $\cfl(2)/n$ by the original definition of $\cfl(2)/n$.
\end{proof}

Let us return to the proof of Proposition \ref{IP-dtime}. It is easy to verify that
$C$ coincides with $(A_0\cup B_0)\cap (A_1\cup B_1)$. Since $\cfl/n$ is closed under union, two sets $A_0\cup B_0$ and $A_1\cup B_1$ also belong to $\cfl/n$.  Claim \ref{CFL(2)-equivalent-def} therefore guarantees that $C$ belongs to  $\cfl(2)/n$.
\end{proof}

An immediate consequence of Proposition \ref{IP-pseudorandom} together with Proposition \ref{IP-dtime} is the  $\cfl/n$-pseudorandomness of the language family $\mathrm{L}\cap\cfl(2)/n$  (and thus $\cfl(2)/n$ alone).

\begin{theorem}\label{CFL(2)-random-CFL/n}
The language family $\mathrm{L}\cap\cfl(2)/n$ is  $\cfl/n$-pseudorandom.
\end{theorem}

If $\mathrm{L}\cap\cfl(2)/n\subseteq \cfl/n$, then  Theorem  \ref{CFL(2)-random-CFL/n} makes $\cfl/n$ become $\cfl/n$-pseudorandom. Obviously, this is absurd because of the self-exclusion property of the $\cfl/n$-pseudorandomness.  Thus, a class separation  holds between $\cfl/n$ and $\mathrm{L}\cap\cfl(2)/n$; moreover, $\cfl(2)\nsubseteq \cfl/n$ holds (as shown in the following corollary). In comparison, it was proven in \cite{Yam08} that $\co\cfl\nsubseteq\cfl/n$. Our separation result is incompatible with  this one and also extends a classical separation result of $\cfl\neq \cfl(2)$ (see, \eg \cite{HMU01}).

\begin{corollary}\label{CFL(2)-vs-CFL/n}
$\mathrm{L}\cap\cfl(2)/n\nsubseteq \cfl/n$. Thus, $\cfl(2)\nsubseteq\cfl/n$.
\end{corollary}

\begin{proof}
As argued earlier, the first separation follows immediately from Theorem  \ref{CFL(2)-random-CFL/n}. The second separation is shown by contradiction.
Let us assume that $\cfl(2)\subseteq \cfl/n$.
Next, we want to assert the following claim.

\begin{claim}
If $\cfl(2)\subseteq \cfl/n$, then $\cfl(2)/n\subseteq \cfl/n$ (and thus $\cfl(2)/n = \cfl/n$).
\end{claim}

\begin{proof}
Take any language $L$ in $\cfl(2)/n$. There exist a language $S\in\cfl(2)$ and a length-preserving advice function $h$ satisfying $L = \{x \mid \track{x}{h(|x|)}\in S\}$. By the premise of the claim, $S$ belongs to $\cfl/n$. Thus, the set $S$ has the form $\{y\mid \track{y}{g(|y|)}\in R\}$ for a ceratin language $R\in\cfl$ and a certain length-preserving advice function $g$. Let us define $f(n) =\track{h(n)}{g(n)}$ for every $n\in\nat$. Moreover, define a new language $T$ as $T=\{\track{x}{z}\mid |x|=|z|, \exists y,u_1,u_2\,[\, z=\track{u_1}{u_2}\, \wedge\, y=\track{x}{u_1}\, \wedge \, \track{y}{u_2}\in R\, ]\}$. It then follows that, for every string $x$, letting $y=\track{x}{h(|x|)}$, $x\in L$ $\Leftrightarrow$ $\track{x}{h(|x|)}\in S$ $\Leftrightarrow$  $\track{y}{g(|y|)}\in R$ $\Leftrightarrow$  $\track{x}{f(|x|)}\in T$.
In conclusion, $L$ belongs to $\cfl/n$.
\end{proof}

By the above claim, our assumption of $\cfl(2)\subseteq \cfl/n$ leads to a containment $\cfl(2)/n\subseteq \cfl/n$, which obviously contradicts the first separation of the corollary. Therefore, the desired separation $\cfl(2)\nsubseteq \cfl/n$ should hold.
\end{proof}

%%%%%%%
\subsection{Efficient Computability of $G$}\label{sec:efficient-computable}

We have already verified the pseudorandomness of the generator $G$,  introduced in Section \ref{sec:random-generator}; however, the proof of Theorem \ref{generator-CFL} has left unproven the efficient computability of $G$. To complete the proof, we wish to discuss the  complexity of computing $G$; in particular,  we shall demonstrate that $G$ actually belongs to both $\mathrm{FL}$ and $\cflmvtwo/n$.

\begin{proposition}\label{G-complexity}
The generator $G$ defined in Section \ref{sec:random-generator} belongs to  $\mathrm{FL}\cap\cflmvtwo/n$.
\end{proposition}

Compared to the proof of $G\in\mathrm{FL}$, the proof of $G\in\cflmvtwo/n$ is much more involved and it is also quite different from the proof of $IP_{3}\in \cfl(2)/n$ (Proposition \ref{IP-dtime}) because we need to ``produce'' $G$'s output strings using only restricted tools (such as,  one-way head moves and push/pop-operations for a stack) provided by npda's.

\begin{proofof}{Proposition \ref{G-complexity}}
It is not difficult to show that $G$ is in $\mathrm{FL}$ by first computing whether $w\in IP_{3}$ using logarithmic space, as done in the proof of  Proposition \ref{IP-dtime}. Once this is done, we determine which case of the definition of $G$ occurs. Finally, we write down an output string according to the chosen case. Clearly, this procedure requires only
logarithmic space.

Next, we shall show that $G$ belongs to $\cflmvtwo/n$. Note that a functional analogue of Claim \ref{CFL(2)-equivalent-def} holds. We describe this as a claim below; however, for readability, we omit the proof of the claim.

\begin{claim}\label{functional-CFL(2)-equiv}
For any multi-valued partial function $g$, $g$ is in $\cflmvtwo/n$ if and only if there are two functions $f,f'\in\cflmv/n$ satisfying  $g(w)=f(w)\cap f'(w)$ for every input $w$.
\end{claim}

Hereafter, we shall define two multi-valued partial functions $f$ and $f'$ in $\cflmv/n$ and prove in Claim \ref{G-vs-f-and-fprime} that $f(w)\cap f'(w) = \{G(w)\}$ holds for every $w\in\Sigma^{\geq9}$. Claim \ref{functional-CFL(2)-equiv} then implies that $G$ is a member of  $\cflmvtwo/n$, completing the proof of Proposition \ref{G-complexity}.

Let $\Sigma=\{0,1\}$ and let $w$ be any input instance in $\Sigma^*$.
If $w\in\Sigma^{\leq 8}$, then we simply set $f(w)=f'(w)=\{G(w)\}$, which obviously implies that $f(w)\cap f'(w) = \{G(w)\}$. Otherwise,
we decompose $w$ into $axbyze$  with $|a|\leq3$, $|x|=n$, $b\in\Sigma$, $|y|=2n-1$, $|z|=n-1$, and $|e|=2$. Similarly to the proof of Proposition \ref{IP-dtime}, we set our advice function $h:\nat\rightarrow\Gamma^*$ to satisfy  $h(4n+|a|) = 2^{|a|}0^n21^{n-1}0^{n}1^{n-1}2^2$, where $\Gamma=\{0,1,2\}$. Let  $x= x_1x_2\cdots x_{n}$ and $z= z_1z_2\cdots z_{n-1}$ with each $x_i,z_i$ in $\Sigma$, and let $y=y_1y_2$ with $|y_1|=|z|$ and $|y_2|=|x|$. For convenience, we set $d'= x\odot y_2^R\;(\mathrm{mod}\;2)$ and $\hat{e}_{d}  = e\oplus (dd')$ for each value $d\in\Sigma$.

Let us begin with defining $f$ by giving a precise description of its underlying npda $M$ that is equipped with a write-only output tape. Note that ``nondeterminism'' of the npda is effectively used in the following description of $M$.
For ease of the description, we assume that $a=\lambda$. Let $\tilde{w} =\track{w}{u}$ be any input string satisfying $|w|=|u|=4n$. For the time being, we further assume that $u$ matches the correct advice string $h(|w|)$.
On this particular input $\tilde{w}=\track{w}{h(|w|)}$,  $M$ initially guesses a value that expresses $y_1\odot z^R\;(\mathrm{mod}\;2)$. Let $d$ denote such a guessed value. In addition, $M$ guesses which case (among Cases 1--3c) of the definition of $G$ in Section \ref{sec:random-generator} occurs. How $M$ behaves after this initial stage  depends on the case guessed during
this stage.

\s

(1) When $M$ guesses ``Case 1'' at the initial stage, $M$ stores $x$ into its stack, remembers $b$,  and copies $xby_1$ onto its output tape.
Using the advice string $h(4n)$ as boundary markers among strings $x$, $y_1$, $y_2$, $z$, and $e$, while reading $y_2$,  $M$ correctly computes $d'$ and $\hat{e}_{d}$. If $d\oplus  d'$ equals $0$, then $M$ rejects  the input $\tilde{w}$ immediately; otherwise, $M$ continues producing an entire string  $xbyz\overline{b}\hat{e}_{d}$ on the output tape using the knowledge of $b$ and $\hat{e}_{d} $. Finally, after scanning the right endmarker $\dollar$, $M$ enters an appropriate accepting state and halts.

(2) If $M$ guesses ``Case 2,'' then $M$ writes down $x$ on the output tape and also places $x$ in the stack. In the case of $b=0$, $M$ rejects the input $\tilde{w}$. Provided that $b=1$, $M$ computes $d'$ and $\hat{e}_{d} $ while reading $y_2$ from the input tape.
If $d\oplus  d'=1$, then $M$ enters a certain rejecting state. Otherwise, it produces $x1yz1\hat{e}_{d} $ on the output tape and then enters an accepting state.

(3a) Assume that ``Case 3a'' is guessed. This is a special case that requires full attention since $M$ is unable to compute the string $\tilde{z}$ (given in the definition of $G$) correctly.  The npda $M$ first writes $x$ onto the output tape and simultaneously stores $x$ into the stack.
If $b=1$, then $M$ instantly rejects the input. Next, let us assume that $b=0$. While writing $x0y$, $M$ computes $d'$ and $\hat{e}_{d}$ and also checks whether $y_1\neq 0^n$ using the boundary markers given by $h(4n)$. Whenever  $y_1=0^n$ occurs, $M$ rejects $\tilde{w}$. If $d\oplus  d'=1$, then $M$ also rejects $\tilde{w}$. Otherwise, $M$ guesses  an index $i\in[n-1]$ and produces $\tilde{z}_{[i]}0\hat{e}_{d}$ on the output tape after $x0y$.
This last guessing process can be done by nondeterministically choosing a step at which $M$ flips a currently-reading bit, provided that there has been no flipping so far.  When $M$ finally terminates in (various) accepting states, its valid outcomes form a set  $\{x0y\tilde{z}_{[i]}\hat{e}_{d} \mid i\in[n-1]\}$ of $n-1$ different strings.

(3b) When ``Case 3b'' is guessed, $M$ further guesses an index $i\in[n]$, writes down  $\tilde{x}_{[n-i+1]}$ on the output tape, and places the string $x' = x_1\cdots x_{n-i}\track{x_{n-i+1}}{1}x_{n-i+2}\cdots x_n$ into the stack.
When $M$ reads $b=0$, it rejects $\tilde{w}$. When $b=1$, on the contrary,  $M$ writes down $0y_1$. Whenever $y_1\neq 0^{n-1}$, $M$ also rejects the input. Using the stored string $x'$ in the stack, $M$ computes $d'$ and $\hat{e}_{d} $ and  checks  whether the symbol $1$ firstly appears at the $i$th bit (which is marked by a special symbol $\track{x_{n-i+1}}{1}$ stored in the stack) of $y_2$. If this is not the case (\eg $y_1\neq 0^{n-1}$), then $M$ instantly enters a rejecting state.
This process eliminates any computation path that has followed an incorrectly guessed index $i$. Moreover, when $d\oplus d'=1$, $M$ also rejects the input. Unless $M$ has already halted, $M$ writes down $y_2z0\hat{e}_{d} $ on the output tape and accepts the input.

(3c) Finally, consider a situation in which ``Case 3c'' is guessed. If $b=1$, then $M$ rejects $\tilde{w}$; otherwise, $M$ computes $d'$ and $\hat{e}_{d}$ exactly. If $d\oplus d'=1$, then $M$ rejects the input. Assume otherwise. While writing down $x1yz1\hat{e}_{d}$, $M$ checks whether $y=0^{2n-1}$. If this is not the case, then $M$ enters a rejecting state.  Otherwise, $M$ enters an accepting state and halts.

\s

In summary of Steps (1)--(3c), the following claim holds for $M$.

\begin{claim}\label{M-algorithm}
When $w$ is of the form $x1yze$ (which corresponds to Steps  (1)--(2)), $M$ always  produces a set $\{x1yz0\hat{e}_{d}, x1yz1\hat{e}_{d} \mid  d\in\Sigma\}$ of output strings; however, when $w$ is of the form $x0yze$ (which corresponds to Steps (1) and (3a)--(3c)), $M$ produces a set  $P_{w}\cup \{x0yz1\hat{e}_{d} \mid  d\in\Sigma\}$ of output strings,  where $P_{w}$ is defined, mostly depending on the value of $y$, as follows. If $w$ satisfies Case 3a, then $P_w$ is set to be $\{x0y\tilde{z}_{[i]}0\hat{e}_{d} \mid i\in[n-1],d\in\Sigma\}$; if $w$ satisfies Case 3b, then $P_w$ equals $ \{\tilde{x}0yz0\hat{e}_{d} \mid d\in\Sigma\}$; if $w$ is of Case 3c, then $P_w$ is $\{x1yz1\hat{e}_{d} \mid d\in \Sigma\}$.
\end{claim}

In a more general case where $\tilde{w}$ is of the form $\track{w}{u}$ with
an arbitrary string  $u\in\Gamma^{|w|}$ not limited to $h(|w|)$,  we need to modify the above-described npda $M$. While scanning the entire input, $M$ additionally checks if $u$ has the form $2^{n_1}0^{n_2}21^{n_3}0^{n_4}1^{n_5}2^2$ for numbers $n_1,n_2,n_3,n_4n,n_5\in[4n]$ with $n_1+n_2+n_3+n_4+n_5+3=4n+|a|$ and $n_1\leq 3$, using only the npda's inner states.  Let $w=axby_1y_2ze$ with $|a|=n_1$, $b\in\Sigma$, $|x|=n_2$, $|y_1|=n_3$, $|y_2|=n_4$, $|z|=n_5$, and $|e|=2$. Moreover, during the computation of $d'=z\odot y_1^R$ described above, $M$ simultaneously checks whether $|z|=|y_1|$. If $M$ detects any inconsistency at any time, then it immediately rejects the input $\tilde{w}$. It is important to note  that, when $u$ is different from $h(|w|)$, $M$ may possibly produce no valid output strings.

Finally, we define $f$ to be a multi-valued partial function whose output is a set of all valid strings produced by $M$ using the advice function $h$; namely, $s\in f(w)$ if and only if $M(\track{w}{h(|w|)})$ produces $s$ in a certain accepting computation path.

Next, we shall define $f'$. In a manner similar to constructing $M$, we define $M'$ by guessing $d'$ and computing $d = x\odot y_2^R\;(\mathrm{mod}\;2)$  accurately. A statement similar to Claim \ref{M-algorithm} also holds for $M'$.
{}From this npda $M'$, the desired function $f'$ can be defined in a manner similar to $f$ using the same advise function $h$. By the behaviors of $M$ and $M'$, both $f$ and $f'$ belong to $\cflmv/n$.
To complete the proof of the proposition, by Claim \ref{functional-CFL(2)-equiv}, what remains to prove is the following claim that $f(w)\cap f'(w)= \{G(w)\}$ holds for every input $w$.

\begin{claim}\label{G-vs-f-and-fprime}
For every string $w\in\Sigma^*$, it holds that $f(w)\cap f'(w) = \{G(w)\}$.
\end{claim}

\begin{proof}
In what follows, it suffices to deal with an arbitrary input instance $w$ of the form $axbyze$ with $|w|\geq9$. For such an input $w$,  we set  $\tilde{w}=\track{w}{h(|w|)}$ as before. Concentrating on $M$, let us consider all accepting  computation paths $p$ of $M(\tilde{w})$ along which all guesses made by $M$ are correct. Note that there is exactly one such accepting computation path.  By Claim \ref{M-algorithm},
$M$ correctly produces $G(w)$ on its output tape as a valid output in this accepting computation path. Therefore,  the set $f(w)$ must contain the string  $G(w)$; namely, $G(w)\in f(w)$. By considering $M'$, we can similarly
obtain $G(w)\in f'(w)$, implying that $G(w)\in f(w)\cap f'(w)$.

Next, we shall prove that $|f(w)\cap f'(w)|\leq 1$.
Let us consider the case where $w$ is of length at least $9$.
First, let $w$ be in the form $ax1yze$ as before.
For simplicity, however, we shall discuss only the case where $a=\lambda$.
By Claim \ref{M-algorithm}, any output string in $f(w)\cap f'(w)$ should  have the form $x1yzb'\hat{e}$, where $b'\in\Sigma$ and $\hat{e}\in\Sigma^2$.
We shall show that $b'$ and $\hat{e}$ are uniquely determined from $w$.
Assume otherwise; that is, $f(w)\cap f'(w)$ contains two different output strings $x1yzb_1e_1$ and $x1yzb_2e_2$.  {}From each $e_j$ ($j\in\{1,2\}$), we can retrieve a two-bit string $d_jd'_j$ satisfying $e_j = e\oplus(d_jd'_j)$ simply by computing $e\oplus e_j$.
Let us target $M$ first. Since $M$ computes $d'=z\odot y_1^R\;(\mathrm{mod}\;2)$ correctly, it should follow that $d'_1=d'_2=d'$. Similarly, since $M'$ correctly computes $d= x\odot y_2^R\;(\mathrm{mod}\;2)$, we obtain $d_1=d_2=d$.  As a consequence, $e_1=e_2$ follows.
Note that, for each index $j\in\{1,2\}$, the value $b_j$ is determined completely from the value $d_jd'_j$ as follows:  $b_j$ must be $0$ if $d_j\oplus d'_j=0$, and $b_j$ must be $1$ otherwise. Since $d_1d'_1 = d_2d'_2$, we obtain $b_1=b_2$, yielding $|f(w)\cap f'(w)|\leq 1$.

In the case of $w=x0yze$, by Claim \ref{M-algorithm}, any output string in $f(w)\cap f'(w)$ must have one of the following three forms: $x0yz1\hat{e}$, $x1yz1\hat{e}$, and  $x'0yz'0\hat{e}$, where  $x'\in\Sigma^n$, $z'\in\Sigma^{n-1}$, and $\hat{e}\in\Sigma^2$.
Assuming $|f(w)\cap f'(w)|\geq2$, we want to draw a contradiction. In what follows, we shall consider only two typical cases since the remaining cases are similar or trivial.

(i) Let us assume that $f(w)\cap f'(w)$ contains two strings $x'_10yz'_10e_1$ and $x'_20yz'_20e_2$. Since these strings are outcomes of $M$ on $w$,
by Claim \ref{M-algorithm}, $M$ must produce either $\{x0y\tilde{z}_{[i]}0\hat{e}_d\mid i\in[n-1], d\in\Sigma\}$ or $\{\tilde{x}0yz0\hat{e}_d\mid d\in\Sigma\}$, but not both. In either case, $x'_1=x'_2\in\{x,\tilde{x}\}$ must hold. Similarly, $M'$ produces either $\{\tilde{x}_{[i]}0yz0\hat{e}_d\mid i\in[n], d\in\Sigma\}$ or $\{x0y\tilde{z}0\hat{e}_d\mid d\in\Sigma\}$ (but not both) and this fact  leads to  $z'_1=z'_2\in\{z,\tilde{z}\}$. In the case where  $x'_1=x$ and $z'_1=\tilde{z}$, since $dd'$ is uniquely determined from $(x,y,\tilde{z})$, it is possible to derive that $e_1=e_2$. Therefore, $|f(w)\cap f'(w)|\leq 1$ follows.
The other cases are similarly treated.

(ii) Next, we assume that there are strings  $x0yz1e_1$ and $x'0yz'0e_2$ in $f(w)\cap f'(w)$. Claim \ref{M-algorithm} indicates that $M$ produces a set  $\{\tilde{x}0yz0\hat{e}_d\mid d\in\Sigma\}\cup \{x0yz1\hat{e}_d\mid d\in\Sigma\}$; thus, $x=\tilde{x}=x'$ follows. This result yields a contradiction because $\tilde{x}$ equals $\tilde{x}_{[2n-i]}$ for a certain index $i\in[n,2n-1]_{\integer}$.

\s

In conclusion, all the cases truly yield the desired inequality $|f(w)\cap f'(w)|\leq 1$.
\end{proof}

Since $f,f'\in\cflmv/n$, Claims \ref{functional-CFL(2)-equiv} and \ref{G-vs-f-and-fprime} imply  that $G$ is indeed a member of $\cflmvtwo/n$. This completes the proof of Proposition \ref{G-complexity}.
\end{proofof}

We remark that the functions $f$ and $f'$ constructed in the above proof are not in $\cflsvtwo/n$ because their underlying npda's $M$ and $M'$ can produce multiple output strings.

%%%%%%%
\subsection{Computational Limitation of Pseudorandom Generators}\label{sec:limitation}

We shall briefly discuss the limitation of the efficiency of pseudorandom generators mapping $\Sigma^*$ to $\Sigma^*$ for an arbitrary alphabet $\Sigma$.
In Sections \ref{sec:random-generator}--\ref{sec:efficient-computable}, we have constructed the pseudorandom generator $G$ designed to fool all languages in $\cfl/n$, which reside in the non-uniform function class $\cflmvtwo/n$.
Naturally, one may ask whether it is possible to find a similar generator that can be computed much more efficiently than $G$ is. In a ``uniform'' setting of computation, however, we shall present a rather negative prospect to this question by exhibiting a computational limitation of pseudorandom generators against the uniform language family $\cfl$.

\begin{theorem}\label{no-generator}
No almost 1-1 pseudorandom generator with stretch factor $n+1$ over a certain  alphabet exists in $\cflmv$ against $\cfl$.
\end{theorem}

To prove Theorem \ref{no-generator}, we first show the computational complexity of the {\em ranges} of single-valued total functions in $\cflmv$ since all pseudorandom generators are, by their definition, single-valued and total.

\begin{lemma}\label{complexity-range-CFL}
Let $f$ be any single-valued total function in $\cflmv$, mapping $\Sigma^*$ to $\Sigma^*$, where $\Sigma$ is an arbitrary alphabet. If $f$ has stretch factor $n+1$, then the set $rang(f)$ belongs to $\cfl$.
\end{lemma}

\begin{proof}
Let $f$ be any generator mapping $\Sigma^*$ to $\Sigma^*$ for a certain alphabet $\Sigma$ and define $S=rang(f)$.
Assuming $f\in\cflmv$, our goal is set to show that $S$ is actually in $\cfl$. Since $f\in\cflmv$, let $N$ be any  npda computing $f$ using an extra  write-only output tape.  We intend to construct a new npda $M$ (with no output tape) that recognizes $S$ in linear time.
Let $y$ be any input of length $n\in\nat^{+}$ to $f$. An underlying idea is that, on input $y$, $M$ guesses a whole input instance $x$ to $f$ and checks whether $f(x)$ equals $y$ using only a single stack with no output tape.
Since $M$ has only a read-only input tape, we need to  simulate $N$ using {\em imaginary} input and output tapes of $M$.
When $N$ reads a new symbol written on its imaginary input tape, $M$ guesses such a symbol (in $\Sigma$) and simulates each of $N$'s moves accurately. As far as $N$'s head keeps scanning the same tape cell, $M$ uses the same symbol without guessing another one. If $N$ writes down symbol $b$ on its imaginary output tape, $M$ first checks whether $b$ appears on a cell at which its head is currently scanning on its own input tape, and then $M$ exactly simulates $N$'s next move. If $b$ does not match the bit written on $M$'s input tape, then $M$ immediately rejects the input $y$; otherwise, $M$ continues its simulation of $N$ step by step. When $N$ halts in an accepting state and $M$ reaches the right endmarker $\dollar$ on its input tape, $M$ accepts the input. In all other cases, $M$ rejects $y$ immediately.

If $y\in S$, then a certain string $x$ makes $N(x)$ produce $y$ on its output tape along a certain accepting computation path, say, $p$. Since $|y|=|x|+1$, $N(x)$ halts along this computation path $p$ in $O(|y|)$ steps. Consider an $M$'s computation path in which $M$ correctly guesses $x$ and simulates $N$ along the path $p$. By following this path faithfully, $M$ finally accepts $y$ in $O(|y|)$ steps. On the contrary, when $y\not\in S$, there is no string $x$  for which $N(x)$ correctly produces $y$ in an accepting computation path. This means that $M$ never accepts $y$ in any computation path of $M$. It is important to note that some of the computation paths of $N$ may not even terminate; thus, we need to modify it so that all computation paths terminate in linear time.

In conclusion, $M$ recognizes $S$. Since $M$ is an npda, $S$ should belong  to $\cfl$.
\end{proof}

The proof of Theorem \ref{no-generator} is now easily described with the help of Lemmas \ref{generator-pseudorandom} and \ref {complexity-range-CFL}.

\begin{proofof}{Theorem \ref{no-generator}}
Let $F$ be any almost 1-1 pseudorandom generator from $\Sigma^*$ to $\Sigma^*$ against $\cfl$ for a certain alphabet $\Sigma$.
To draw a contradiction, we assume that $F$ belongs to $\cflmv$. By Lemma \ref{generator-pseudorandom}, the set $rang(F)$ is  $\cfl$-pseudorandom, implying that $rang(F)\not\in\cfl$, because of the self-exclusion property of the $\cfl$-pseudorandomness (namely, no language in $\cfl$ is $\cfl$-pseudorandom). On the contrary, Lemma \ref{complexity-range-CFL} leads to another conclusion that $rang(F)$ is in $\cfl$. These two consequences are contradictory; therefore, $F$ cannot be in $\cflmv$.
\end{proofof}

%%%%%%%%%%%%%%%%%%%%%%%%%%%%%%%%%%%%%%%%%
%%%%%%%%%%%%%%%%%%%%%%%%%%%%%%%%%%%%%%%%%
\section{Swapping Property Lemma}\label{sec:swapping}

The rest of the paper will be devoted to prove Proposition \ref{IP-pseudorandom}, whose proof relies on an analysis of behaviors of advised context-free languages. Prior to the actual proof of the proposition, we intend to examine those behaviors extensively.
In particular, we shall be focused on one of the essential structural properties of the advised context-free languages, which is similar in nature to a {\em swapping property of advised regular languages} \cite{Yam11},  originating in the so-called the {\em swapping lemma for regular languages}\footnote{[Swapping Lemma for Regular Languages] Let $L$ be any infinite language over alphabet $\Sigma$ with $|\Sigma|\geq2$. There exists a positive integer $m$ (called the swapping-lemma constant) such that, for any integer $n\geq1$, any subset $S$ of $L\cap\Sigma^n$ with $|S|>m$, the following condition holds: for any integer $i\in[0,n]_{\integer}$, there exists two strings $x=x_1x_2$ and $y=y_1y_2$ in $S$ with $|x_1|=|y_1|=i$ and $|x_2|=|y_2|$ for which (i) $x\neq y$, (ii) $y_1x_2\in L$, and (iii) $x_1y_2\in L$. } \cite{Yam08}.

Our intended swapping property roughly states that, given a language $L$ in $\cfl/n$,  any long string $w$ in $L$ can be decomposed into $xyz$ in such a way that, under an appropriate condition, if two decompositions, say, $x_1y_1z_1$ and $x_2y_2z_2$ belong to $L$ then the strings $x_1y_2z_1$ and $x_2y_1z_2$ obtained by swapping their middle portions also belong to $L$. A basic form of this fundamental property appeared implicitly  in the proof of the {\em swapping lemma for context-free languages} \cite{Yam08}.
For languages in $\reg/n$, a more useful formulation was given explicitly in \cite[Lemma 5.5]{Yam11}. Here, we intend to shall give a full formulation of the desired swapping property for  languages in $\cfl/n$.
Let us describe our swapping property and give its proof by utilizing an extensive analysis conducted  in \cite{Yam08} for context-free languages.
In what follows, the notation $\Sigma$ is used again
to denote an arbitrary alphabet of cardinality at least $2$. For clarity, we intentionally express  $xz$ and $(xz,y)$ as $(x,z)$ and $(x,z,y)$, respectively, in the lemma and throughout this section.

\begin{lemma}\label{swapping-property}{\rm [Swapping Property Lemma]}\hs{1}
Let $\Sigma$ be any input alphabet with $|\Sigma|\geq2$ and let $L$ be any language over $\Sigma$. If $L\in \cfl/n$, then there exists another alphabet $\Gamma$ that satisfies the following statement. For any triplet $(j_0,k_0,n)$ of integers satisfying $j_0\geq2$ and $2j_0\leq k_0 < n$, there always exist two finite series $\{A_{e}\}_{e\in\Delta_{j_0,k_0,n}}$ and $\{B_{e}\}_{e\in \Delta_{j_0,k_0,n}}$ that meet the four conditions described below,
where $\Delta_{j_0,k_0,n}$ denotes $\{(i,j,u,v)\mid u,v\in\Gamma, i\in[0,n]_{\integer},j\in[j_0,k_0]_{\integer},i+j\leq n\}$.

\begin{itemize}\vs{-2}
  \setlength{\topsep}{-2mm}%
  \setlength{\itemsep}{0mm}% original = 1mm
  \setlength{\parskip}{0cm}%

\item[(1)] For any index tuple $e=(i,j,u,v)\in\Delta_{j_0,k,n}$, it holds that  $A_{e}\subseteq \Sigma^{i}\times\Sigma^{n-i-j}$ and $B_{e}\subseteq \Sigma^{j}$.

\item[(2)] For every string $w\in\Sigma^n$ with $|w|\geq4$, $w$ is in $L$ if and only if there exist an index  $e=(i,j,u,v)\in \Delta_{j_0,k_0,n}$ and three strings $x\in\Sigma^{i}$,  $y\in\Sigma^{j}$, and $z\in\Sigma^{n-i-j}$ for which $w=xyz$, $(x,z)\in A_{e}$, and $y\in B_{e}$.

\item[(3)] \sloppy (swapping property) For every index $e\in\Delta_{j_0,k_0,n}$ and any six strings $x_1,x_2,y_1,y_2,z_1,z_2\in\Sigma^*$, if $(x_1,z_1,y_1),(x_2,z_2,y_2)\in A_{e}\times B_{e}$, then $(x_1,z_1,y_2),(x_2,z_2,y_1)\in A_{e}\times B_{e}$.

\item[(4)] (disjointness) All product sets in $\{A_e\times B_e\mid e\in \Delta_{j_0,k_0,n}\}$ are mutually disjoint.
\end{itemize}
\end{lemma}

\begin{figure}[t]
\begin{center}
\includegraphics*[width=13.0cm]{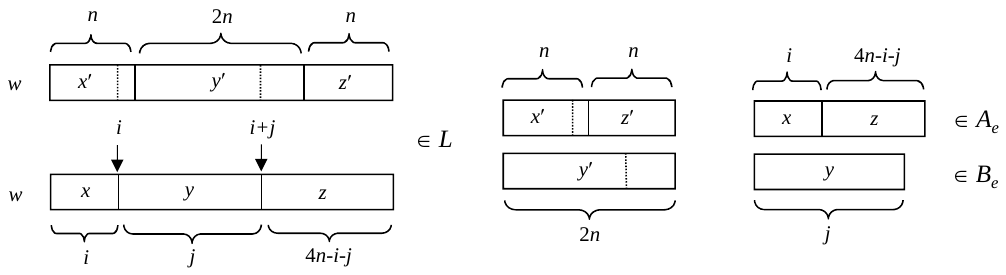}
\caption{An example of input string $w=x'y'z'$ of length $4n$ given to $L$ with $|x'|=|z'|=n$ and $|y'|=2n$.
For appropriate indices $j_0$, $k_0,$ and $e=(i,j,u,v) \in \Delta_{j_0,k_0,4n}$, the same string
$w$ can be  decomposed into $xyz$ with $|x|=i$, $|y|=j$, and $|z|=4n-i-j$,
and two strings $xz$ and $y$ respectively belong to
two sets $A_e$ and $B_e$ of Lemma \ref{swapping-property} so that
$w\in L$ iff $(x,z,y)\in A_e\times B_e$.}\label{fig:IP-string}
\end{center}
\end{figure}

In Section \ref{sec:pseudo-IP*}, we shall apply Lemma \ref{swapping-property} to prove the $\cfl/n$-pseudorandomness of $IP_3$. For this proof, we need to cope with any language $L$ in $\cfl/n$ and any given input string $w$ of length, particularly, $4n$.
It follows from Condition (2) of Lemma \ref{swapping-property} that, for every appropriately chosen number  $n$ in $\nat$, the set  $L\cap \Sigma^{4n}$ is expressed as $\{xyz\mid (x,z,y)\in A_e\times B_e,e\in\Delta_{j_0,k_0,4n}\}$.
Figure~\ref{fig:IP-string} illustrates this situation.

%%%%%
%%%%%
%%%%%
%%%%%

The proof of Lemma \ref{swapping-property} will be given in Sections \ref{sec:features}--\ref{sec:four-conditions}.
As a corollary of Lemma \ref{swapping-property}, the swapping lemma for context-free languages \cite{Yam08} follows easily.
For each fixed subset $S$ of $\Sigma^n$, any two indices
$i\in[0,n]_{\integer}$ and $j\in[1,n]_{\integer}$ with $i+j\leq n$, and any string $u\in\Sigma^{j}$, the notation $S_{i,u}$ denotes the set $\{x\in S\mid u = midd_{i,i+j}(x)\}$. It thus follows that $S = \bigcup_{u\in\Sigma^j}S_{i,u}$ for each fixed index $j\in[1,n-i]_{\integer}$.

\begin{corollary}\label{swapping-lemma-CFL}{\rm
[Swapping Lemma for Context-Free Languages] \cite{Yam08}}\hs{1}
Let $L$ be any infinite context-free language over  an
alphabet $\Sigma$ with $|\Sigma|\geq 2$. There is a positive number $m$ that satisfies the following. Let $n$ be any positive number at least $2$,
let $S$ be any subset of
$L\cap\Sigma^{n}$, and let  $j_0,k_0\in[2,n-1]_{\integer}$ be any two indices satisfying that $k_0 \geq 2j_0$ and $|S_{i,u}|< |S|/m(k_0-j_0+1)(n-j_0+1)$ for any index $i\in[0,n-j_0]_{\integer}$ and any string $u\in\Sigma^{j_0}$.
There exist two indices $i\in[0,n]_{\integer}$ and $j\in[j_0,k_0]_{\integer}$ with $i+j\leq n$ and two strings $x =x_1x_2x_3$ and $y=y_1y_2y_3$ in $S$ with $|x_1|=|y_1|=i$, $|x_2|=|y_2|=j$, and $|x_3|=|y_3|$ such that
(i) $x_2\neq y_2$, (ii) $x_1y_2x_3\in L$, and
(iii) $y_1x_2y_3\in L$.
\end{corollary}

\begin{proofsketch}
Let $L$ be any infinite language in $\cfl$ and take $\Gamma$, $\{A_e\}_{e\in\Delta_{j,k,n}}$, and $\{B_e\}_{e\in\Delta_{j,k,n}}$ that meet Conditions (1)-(4) of Lemma \ref{swapping-property} for all appropriate parameters $(j,k,n)$. Set $m=|\Gamma|^2$ and assume that the conclusion of the corollary fails for this $m$ and parameters $(j_0,k_0,n,S)$. For simplicity, set $\Delta=\Delta_{j_0,k_0,n}$. Note that $|\Delta|=(k_0-j_0+1)(n+1-\frac{j_0+k_0}{2})|\Gamma|^2 \leq m(k_0-j_0+1)(n-j_0+1)$. Assuming an appropriate order for $\Delta$, for each $x\in S$, we denote by $e(x)$ the {\em minimal} element $(i,j,a,b)\in \Delta$ satisfying Condition (2) of Lemma \ref{swapping-property}. Moreover, we set $D_{i,j,a,b} = \{x\in S\mid e(x) = (i,j,a,b)\}$.

Since $e$ is a map from $S$ to $\Delta$, choose an element $e'=(i,j,a,b)\in\Delta$ satisfying $|D_{e'}|\geq |S|/|\Delta|$.
For any string $u\in\Sigma^j$,  it follows from the premise of the corollary that  $|S_{i,u}|< \frac{|S|}{m(k-j_0+1)(n-j_0+1)}\leq \frac{|S|}{|\Delta|}\leq |D_{e'}|$.
Thus, there are four strings $x,y\in S$ and $u,v\in\Sigma^j$ for which  $u\neq v$, $x\in S_{i,u}$, $y\in S_{i,v}$, and $e(x)=e(y)=e'$. Write $x=x_1x_2x_3$ and $y=y_1y_2y_3$, where $|x_1|=|y_1|=i$, $x_1=u$,
and $y_2=v$. By Condition (2), $A_{e'}\times B_{e'}$ must contain both $(x_1,x_3,x_2)$ and $(y_1,y_3,y_2)$. However, Condition (3) implies that $(x_1,x_3,y_2),(y_1,y_3,x_2)\in A_{e'}\times B_{e'}$, in other words, $x_1y_2x_3,y_1x_2y_3\in L$. This is obviously a contradiction; therefore, the corollary holds.
\end{proofsketch}

%%%%
%%%%
\subsection{Structural Features of Npda's}\label{sec:features}

Let us start the proof of Lemma \ref{swapping-property}. Our proof will use
certain structural features of npda's, which were first explored in the proof of the swapping lemma for context-free languages, given in \cite{Yam08}.
Since our proof is founded on such features, it is necessary for us to
review a key lemma (Lemma \ref{height-interval}) of \cite{Yam08} first.

As a starter, we take an arbitrary advised context-free language $L$ over  alphabet $\Sigma$ satisfying $|\Sigma|\geq2$. Assuming that $L$ is in $\cfl/n$, we choose a context-free language $S$, an advice alphabet $\Theta$, and a length-preserving advice function $h:\nat\rightarrow \Theta^*$ satisfying $L=\{x\in\Sigma^*\mid \track{x}{h(|x|)}\in S\}$. For convenience, let  $\Sigma_{\Theta}$ indicate an {\em induced alphabet} $\{\track{w}{x}\mid w\in\Sigma,x\in\Theta\}$ from $\Sigma$ and $\Theta$, and assume that $S\subseteq (\Sigma_{\Theta})^*$.  Since $n\geq 2$,
it is harmless to assume further that $L$ (as well as $S$) contains no empty string $\lambda$.

Since $S\in\cfl$, $S$ is recognized by a certain npda, say, $M$.
To make our later proof simple, we demand that $M$ should have a specific simple form, which we shall explain in the following.
First, we  consider a
context-free grammar $G=(V,T,S_0,P)$ that generates $S$ with $T= \Sigma_{\Theta}$,
where $V$ is a set of variables, $T$ is a set of terminal symbols, $S_0\in V$ is the start variable, and $P$ is a set of productions. We assume that $G$ is in {\em Greibach normal form}; that is, $P$ consists of the production rules of the form $A\,\rightarrow au$, where $A\in V$, $a\in\Sigma_{\Theta}$, and $u\in V^*$.

Closely associated with the grammar $G$, we want to construct an npda $M$ of the form  $(Q,\Sigma_{\Theta},\Gamma,\delta,q_0,Z_0,Q_{acc},Q_{rej})$, where   $Q_{acc}=\{q_{acc}\}$, $Q_{rej}=\{q_{rej}\}$, $Q=\{q_0,q_1\}\cup Q_{acc}\cup Q_{rej}$, and $\Gamma = V\cup\{Z_0\}$ with $Z_0\notin V$. The transition function $\delta$ will be given later.
In this section, we shall deal only with inputs of the form $\cent x \dollar$, where $x\in(\Sigma_{\Theta})^*$, by treating the endmarkers as an integrated part of the input. Notice that $|\cent x \dollar|=|x|+2$. For convenience, every tape cell is indexed with integers and the left endmarker $\cent$ is always written in the $0$th cell. The original input string $x$ of length $n$ is written in the cells indexed between $1$ and $n$ and the right endmarker $\dollar$ is written in the $n+1$st cell.

When we express the content of the stack of $M$ as a series $s = s_1s_2s_3\cdots s_m$ of stack symbols from $\Gamma$, we understand that the leftmost symbol $s_1$ is located at the top of the stack and the $s_m$ is at the bottom of the stack. We then define the transition function $\delta$ as follows:
\begin{enumerate}\vs{-1}
  \setlength{\topsep}{-2mm}%
  \setlength{\itemsep}{0mm}% original = 1mm
  \setlength{\parskip}{0cm}%

\item[1.] $\delta(q_0,\cent,Z_0) = \{ (q_1,S_0Z_0) \}$;

\item[2.] $\delta(q_1,a,A) = \{ (q_1,u) \mid u\in V^{*}, \text{$P$ contains } A\rightarrow au \}$ for every $a\in\Sigma_{\Theta}$ and $A\in V$; and

\item[3.] $\delta(q_1,\dollar,Z_0) = \{(q_{acc},Z_0)\}$.
\end{enumerate}\vs{-1}
It is important to note that the npda $M$ is always in the inner state $q_1$ while the tape head scans any cell located between $1$ and $n$.  Along each accepting computation path, say, $p$ of $M$ on any input, the stack of $M$ never becomes empty (except for $Z_0$) because of the form of production rules in $P$. After the tape head of $M$ scans $\dollar$ along the computation path $p$, the stack must be empty (except for $Z_0$). Therefore, we further
demand that $\delta$ should satisfy the  following requirement.
\begin{enumerate}\vs{-1}
  \setlength{\topsep}{-2mm}%
  \setlength{\itemsep}{0mm}% original = 1mm
  \setlength{\parskip}{0cm}%

\item[4.] For any symbol $a\in\Sigma_{\Theta}$, $\delta(q_1,a,Z_0) = \{(q_{rej},Z_0)\}$.

\item[5.] For every stack symbol $A\in V$, $\delta(q_1,\dollar,A) = \{(q_{rej},A)\}$.
\end{enumerate}\vs{-1}

Additionally, we modify the above npda $M$ and force its stack to increase in size by {\em at most two} by encoding several consecutive stack symbols (except for $Z_0$) into one new stack symbol.
For instance, provided that the original npda $M$ increases its stack size by at most $3$, we introduce a new stack alphabet $\Gamma'$ consisting of $(v_1)$, $(v_1v_2)$, and $(v_1v_2v_3)$, where $v_1,v_2,v_3\in\Gamma$. A new transition $\delta'$ is defined as follows. Initially, we define $\delta'(q_0,\cent,Z'_0)=\{(q_1,S'_0Z'_0)\}$, where $S'_0=(S_0)$ and $Z'_0=(Z_0)$. Consider the case where the top of a new stack contains a new stack symbol $(v_1v_2v_3)$, which indicates that the top three stack symbols of the original computation are $v_1v_2v_3$. If $M$ applies a transition of the form $(q_1,w_1w_2w_3) \in\delta(q_1,a,v_1)$, then we instead apply
$(q_1,(w_1w_2)(w_3v_2v_3)) \in\delta'(q_1,a,(v_1v_2v_3))$. In the case of $(q_1,\lambda)\in\delta(q_1,a,v_1)$, we apply $(q_1,(v_2v_3))\in\delta'(q_1,a,(v_1v_2v_3))$.
The other cases of $\delta'$ are similarly defined.
For more details, refer to, \eg \cite{HMU01}. For brevity, we shall express $\Gamma'$ as $\Gamma$.
Overall, we can demand the following extra requirement for $M$.
\begin{enumerate}\vs{-1}
\item[6.] For any $a\in\Sigma_{\Theta}$, any $v\in\Gamma$, and any $w\in\Gamma^*$, if $(q_1,w)\in \delta(q_1,a,v)$, then $|w|\leq 2$.
\end{enumerate}\vs{-1}

Hereafter, we assume that our npda $M$ always satisfies  the aforementioned  five  conditions. For each string $x\in S$, we write $ACC(x)$ for the set of all accepting computation paths of $M$ on the input $x$. For simplicity, we write $ACC_n$ to express the union $\bigcup_{x\in S\cap(\Sigma_{\Theta})^n}ACC(x)$.

%%%
%%%
\begin{figure}[t]
\begin{center}
\includegraphics*[width=12.0cm]{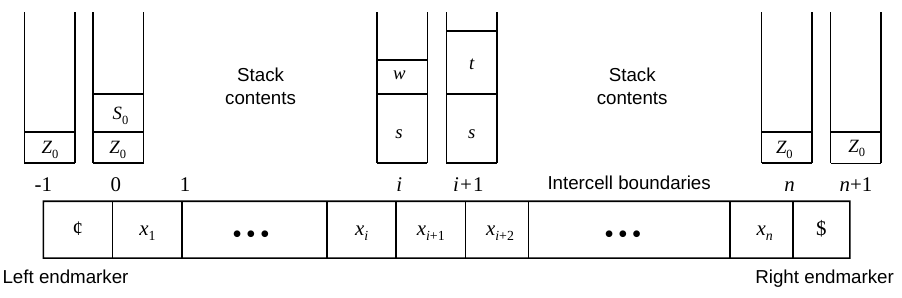}
\caption{An example of intercell boundaries and transitions of
stack contents at those intercell boundaries. Lower-case letters $s,t,w$ refer to series of stack symbols and $x_1,x_2,\ldots,x_n$ is an input string.}\label{fig:intercell-boundary}
\end{center}
\end{figure}
%%%
%%%

An {\em intercell boundary} $i$ refers to a boundary or a border between two adjacent cells---the $i$th cell and the $i+1$st cell---in our npda's input tape.
We sometimes call the intercell boundary $-1$ the {\em initial intercell boundary} and the intercell boundary $n+1$ the {\em final intercell boundary}.
Meanwhile, we fix a string $x$ in $S\cap(\Sigma_{\Theta})^n$ and a computation path $p$ of $M$ in $ACC(x)$.
Along this accepting computation path $p$, we assign
to intercell boundary $i$
a stack content produced after scanning the $i$th cell but before scanning the $i+1$st cell. For convenience, such a stack content is referred to as the ``stack content at intercell boundary $i$.'' For instance, the stack contents at the initial and final intercell boundaries are both $Z_0$,   independent of the choice of accepting computation paths.
Figure \ref{fig:intercell-boundary} illustrates intercell boundaries and  transitions of stack contents at those intercell boundaries.

We define the {\em basis interval} to be $I_0=[-1,n+1]_{\integer}$.
Any accepting computation path of the npda $M$ can generate a certain length-$(n+2)$ series $(s_{-1},s_0,s_1,\ldots,s_n,s_{n+1})$ of stack contents, where  $s_{-1}= s_n = s_{n+1} =Z_0$ and $s_0=S_0Z_0$.
For any subinterval $I=[i_0,i_1]_{\integer}$ of $I_0$, let the {\em size of $I$} be $|I|=i_1-i_0$. We call a subsequence $\gamma = (s_{i_0},s_{i_0+1},\ldots,s_{i_1})$
a {\em stack transition} associated with this interval $I$.
The {\em height} at intercell boundary $b$ of $\gamma$
is the length $|s_{b}|$ of the stack content $s_b$ at $b$.
Since $Z_0$ cannot be removed, the minimal height must be $1$.
An {\em ideal} stack transition $\gamma$ associated with an interval $[i_0,i_1]_{\integer}$ should satisfy that (a) both of the intercell boundaries $i_0$ and $i_1$ have the same height $\ell$ and (b) all heights within this interval are more than or equal to $\ell$.

Take  any subinterval $I=[i_0,i_1]_{\integer}$ of $I_0$ and let $\gamma= (s_{i_0},s_{i_0+1},\ldots,s_{i_1})$ be any ideal stack transition with $I$. For every possible height $\ell$, we define the {\em minimal width}, denoted $minwid_{I}(\ell)$ (resp., the {\em maximal width}, denoted $maxwid_{I}(\ell)$), to be the minimal size (resp., maximal size) $|I'|$ for which (i) $I'=[i'_0,i'_1]_{\integer}\subseteq I$, (ii) $\gamma$ has height $\ell$ at both intercell boundaries $i'_0$ and $i'_1$, and (iii) at no intercell boundary $i\in I'$, $\gamma$ has height less than $\ell$. Such a pair $(i'_0,i'_1)$ naturally induces a subsequence $\gamma'=(s_{i'_0},s_{i'_0+1},\ldots,s_{i'_1})$ of $\gamma$. For convenience, we say that $\gamma'$ (as well as $I'$) {\em realizes} the minimal width $minwid_{I}(\ell)$ (resp., maximal width $maxwid_{I}(\ell)$).

Finally, we come to the point of describing a key lemma, given implicitly in \cite{Yam08}, which holds for any accepting computation path $p$ of $M$.
For completeness, we include the proof of the lemma because the proof itself is interesting in its own right.

\begin{lemma}{\rm \cite{Yam08}}\label{height-interval}
Let $M$ be any npda that satisfies Conditions 1--6 given earlier. Let $x$ be any string of length $n$ accepted by $M$.  Assume that $j_0\geq2$ and $2j_0\leq k_0 \leq n$. Along any computation path $p\in ACC(x)$, for any interval $I =[i_0,i_1]_{\integer} \subseteq I_0$ with $|I|\geq k_0$ and for any ideal stack transition $\gamma$ associated with the interval $I$ having height $\ell_0$ at the two
intercell boundaries $i_0$ and $i_1$, there exist a subinterval $I'=[i'_0,i'_1]_{\integer}$ of $I$ and a height $\ell\in[n]$
such that $\gamma$ has height $\ell$ at both intercell boundaries $i'_0$ and $i'_1$, $j_0\leq |I'|\leq k_0$, and $minwid_{I}(\ell)\leq |I'| \leq maxwid_{I}(\ell)$.
\end{lemma}

\begin{proof}
Fix ten parameters $(n,x,p,\gamma,i_0,i_1,j_0,k_0,\ell_0,I)$ given in the premise of the lemma. Recall that $\gamma$ is of the form $(s_{i_0},s_{i_0+1},\ldots,s_{i_1})$ associated with $I$.
Let us introduce several terminologies necessary to go through this proof.
We say that $\gamma$ has a {\em peak at $i$} if $|s_{i-1}|<|s_i|$ and $|s_{i+1}|<|s_i|$. Moreover, $\gamma$ has a {\em flat peak in $(i'_0,i'_1)$} if $|s_{i'_0-1}|<|s_{i'_0}|=|s_{i'_0+1}|=\cdots = |s_{i'_1}|$ and $|s_{i'_1+1}|<|s_{i'_1}|$.
On the contrary, we say that $\gamma$ has a {\em base at $i$} if
$|s_{i-1}|>|s_i|$ and $|s_{i+1}|>|s_i|$; $\gamma$ has a {\em flat base in $(i'_0,i'_1)$} if $|s_{i'_0-1}|>|s_{i'_0}|=|s_{i'_0+1}|=\cdots = |s_{i'_1}|$ and $|s_{i'_1+1}|>|s_{i'_1}|$. Figure \ref{fig:stack-transition} provides an illustration of (flat) peaks and (flat) bases.

%%%
%%%
\begin{figure}[t]
\begin{center}
\includegraphics*[width=9.5cm]{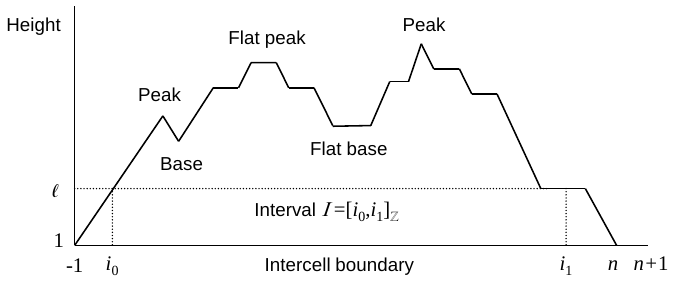}
\caption{An example of (ideal) track transitions associated with interval $I=[i_0,i_1]_{\integer}$ and height $\ell$}\label{fig:stack-transition}
\end{center}
\end{figure}
%%%
%%%

We wish to prove the lemma by induction on the number of peaks or flat peaks along the given accepting computation path $p$ of $M$ on $x$.

\s

(Basis Step) Assume that the ideal stack transition $\gamma$ with $I=[i_0,i_1]_{\integer}$ has either one peak or one flat peak
and that $\gamma$ has no base or flat base.
Let us consider the first case where there is a unique peak.
Let $\ell_1$ be the height of such a peak.
Clearly, we obtain $minwid_{I}(\ell-1)=0$.
Since $M$ satisfies Condition 6, it follows that
\begin{equation*}
minwid_{I}(\ell) = maxwid_{I}(\ell+1) +2
\end{equation*}
for any height $\ell$ with $\ell_0\leq \ell<\ell_1$.

We first assume that $minwid_{I}(\ell_0)\leq j_0$. Define $\tilde{I}=[i'_0,i'_1]_{\integer}$ to be any subinterval of $I$ that realizes $minwid_{I}(\ell_0)$. Note that $maxwid_{I}(\ell_0)=|I|\geq k_0$ holds. In this case, we set $\ell=\ell_0$ and choose any interval $I'$ so that $\tilde{I}\subseteq I'$ and $|I'|=j_0$. Obviously, it follows that $minwid_{I}(\ell)\leq |I'|\leq maxwid_{I}(\ell)$ and $j_0\leq |I'|\leq k_0$, as requested.

Next, we assume that $minwid_{I}(\ell_0)>j_0$. Let  $\ell'$ denote the maximal height in $[\ell_0,\ell_1-1]_{\integer}$ satisfying that $minwid_{I}(\ell'+1)\leq j_0 < minwid_{I}(\ell')$.
Let $I_{min}=[i'_0,i'_1]_{\integer}$ be a subinterval of $I$
that realizes $minwid_{I}(\ell'+1)$. Similarly, let $I_{max} = [i''_0,i''_1]_{\integer}$ express a subinterval of $I$ that realizes $maxwid_{I}(\ell'+1)$.
If $j_0= minwid_{I}(\ell'+1)$, then we choose $I_{min}$ as
the desired interval $I'$ and  $\ell'+1$ as the height $\ell$ for the lemma.
If $j_0\leq maxwid_{I}(\ell'+1)$, then we pick
an interval $I'$ satisfying that $I_{min}\subseteq I'\subseteq I_{max}$ and $|I'|=j_0$. We also define $\ell=\ell'+1$ for the lemma.
The remaining case to consider is that $maxwid_{I}(\ell'+1) < j_0 < minwid_{I}(\ell')$. In this particular case, it follows that
\[
j_0< minwid_{I}(\ell') = maxwid_{I}(\ell'+1)+2 < j_0 + 2 \leq 2j_0 \leq k_0
\]
since $j_0\geq 2$. For any subinterval $I'_{min}$ of $I$ that realizes $minwid_{I}(\ell')$, it follows that $j_0<|I'_{min}|<k_0$. It is thus enough to define $I'=I'_{min}$ and $\ell= \ell'$ for the lemma.

Let us consider the second case where there is a unique flat peak in $(i_2,i_3)$ with height $\ell_1$.
If $i_3-i_2 \geq j_0$, then we define $I'=[i_2,i_2+j_0]_{\integer}$ and $\ell=\ell_1$ for the lemma. The other case where $i_3-i_2 < j_0$ is similar in essence to the ``peak'' case discussed above.

\s

(Induction Step) Let $c>1$ and consider the case where $\gamma$ has  $c$ peaks and/or flat peaks. Unlike the basis step, we need to consider bases and flat bases as well. Choose the {\em lowest} base or flat base within this interval.
In case of more than one such base and/or flat base, we always choose the leftmost one.

Let us consider the first case where
there is the lowest base at $i_2$. Let $\ell_2$ denote the height at $i_2$. Since $\gamma$ is an ideal stack transition, $\ell_2\geq \ell_0$ follows.
Let  $I^*=[i'_0,i'_1]_{\integer}$ be the {\em largest} interval for which the heights at both $i'_0$ and $i'_1$ both equal $\ell_2$.
The choice of $I^*$ implies that $i_2\in I^*$ and $|I^*| = maxwid_{I}(\ell_2)$.
If $j_0\leq |I^*|\leq k_0$, then we set $I'=I^*$ and $\ell=\ell_2$ for the lemma.
If $|I^*|<j_0$, then a similar argument used for the basis step proves the lemma.
Next, assume that $|I^*|> k_0$. Let us split $I^*$ into two subintervals $I_1=[i'_0,i_2]_{\integer}$ and $I_2 = [i_2,i'_1]_{\integer}$.
Since $k_0\geq 2j_0$, either one of of $I_1$ and $I_2$ has size more than $j_0$. We pick such an interval, say $I_3$.  Let $\gamma'$ denote a unique subsequence of $\gamma$ associated with the interval $I_3$.
If $|I_3|\leq k_0$,
then we choose $I'=I_3$ and $\ell=\ell_2$ for the lemma.
Let us assume that $|I_3|>k_0$. By the choice of $I_3$,
$\gamma'$ is an ideal stack transition. Since $\gamma'$ has fewer than $c$ peaks and/or flat peaks, we can apply the induction hypothesis to obtain the lemma.

Consider the second case where there is the lowest flat base in $(i_2,i_3)$.
We set $I^*=[i'_0,i'_1]_{\integer}$ as in the first case so that $[i_2,i_3]_{\integer}\subseteq I^*$.  Unlike the first case, nevertheless,
we need to split $I^*$ into three intervals $I_1=[i'_0,i_2]_{\integer}$, $I_2=[i_2,i_3]_{\integer}$, and $I_3=[i_3,i'_1]_{\integer}$.
If either $|I^*|<j_0$ or $j_0\leq |I^*|\leq k_0$ holds, then it suffices to apply a similar argument used for the previous case. Finally, we examine the case of $|I^*|>k_0$. Since $k_0\geq 2j_0$, either one of the two intervals $I_1\cup I_2$  and $I_3$ has size more than $j_0$.
We pick such an interval. The rest of our argument is similar to the one for the previous case.
\end{proof}

%%%
%%%
\subsection{Four Conditions of the Lemma}\label{sec:four-conditions}

Following the previous subsection, we continue the proof of Lemma \ref{swapping-property}. Recall that $M =(Q,\Sigma_{\Theta},\Gamma,\delta,q_0,Z_0,Q_{acc},Q_{rej})$ is our target npda and satisfies
Condition 1--6 given in the previous subsection. Our goal is to define $A_e$'s and $B_e$'s so that they satisfy Conditions (1)--(4) of Lemma \ref{swapping-property}.

Hereafter, we arbitrarily fix a length $n\in\nat$ and a pair $(j_0,k_0)$ that satisfies $4\leq 2j_0\leq k_0 < n$. Recall the index set $\Delta_{j_0,k_0,n}=\{(i,j,u,v)\mid  u,v\in\Gamma, i\in[0,n]_{\integer},j\in[j_0,k_0]_{\integer},i+j\leq n \}$ given in the premise of Lemma \ref{swapping-property}. Notice that $|\Delta_{j_0,k_0,n}| \leq (n+1)^2|\Gamma|^2$ holds.
We then apply Lemma \ref{height-interval} to obtain the following claim concerning stack contents of  $M$.

\begin{claim}\label{interpreted-lemma}
For every string $w$ in $S\cap (\Sigma_{\Theta})^n$, there exist an index   $(i,j,u,v)\in\Delta_{j_0,k_0,n}$, four strings $x,y,z\in(\Sigma_{\Theta})^*$ and  $s\in\Gamma^*$, and a computation path
$p\in ACC(w)$ such that
(i) $w=xyz$ with $|x|=i$ and $|y|=j$ and (ii) along the computation path $p$, $M$ produces stack content $us$ after reading $\cent x$ and stack content $vs$ after reading $y$, and  no symbol in $s$ is ever accessed by $M$ while  reading $y$. We call this $s$ a {\em rooted stack content}.
\end{claim}

\begin{proof}
Let $w$ be any input string in $(\Sigma_{\Theta})^n$ that is accepted by $M$. We choose $i_0=0$ and $i_1=n-1$ and consider the interval $I=[i_0,i_1]_{\integer} \subseteq I_0$. Choose any ideal track transition $\gamma$ made by $M$ along a certain computation path in $ACC(w)$. By applying  Lemma \ref{height-interval}, we obtain a subinterval $I'=[i'_0,i'_1]_{\integer}$ of $I$ and a height $\ell\in[n]$ such that  $j_0\leq |I'|\leq k_0$, $minwid_{I}(\ell)\leq |I'| \leq maxwid_{I}(\ell)$, and $\gamma$ has height $\ell$ at both intercell boundaries $i'_0$ and $i'_1$.  Here, we set $i=i'_0$ and $j=|I'|$ and decompose $w$ into $w=xyz$ with $|x|=i$ and $|y|=j$. Let us assume that $\gamma$ has stack content $us$ of length $\ell$ at the intercell boundary $i'_0$ (\ie just after reading $\cent x$) and similarly stack content $vs'$ of length $\ell$ at $i'_1$ (\ie just after reading $y$) for certain elements $u,v\in\Gamma$ and $s,s'\in\Gamma^*$.
Notice that $|s|=|s'|$ because $|u|=|v|=1$. Since $minwid_{I}(\ell)\leq |I'| \leq maxwid_{I}(\ell)$, $\gamma$ never has height less than $\ell$ at any cell number between $i'_0$ and $i'_1$; namely, $M$ accesses no symbol  inside $s$. Hence, $s$ must be a rooted stack content. From this fact, we derive that $s'$ coincides with $s$. Thus, $(i,j,u,v)$ falls into  $\Delta_{j_0,k_0,n}$. In conclusion, Claim \ref{interpreted-lemma} should be true.
\end{proof}

Let us return to our proof of Lemma \ref{swapping-property}. To improve  the readability, we shall define two ``temporary'' series $\{A_{e}\}_{e\in\Delta_{j_0,k_0,n}}$ and $\{B_{e}\}_{e\in\Delta_{j_0,k_0,n}}$ and then verify Conditions (1)--(3) of the lemma. Later in this subsection, we shall modify them appropriately to further satisfy Condition (4) (as well as Conditions (1)--(3)).
Given every index tuple $(i,j,u,v)\in \Delta_{j_0,k_0,n}$, we shall define three sets $T^{(1)}_{i,u}$, $T^{(2)}_{i,j,v}$, and $T^{(3)}_{j,u,v}$.
Recall that $ACC_n$ is shorthand for the union $\bigcup_{x\in S\cap(\Sigma_{\Theta})^n}ACC(x)$. Assume that $h(n)$ has the form $h_1h_2h_3$ with $|h_1|=i$ and $|h_2|=j$.  Remember that $M$ stays in inner state $q_1$ except for the first and final steps.  Since $n$ is fixed, we often omit ``$n$'' in the rest of the proof.

\begin{itemize}\vs{-1}
  \setlength{\topsep}{-2mm}%
  \setlength{\itemsep}{0mm}% original = 1mm
  \setlength{\parskip}{0cm}%

\item Let $T^{(1)}_{i,u}$ be a collection of all triplets  $(\track{x}{h_1},s,p)$ with $x\in\Sigma^i$, $s\in\Gamma^*$, and $p\in ACC_n$  such that,  along the computation path $p$, $M$ produces $us$ in the stack after reading $\cent \track{x}{h_1}$.

\item Let $T^{(2)}_{i,j,v}$ be a collection of all triplets  $(\track{z}{h_3},s,p)$ with $z\in\Sigma^{n-i-j}$, $s\in\Gamma^*$, and $p\in ACC_n$  such that,  along the computation path $p$,  $M$ is in the inner state $q_1$ with stack content $vs$ before reading $\track{z}{h_3}\dollar$ and $M$ enters the unique accepting state $q_{acc}$ after reading $\track{z}{h_3}\dollar$.

\item Let $T^{(3)}_{j,u,v}$ be a collection of all triplets  $(\track{y}{h_2},s,p)$ with $y\in\Sigma^{j}$, $s\in\Gamma^*$, and $p\in ACC_n$  such that,  along the computation path $p$, $M$ is in the inner state $q_1$ with stack content $us$ before reading $\track{y}{h_2}$ and $M$ produces stack content $vs$ after reading $\track{y}{h_2}$, provided that $s$ is a rooted stack content (\ie $M$ does not access any symbol in $s$ while reading $\track{y}{h_2}$).
\end{itemize}\vs{-1}

\n Given each index $e=(i,j,u,v)$ in $\Delta_{j_0,k_0,n}$, the desired sets $A_{e}$ and $B_{e}$ are defined as follows.

\begin{itemize}\vs{-1}
  \setlength{\topsep}{-2mm}%
  \setlength{\itemsep}{0mm}% original = 1mm
  \setlength{\parskip}{0cm}%

\item $A_{e} = \{ (x,z)\in\Sigma^{i}\times\Sigma^{n-i-j} \mid \exists s\in\Gamma^*\, \exists p\in ACC_n [\, (\track{x}{h_1},s,p)\in T^{(1)}_{i,u}\,\wedge\, (\track{z}{h_3},s,p)\in T^{(2)}_{i,j,v}, ] \}$.

\item $B_{e} = \{ y\in\Sigma^j\mid \exists s\in\Gamma^*\, \exists p\in  ACC_n[\, (\track{y}{h_2},s,p)\in T^{(3)}_{j,u,v}\,]  \}$.
\end{itemize}\vs{-1}

\n Next, we wish to argue that the series $\{A_e\times B_e\}_{e\in\Delta_{j_0,k_0,n}}$ satisfies Conditions (1)--(3) of Lemma \ref{swapping-property}.

\s

(1) Clearly, for every $e\in D$, we obtain  $A_{e}\subseteq \Sigma^{i}\times \Sigma^{n-i-j}$ and $B_{e}\subseteq \Sigma^j$, and thus Condition (1) follows instantly.

\s

(2) Using Claim \ref{interpreted-lemma}, we want to show Condition (2).
Let $w$ be any string in $L\cap \Sigma^n$; that is, $\track{w}{h(n)}\in S\cap (\Sigma_{\Theta})^n$.
Conditions (i)--(ii) of Claim \ref{interpreted-lemma} imply the existence of an index $e=(i,j,u,v)\in\Delta_{j_0,k_0,n}$, four strings $x\in\Sigma^i$, $y\in\Sigma^j$, $z\in\Sigma^{n-i-j}$, $s\in\Gamma^*$,
 and a computation path $p\in ACC_n$ satisfying the following membership relations:
$(\track{x}{h_1},s,p)\in T^{(1)}_{i,u}$, $(\track{z}{h_3},s,p)\in T^{(2)}_{i,j,v}$, and $(\track{y}{h_2},s,p)\in T^{(3)}_{j,u,v}$, provided that  $h(n)$ has the form $h_1h_2h_3$ with $|h_1|=i$ and $|h_2|=j$. From those relations,  we obtain both  $(x,z)\in A_e$ and $y\in B_e$, as requested.

Conversely, assume that $w=xyz$ and $(x,z,y)\in A_e\times B_e$ for a certain  index $e=(i,j,u,v)\in\Delta_{j_0,k_0,n}$ and three strings $x,y,z\in\Sigma^*$. By the definitions of $A_e$ and $B_e$, this assumption indicates the existence of two stack contents $s,s'\in\Gamma^*$ and two computation paths $p,p'\in ACC_n$ for which $(\track{x}{h_1},s,p)\in T^{(1)}_{i,u}$, $(\track{z}{h_3},s,p)\in T^{(2)}_{i,j,v}$, and $(\track{y}{h_2},s',p')\in T^{(3)}_{j,u,v}$. Since $(s',p')$ may be in general different from $(s,p)$, we cannot immediately conclude the
acceptance of the input $\track{xyz}{h_1h_2h_3}$ ($=\track{w}{h(n)}$) by $M$. We thus need the following claim.

\begin{claim}\label{another-stack-path}
For any $s,s'\in\Gamma^*$ and $p,p'\in ACC_n$, if three conditions $(\track{x}{h_1},s,p)\in T^{(1)}_{i,u}$, $(\track{z}{h_3},s,p)\in T^{(2)}_{i,j,v}$, and $(\track{y}{h_2},s',p')\in T^{(3)}_{j,u,v}$ hold, then there exists another computation path $r\in ACC_n$ for which   $(\track{x}{h_1},s,r)\in T^{(1)}_{i,u}$, $(\track{z}{h_3},s,r)\in T^{(2)}_{i,j,v}$, and $(\track{y}{h_2},s,r)\in T^{(3)}_{j,u,v}$ hold.
\end{claim}

\begin{proof}
From the computation paths $p$ and $p'$ given in the premise of the claim, we want to  find a new computation path $r$, along which $M$ behaves as follows.
In the first stage, following the computation path $p$, $M$ produces $us$ in its stack after reading $\cent\track{x}{h_1}$. In the second stage, $M$ starts with the current configuration and follows the computation path $p'$ by  imagining that $s$ stored in the stack is $s'$. This switch of computation paths is possible because, along the computation path $p'$, $s'$ is a rooted stack content and thus $M$ accesses no symbol in $s'$ while reading $\track{y}{h_2}$. After  reading $\track{y}{h_2}$, the stack holds  $vs$.  In the third stage, $M$ starts with the current configuration, follows the computation path $p$ again, and finally enters an appropriate accepting state  after reading $\track{z}{h_3}\dollar$. The resulted computation path $r$ is truly an accepting  computation path,  and thus $M$ accepts the input $\track{xyz}{h_1h_2h_3}$ ($=\track{w}{h(n)}$). The behavior of $M$ along  $r$ obviously satisfy the desired conditions, namely, $(\track{x}{h_1},s,r)\in T^{(1)}_{i,u}$, $(\track{z}{h_3},s,r)\in T^{(2)}_{i,j,v}$, and $(\track{y}{h_2},s,r)\in T^{(3)}_{j,u,v}$.
\end{proof}

Claim \ref{another-stack-path} helps us choose another computation path $r\in ACC_n$ that meets the following three conditions:  $(\track{x}{h_1},s,r)\in T^{(1)}_{i,u}$, $(\track{z}{h_3},s,r)\in T^{(2)}_{i,j,v}$, and $(\track{y}{h_2},s,r)\in T^{(3)}_{j,u,v}$.  These conditions altogether indicate that $M$ accepts the input $\track{w}{h(|w|)}$. In conclusion, $w$ indeed belongs to  $L$.

\s

(3) Next, we shall discuss Condition (3). Let us take two arbitrary triplets $(x_1,z_1,y_1),(x_2,z_2,y_2)\in A_{e}\times B_{e}$ satisfying  $|x_1|=|x_2|$, $|y_1|=|y_2|$, and $|z_1|=|z_2|$. For each index  $b\in\{1,2\}$, there are two stack contents $s_b,s'_b\in\Gamma^*$ and two computation paths $p_b,p'_b\in ACC_n$ for which   $(\track{x_{b}}{h_1},s_b,p_b)\in T^{(1)}_{i,u}$, $(\track{z_b}{h_3},s_b,p_b)\in T^{(2)}_{i,j,v}$, and $(\track{y_b}{h_2},s'_b,p'_b)\in T^{(3)}_{j,u,v}$. Of those six conditions, we particularly select
$(\track{x_1}{h_1},s_1,p_1)\in T^{(1)}_{i,u}$, $(\track{z_1}{h_3},s_1,p_1)\in T^{(2)}_{i,j,v}$, and $(\track{y_2}{h_2},s'_2,p'_2)\in T^{(3)}_{j,u,v}$.
Claim \ref{another-stack-path} then provides a computation path $r\in ACC_n$ for which  $(\track{x_1}{h_1},s_1,r)\in T^{(1)}_{i,u}$, $(\track{z_1}{h_3},s_1,r)\in T^{(2)}_{i,j,v}$, and $(\track{y_2}{h_2},s_1,r)\in T^{(3)}_{j,u,v}$. Obviously, from these three  conditions, it follows that $(x_1,z_1,y_2)$ belongs to $A_{e}\times B_{e}$. Similarly, we obtain $(x_2,z_2,y_1)\in A_{e}\times B_{e}$, leading to Condition (3).

\s

(4) Finally, we shall prove Condition (4). Up to this point, we have proven that
the two series  $\{A_e\}_{e\in \Delta_{j_0,k_0,n}}$ and $\{B_e\}_{e\in \Delta_{j_0,k_0,n}}$ satisfy Conditions (1)--(3). Unfortunately,
all product sets in $\{A_e\times B_e\mid e\in \Delta_{j_0,k_0,n}\}$ are not guaranteed to be mutually disjoint.
To amend this drawback, we shall slightly modify the two series and make them satisfy the required disjointness. In what follows, we assume a (lexicographic) linear order $<$ among all indices in $\Delta_{j_0,k_0,n}$. Here, let us define two additional sets $A'_e$ and $B'_e$ as follows.
\begin{itemize}\vs{-1}
\item $A'_e = \{(x,z)\in A_e\mid \forall d\in\Delta_{j_0,k_0,n} [\,d<e\rightarrow (x,z)\not\in A_{d}\,]\}$.
\vs{-2}
\item $B'_e = \{y \in B_e\mid \forall d\in\Delta_{j_0,k_0,n} [\,d<e\rightarrow (x,z)\not\in B_{d}\,]\}$.
\end{itemize}
Note that, if $A'_{e_1}\cap A'_{e_2}\neq\setempty$ holds for two indices $e_1,e_2\in\Delta_{j_0}$, then  the above definition of $\{A'_e\}_{e\in\Delta_{j_0,k_0,n}}$ leads to  $e_1=e_2$. Similarly, $B'_{e_1}\cap B'_{e_2}\neq\setempty$ yields $e_1=e_2$.
Assuming that $(A'_{e_1}\times B'_{e_1})\cap (A'_{e_2}\times B'_{e_2})\neq\setempty$, we choose a triplet $(x,z,y)$ in $(A'_{e_1}\times B'_{e_1})\cap (A'_{e_2}\times B'_{e_2})$. For those strings $x,y,z$, it follows that  $(x,z)\in A'_{e_1}\cap A'_{e_2}$ and $y\in B'_{e_1}\cap B'_{e_2}$. By the above-mentioned property of $\{A'_e\}_{e\in \Delta_{j_0,k_0,n}}$ and $\{B'_e\}_{e\in \Delta_{j_0,k_0,n}}$, we obtain  $e_1=e_2$. Therefore, all product sets in $\{A'_e\times B'_e\}_{e\in\Delta_{j_0,k_0,n}}$ are mutually disjoint. It is worth mentioning that Condition (1)--(3) also hold for the series $\{A'_e\}_{e\in\Delta_{j_0,k_0,n}}$ and $\{B'_e\}_{e\in\Delta_{j_0,k_0,n}}$.

\s

Since all four conditions are properly met, the proof of Lemma \ref{swapping-property} is finally completed. We are now well-prepared for proving Proposition \ref{IP-pseudorandom} in the next section.

%%%%%%%%%%%%%%%%%%%%%%%%%%%%%%%%%%%
%%%%%%%%%%%%%%%%%%%%%%%%%%%%%%%%%%%
\section{Proof of Proposition \ref{IP-pseudorandom}}\label{sec:proof-prop}

In order to complete the proof of Theorem \ref{generator-CFL}, there is still a missing proof of Proposition \ref{IP-pseudorandom}, which states that the language $IP_{3}$ is indeed  $\cfl/n$-pseudorandom. Exclusively in this section, we shall present its proof in details. Firstly, we shall present in Section \ref{sec:pseudo-IP*} a key lemma  (Lemma \ref{disc_M(T)-bound}) regarding a discrepancy upper bound of an arbitrary set with respect to a special function.  With the help of this key lemma, we shall prove the desired proposition by directly applying the swapping property lemma (Lemma \ref{swapping-property}). Secondly, we shall verify in Section \ref{sec:discrepancy} the key lemma by studying five different situations separately,  depending on  characteristic behaviors of a target npda.

%%%%%
\subsection{Pseudorandomness of $IP_{3}$}\label{sec:pseudo-IP*}

To prove that $IP_{3}$ is $\cfl/n$-pseudorandom, let us fix an arbitrary  language $S$  in $\cfl/n$  over the binary alphabet $\Sigma=\{0,1\}$.
To achieve our goal, it suffices by Lemma \ref{dense-dense-lemma} to prove  that the function $\ell''(n) = \frac{|dense(IP_{3}\cap S)(n) - dense(\overline{IP_{3}}\cap S)(n)}{2^{n}}$ is negligible. Let us start proving the negligibility of $\ell''(n)$.

Let  $p$ be any positive polynomial and assume without loss of generality  that $p$ is {\em strictly increasing} (\ie $m<n$ implies $p(m)<p(n)$). We choose a constant $c\in\nat^{+}$ that forces $p(4n+3)\leq 3^cp(4n)$ to hold for every number $n\in\nat^{+}$. Such a constant actually exists because $p$ is an increasing positive polynomial. Let us fix an arbitrary number $n\in\nat^{+}$ for which $(4n+1)^2|\Gamma|^2<2^{n/4}$ is satisfied.  Those two conditions regarding $c$ and $n$ will be used later to obtain an inequality $\ell''(4n)\leq 1/p(4n)$.

We shall consider the basic case of $|axyz|=4n$ with $a=\lambda$. It suffices for us  to concentrate on $n\geq4$.
For notational convenience, we abbreviate the sets $S\cap IP_{3}\cap\Sigma^{4n}$  and $S\cap \overline{IP_{3}}\cap\Sigma^{4n}$ respectively as $U_1$ and $U_0$. Since
\[
2^{4n}\cdot \ell''(4n) = |dense(IP_{3}\cap S)(4n) - dense(\overline{IP_{3}}\cap S)(4n)| =  ||U_1| - |U_0| |,
\]
our first goal is to verify that $||U_1| - |U_0||\leq 2^{4n}/3^{c}p(4n)$.

Initially, we set our magic numbers $j_0$ and $k_0$ as $j_0 = \floors{4n/3}$ and $k_0 = 2j_0$ and we then apply Lemma \ref{swapping-property} with these numbers.
Unlike Section \ref{sec:swapping}, here we use ``$4n$''  in place of ``$n$'' as the length of our input strings. To simplify our notation further, we intend to write $\Delta_{j_0}$ for $\Delta_{j_0,k_0,4n}$ by simply dropping the subscripts ``$k_0$'' and ``$4n$.''
Since $|\Delta_{j_0}|\leq (4n+1)^2|\Gamma|^2$ holds as shown in Section \ref{sec:swapping}, we then obtain $|\Delta_{j_0}|<2^{n/4}$ by our choice of $n$. With respect to $\Delta_{j_0}$, Lemma \ref{swapping-property} provides two useful series $\{A_{e}\}_{e\in\Delta_{j_0}}$ and $\{B_{e}\}_{e\in\Delta_{j_0}}$.

To estimate the value $||U_1| - |U_0||$, we want to decompose $S\cap\Sigma^{4n}$ into a certain finite series $\{S_e\}_{e\in \Delta_{j_0}}$ of sets. Associated with each index $e=(i,j,u,v)\in \Delta_{j_0}$, we set $S_e$ to
be a collection of all strings $w$ in $S\cap \Sigma^{4n}$ that satisfy  both $(x',z')\in A_e$ and $y'\in B_e$, where $x'= pref_{i}(w)$, $y'= midd_{i,i+j}(w)$, and $z'=suf_{4n-i-j}(w)$. As the following statement shows, it is enough to concentrate  on the value $||U_1\cap S_{e}| - |U_0\cap S_{e}||$ for each index $e\in\Delta_{j_0}$.

\begin{claim}
$||U_1| - |U_0|| \leq \sum_{e\in \Delta_{j_0}}||U_1\cap S_{e}| - |U_0\cap S_{e}||$.
\end{claim}

The above claim can be proven in the following fashion. Based on the definition of $S_e$,  the equality $S\cap\Sigma^{4n} = \bigcup_{e\in \Delta_{j_0}}S_e$ follows instantly from Lemma \ref{swapping-property}(2). Moreover, since all sets in $\{A_e\times B_e\}_{e\in\Delta_{j_0}}$ are mutually disjoint by Lemma \ref{swapping-property}(4), so are  all sets in  $\{S_e\}_{e\in\Delta_{j_0}}$. It thus follows that
$|U_b| = \sum_{e\in\Delta_{j_0}}|U_b\cap S_e|$ for each index $b\in\{0,1\}$. This equality leads to
\begin{equation}\label{eqn:U_1-vs-U_0}
\left||U_1| - |U_0|\right|
= \left| \sum_{e\in\Delta_{j_0}}\left| U_1\cap S_{e}\right| - \sum_{e\in \Delta_{j_0}}\left| U_0\cap S_{e}\right| \right|
\leq \sum_{e\in \Delta_{j_0}}\left| |U_1\cap S_{e}| - |U_0\cap S_{e}|\right|.
\end{equation}
A crude upper-bound of the term $\left||U_1\cap S_e| - |U_0\cap S_e|\right|$ in Eq.(\ref{eqn:U_1-vs-U_0}) is given by the following lemma, whose proof  will be given later for readability.

\begin{lemma}\label{simple-upper-bound}
For every index $e\in\Delta_{j_0}$, it holds that $||U_1\cap S_{e}| - |U_0\cap S_{e}||\leq 2^{7n/2}$.
\end{lemma}

Given an index $e\in\Delta_{j_0}$, we abbreviate as $m(e)$ the target value $||U_1\cap S_e| - |U_0\cap S_e||$.  Lemma \ref{simple-upper-bound} then states that  $m(e) \leq 2^{7n/2}$  for every index $e\in\Delta_{j_0}$.
Since  $|\Delta_{j_0}|< 2^{n/4}$, it follows from the lemma that
\[
||U_1|-|U_0|| \leq \sum_{e\in \Delta_{j_0}} m(e) \leq |\Delta_{j_0}|\cdot \max_{e\in\Delta_{j_0}}\{m(e)\} \\
\leq   2^{n/4}\cdot 2^{7n/2} = 2^{15n/4}.
\]
By $3^cp(4n)<2^{n/4}$, the last term $2^{15n/4}$ is further
bounded from above by ${2^{4n}}/3^{c}p(4n)$, from which  we immediately conclude that $||U_1|-|U_0|| \leq 2^{4n}/3^{c}p(4n)$. In other words, it holds that   $2^{4n} \ell''(4n) \leq 2^{4n}/3^{c}p(4n)$, or equivalently,
$\ell''(4n)\leq 1/3^{c}p(4n)$. This consequence will be used again for the next general case.  As a result, we reach the desired bound of  $\ell''(4n)\leq 1/p(4n)$ because of $c\geq1$.

In the previous basic case, we have assumed that $a=\lambda$. Here, we
want to consider a general case of $a\in\Sigma^{\leq3}$ and $|xyz|=4n$. Let $d\in[0,3]_{\integer}$, representing the length of $a$, and define a restriction $S'_a$ of $S$ for each $a\in\Sigma^{\leq 3}$ to be $S'_a = \{xyz\mid axyz\in S, |xyz|=0\;(\mathrm{mod}\;4)\}$.  Recall that the notation $aS'_a$ expresses the concatenation set $\{aw\mid w\in S'_a\}$. By  the definition of $IP_{3}$, it is not difficult to show that $dense(IP_{3}\cap aS'_a)(4n+d) = dense(IP_{3}\cap S'_a)(4n)$
and $dense(\overline{IP_{3}}\cap aS'_a)(4n+d) = dense(\overline{IP_{3}}\cap S'_a)(4n)$. {}From $S\cap\Sigma^{4n+d} = \left(\bigcup_{a\in\Sigma^{d}}aS'_a\right)\cap\Sigma^{4n+d} = \bigcup_{a\in\Sigma^d}\left( aS'_a\cap\Sigma^{4n+d}\right)$, we can deduce
\begin{eqnarray*}
2^{4n+d}\cdot \ell''(4n+d) &=& \left|dense(IP_{3}\cap S)(4n+d) - dense(\overline{IP_{3}}\cap S)(4n+d)\right| \\
&\leq& \sum_{a\in\Sigma^{d}}\left|dense(IP_{3}\cap S'_a)(4n) - dense(\overline{IP_{3}}\cap S'_a)(4n)\right|.
\end{eqnarray*}
As shown in the basic case, it must hold that $|dense(IP_{3}\cap S'_a)(4n) - dense(\overline{IP_{3}}\cap S'_a)(4n)| < 2^{4n}/3^cp(4n)$.
This inequality guides us to a bound:
\begin{equation}\label{ell''-bound}
2^{4n+d}\cdot \ell''(4n+d)  \leq \left|\Sigma^d\right| \cdot \frac{2^{4n}}{3^{c}p(4n)} = \frac{2^{4n+d}}{3^{c}p(4n)}.
\end{equation}
By our assumption $p(4n+d)\leq p(4n+3)\leq 3^cp(4n)$, it therefore follows from Eq.(\ref{ell''-bound}) that $2^{4n+d}\ell''(4n+d) \leq 2^{4n+d}/p(4n+d)$, or equivalently,  $\ell''(4n+d)\leq 1/p(4n+d)$.

Since $d$ is arbitrary, the inequality  $\ell''(n)\leq 1/p(n)$ is satisfied   for any number $n\in\nat^{+}$.
Since $p$ is also arbitrary, $\ell''(n)$ should be a negligible function. Overall, we can conclude that $IP_{3}$ is indeed  $\cfl/n$-pseudorandom; thus, the proof of Proposition \ref{IP-pseudorandom} is finally completed.

\ms

Note that the aforementioned proof of Proposition \ref{IP-pseudorandom} requires Lemma \ref{simple-upper-bound} to be true.
Henceforth, we shall aim at proving this lemma using a well-known discrepancy upper bound of an inner-product-modulo-two function. To explain this bound, let us introduce a critical notion of {\em discrepancy}.
For convenience, we switch our values $\{0,1\}$ to $\{1,-1\}$ and define our {\em (binary) inner-product-modulo-two function} $f$ as $f(x,y) = (-1)^{x\odot y}$. Now, the discrepancy of a set $T\subseteq\Sigma^{2n}\times\Sigma^{2n}$ with respect to $f$ is then defined as follows.

\begin{definition}
For any set $T\subseteq\Sigma^{2n}\times\Sigma^{2n}$, the {\em discrepancy} of $T$ with respect to $f$ is $Disc_{f}(T) = 2^{-4n}\left| \sum_{(x,y)\in T} f(x,y) \right|$.
\end{definition}

We shall utilize the following technical but crucial lemma, which gives an upper bound of the discrepancy of a particular set $T^{(i,j)}_{A,B}$ induced from pair $(A,B)$. By our choice of $j_0=\floors{4n/3}$ and $k_0=2j_0$, any index $(i,j,u,v)$ in $\Delta_{j_0,k_0,4n}$ must satisfy  $i\in[0,4n]_{\integer}$, $j\in[\floors{4n/3},\ceilings{8n/3}]_{\integer}$, and $i+j\leq 4n$, from which we obtain, in particular,  $i\in[0,3n]_{\integer}$ and $j\in[n,4n]_{\integer}$. In the following lemma, we shall use those relaxed conditions.
Here, we remark that, being consistent with a later application of this lemma to $IP_{3}$, we shall describe this lemma using the {\em reverse} of $y$,   instead of $y$ itself.

\begin{lemma}\label{disc_M(T)-bound}{\em [Key Lemma]}
Let $n$ be any number in $\nat$ at least $4$. Let $i\in[0,3n]_{\integer}$ and $j\in[n,4n]_{\integer}$ with $i+j\leq 4n$. Let $A\subseteq \Sigma^{i}\times\Sigma^{4n-i-j}$ and $B\subseteq \Sigma^{j}$. Define $T^{(i,j)}_{A,B}$ to be a set of all pairs $(xz,y)$ with $x,z\in\Sigma^n$ and $y\in\Sigma^{2n}$ such that there exist three strings $p,q,r\in\Sigma^*$ satisfying $xy^Rz = pqr$,  $(p,r)\in A$,  and $q\in B$. It then holds that $\ell= Disc_{f}(T^{(i,j)}_{A,B})\leq 2^{-n/2}$.
\end{lemma}

Meanwhile, we postpone the proof of Lemma \ref{disc_M(T)-bound} until Section \ref{sec:discrepancy} and we continue the proof of Lemma \ref{simple-upper-bound}.
Let us examine three major cases first and give their associated discrepancy upper bounds, which can be derived by an dexterous application of Lemma \ref{disc_M(T)-bound}.

%%%%
%%%%

Remember that $n$ has been fixed and,  for each index $e\in\Delta_{j_0}$,  $S_e$ expresses the set $\{x'y'z'\in\Sigma^{4n}\mid (x',z')\in A_e, y'\in B_e\}$.
To estimate the value $m(e)=||U_1\cap S_e|-|U_0\cap S_e||$, we shall  consider the corresponding set $T_e = T^{(i,j)}_{A_e,B_e}$, as in Lemma \ref{disc_M(T)-bound}, which is defined from $(A_e,B_e)$ as $T_e =\{(xz,y^R)\mid x,z\in\Sigma^n, y\in\Sigma^{2n}, \exists p,q,r\,[\,xyz=pqr\wedge (p,r)\in A_e\wedge q\in B_e\,]\}$, where we intentionally swap the roles of $y$ and $y^R$ to improve the readability.    The following claim will establish a bridge between $S_e$ and $T_e$. Remember that, since $n\geq4$, the choice of $j_0$ and $k_0$ implies that $n<j_0<2n<k_0<3n$.

\begin{claim}\label{T_e-S}
For any index $e\in\Delta_{j_0}$ and any three strings $x,z\in\Sigma^n$ and $y\in\Sigma^{2n}$, the following relationship holds: $xyz\in S_e$ if and only if  $(xz,y^R)\in T_e$.
\end{claim}

\begin{proof}
Let $e$ denote an arbitrary tuple $(i,j,u,v)$ in $\Delta_{j_0}$ and choose arbitrary strings $x,y,z\in\Sigma^*$ with $|x|=|z|=n$ and $|y|=2n$ for $n\geq4$. Define $w=xyz$. Let us consider $S_e$ and $T_e$ defined earlier.

\s

(Only If--part) Assume that $w\in S_e$. There exists a pair $(i,j)$ with $0\leq i\leq 3n$, $j_0\leq j\leq k_0$, and $i+j\leq 4n$ such that three strings $x'= pref_{i}(w)$, $y'= midd_{i,i+j}(w)$, and $z'=suf_{4n-i-j}(w)$
make the pair $(x'z',y')$ fall into $A_e\times B_e$.
Here, we shall consider only the case where $0\leq i\leq n$ and $n<  i+j\leq 3n$ because the other cases can be proven quite similarly.
Let us express $x$ and $y$ as $x=x_1x_2$ and $y=y_1y_2$ using  four strings $x_1,x_2,y_1,y_2$ that satisfy the condition: (*) $x'=x_1$, $y'=x_2y_1$, and $z'=y_2z$.
Note that $(x_1y_2z,x_2y_1)$ belongs to  $A_e\times B_e$. {}From the definition of $T_{e}$, we can deduce   $(x_1x_2z,y_2^Ry_1^R)\in T_{e}$, which is obviously equivalent to  $(xz,y^R)\in T_e$.

\s

(If--part) Assume that $(xz,y^R)\in T_e$. Take a pair $(i,j)$ with  $0\leq i\leq 3n$, $j_0\leq j\leq k_0$, and $i+j\leq 4n$ satisfying $xyz = x'y'z'$ for three strings $x'= pref_{i}(w)$, $y'= midd_{i,i+j}(w)$, and $z'=suf_{4n-i-j}(w)$. As before, we shall study only the case where $0\leq i\leq n$ and $n<  i+j\leq 3n$. Decompose $x$ and $y$ into $x=x_1x_2$ and $y=y_1y_2$, respectively, to satisfy Condition (*). From $(xz,y^R)\in T_e$, it follows that  $(x_1x_2z,y_2^Ry_1^R)$ is a member of $T_{e}$. This means  that $(x_1y_2z,x_2y_1)$ is contained in $A_e\times B_e$. Since this containment
is further equivalent to  $(x'z',y')\in A_e\times B_e$,  we conclude that $x'y'z'\in S_e$; therefore, we obtain $xyz\in S_e$.
\end{proof}

Let us return to the proof of Lemma \ref{simple-upper-bound}. Using Claim \ref{T_e-S}, we shall connect the value $m(e)$ to the discrepancy of $T_e$.

\begin{claim}\label{u0-u1-disc}
For each index $e$ in $\Delta_{j_0}$, it holds that $m(e)  = 2^{4n}Disc_{f}(T_{e})$.
\end{claim}

\begin{proof}
Let $e$ be any index in $\Delta_{j_0}$. We shall use the following close relationship between  the inner-product-modulo-two function $f$ and $IP_{3}$:
for any strings $x,z\in\Sigma^{n}$ and $y\in\Sigma^{2n}$, it holds that   $f(xz,y^R)=1$ if and only if $xyz\in IP_{3}$. By a direct translation between $S_e$ and $T_e$ given in Claim \ref{T_e-S}, it immediately follows that, for each  index $b\in\{0,1\}$,
\[
|U_b\cap S_e| = |\{(xz,y^R)\in T_e\mid f(xz,y^R)=b\}| = |T_{e}\cap f^{-1}(b)|.
\]
Using these equalities, we calculate the value $2^{4n}Disc_{f}(T_{e})$ as follows:
\begin{eqnarray*}
2^{4n}\cdot Disc_{f}(T_{e}) &=& \left|\sum_{(x,y)\in T_e}f(x,y)\right|
\;\;=\;\; \left|\sum_{(x,y)\in T_e\cap f^{-1}(1)} 1 + \sum_{(x,y)\in  T_e\cap f^{-1}(0)}(- 1)\right| \\
&=& \left||T_{e}\cap f^{-1}(1)| - |T_{e}\cap f^{-1}(0)|\right|
\;\;=\;\; ||U_0\cap S_e| - |U_1\cap S_e||.
\end{eqnarray*}
This result clearly establishes the desired conclusion of the claim.
\end{proof}

For each index  $e=(i,j,u,v)$, by applying Lemma \ref{disc_M(T)-bound} to $(i,j,n,A_e,B_e)$, we immediately obtain a useful bound
$Disc_{f}(T_{e})\leq 2^{-n/2}$.
{}From this bound and also by Claim \ref{u0-u1-disc}, it follows that
\begin{equation*}
m(e) = 2^{4n}\cdot  Disc_{f}(T_e) \leq 2^{4n}\cdot 2^{-n/2}
= 2^{7n/2}.
\end{equation*}
Therefore, we obtain the desired inequality $m(e)\leq 2^{7n/2}$ and, in the end, we have finished the proof of Lemma \ref{simple-upper-bound}, which leads to Proposition \ref{IP-pseudorandom}. The remaining proof of Lemma  \ref{disc_M(T)-bound} will be proven in the next subsection.

%%%%%%%%%%%%%%%%%%%
\subsection{Discrepancy Upper Bounds}\label{sec:discrepancy}

In Section \ref{sec:pseudo-IP*}, we have started proving Proposition \ref{IP-pseudorandom} with the help of our key lemma, Lemma \ref{disc_M(T)-bound}, which have been left unproven. We are now ready to verify this yet-proven lemma and complete the entire proof of the first main theorem, Theorem \ref{generator-CFL}.

To prove Lemma \ref{disc_M(T)-bound}, let us assume that $n$ is an arbitrary integer  with $n\geq4$, $i$ is in $[0,3n]_{\integer}$, and $j$ is in $[n,4n]_{\integer}$ satisfying $i+j\leq 4n$. Moreover, let $A\subseteq \Sigma^i\times \Sigma^{4n-i-j}$ ($=\Sigma^{4n-j}$) and $B\subseteq \Sigma^{j}$. {}From this pair $(A,B)$, its associated set $T^{(i,j)}_{A,B}$ can be introduced as in Lemma \ref{disc_M(T)-bound}. In what follows, for readability, we shall write $T$ for $T^{(i,j)}_{A,B}$ since $i$, $j$, $A$, and $B$ are all fixed throughout this proof. Our goal is to show that the discrepancy $\ell= Disc_{f}(T)$ is upper-bounded by $2^{-n/2}$.

In this proof, there are four separate cases to examine, depending on the conditions of the given pair $(i,j)$. Let us begin with the first case, which deals with the most fundamental situation. Since Case 1 showcases a core of our argument, we wish to detail this case here.

\s
\n{\bf Case 1:} As the first case, we shall consider the case where the pair $(i,j)$ satisfies that $0\leq i\leq n$ and $2n\leq i+j\leq 3n$. Depending on the value of $2i+j$, we shall further argue two separate subcases.

\begin{figure}[t]
\begin{center}
\includegraphics*[width=13.0cm]{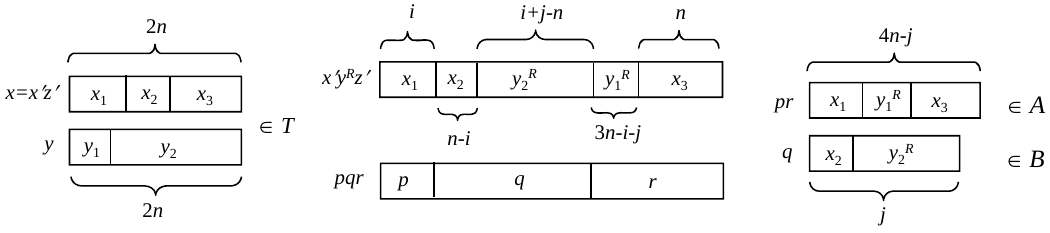}
\caption{A relationship between $T$ and $(A,B)$ in Subcase 1 of Case 1. String $x$ is of the form $x'z'$ and $(x,y)\in T$. Moreover, $pqr$ equals $x'y^Rz'$  and $(pr,q)\in A\times B$.}\label{fig:case-1-subcase-1}
\end{center}
\end{figure}

\s
\n{\bf Subcase 1:} Assume that $2i+j\geq 3n$.
First, let us state the precise definition of $T$ under the current assumption. Each element $(x,y)\in\Sigma^{2n}\times\Sigma^{2n}$ in $T$ should satisfy the following condition: (*)
there exist six strings $x_1,x_2,x_3,y_1,y_2\in\Sigma^*$ with  $x=x_1x_2x_3$, $y=y_1y_2$,  $|x_1|=i$, $|x_2|=n-i$,  $|x_3|=n$, $|y_1|=3n-i-j$, and $|y_2|=i+j-n$ for which $x_1y_1^Rx_3\in A$ and $x_2y_2^R\in B$ hold. This condition (*) is illustrated in Figure~\ref{fig:case-1-subcase-1}.

%%%%%
%%%%%

%%%%%
%%%%%

Next, we choose an arbitrary pair $(x,y)\in T$ and consider their decompositions, $x=x_1x_2x_3$ and $y=y_1y_2$, whose components satisfy the condition (*).
To estimate the value $\ell= Disc_{f}(T)$, we want to use the following simple upper bound of the discrepancy of a product set. How to obtain this bound is  demonstrated in, \eg \cite[Example 13.15]{AB09}.

\begin{lemma}\label{discrepancy-bound}
For any two sets $P,Q\subseteq \Sigma^{2n}$, it holds that $Disc_{f}(P\times Q)\leq  2^{-3n}\sqrt{|P||Q|}$.
\end{lemma}

Unfortunately, we are unable to apply Lemma \ref{discrepancy-bound} directly to $T$ because $T$ is not of the form $P\times Q$ , and thus we need to seek a slightly different way of viewing $T$. One simple way is to view $T$ as a {\em union} of product sets whose discrepancy can be easily estimated by Lemma \ref{discrepancy-bound}. To implement this idea,
we introduce an index set $D = \{(x_2,y_1)\mid x_2\in\Sigma^{n-i},y_1\in\Sigma^{3n-i-j}\}$, which immediately implies
$|D|= 2^{4n-2i-j}$. Fixing each index pair $(a,b)$ in $D$, we further introduce two new subsets $P_{a,b}$ and $Q_{a,b}$ of $\Sigma^{2n}$
as follows.
\begin{itemize}\vs{-1}
\item $P_{a,b}=\{ x_1ax_3 \mid  x_1\in\Sigma^i, x_3\in\Sigma^n,
\exists y_2\in\Sigma^{i+j-n}\, [\, (x_1ax_3, by_2)\in T\,]\}$.
\vs{-2}
\item $Q_{a,b}=\{by_2 \mid y_2\in\Sigma^{i+j-n},
\exists x_1\in\Sigma^{i}\; \exists x_3\in\Sigma^{n}\, [\, (x_1ax_3, by_2)\in T\,]\}$.
\end{itemize}\vs{-1}
Notice that $|P_{a,b}|\leq 2^{n+i}$ since $a$ is fixed. Similarly, since $b$ is fixed, we obtain  $|Q_{a,b}|\leq 2^{i+j-2n}$.  In conclusion, it holds that
$|P_{e}||Q_{e}|\leq 2^{n+i}\cdot 2^{i+j-n} = 2^{2i+j}$.

We further show two useful properties of each product set
$P_{a,b}\times Q_{a,b}$.

\begin{claim}\label{T-equal-P-times-Q}
\begin{enumerate}
\item All product sets in $\{P_{e}\times Q_{e}\}_{e\in D}$ are mutually disjoint.
\vs{-2}
\item $T= \bigcup_{e\in D} (P_{e}\times Q_{e})$.
\end{enumerate}
\end{claim}

\begin{proof}
(1) We want to prove this statement by contradiction. To draw a contradiction, assume that there are two distinct pairs $(a,b),(a',b')\in D$ and an element $(x,y)\in\Sigma^{2n}\times\Sigma^{2n}$ in both $P_{a,b}\times Q_{a,b}$
and $P_{a',b'}\times Q_{a',b'}$. In other words, it holds that
$x\in P_{a,b}\cap P_{a',b'}$ and $y\in Q_{a,b}\cap Q_{a',b'}$.
{}From these membership relations, we obtain $x=x_1ax_3=x'_1a'x'_3$ and $y=by_2=b'y'_2$
for certain strings $x_1,x_3,x'_1,x'_3,y_2,y'_2$ of appropriate lengths. Since $|x_1|=|x'_1|$ and $|y_2|=|y'_2|$, it is obvious that  $a=a'$ and $b=b'$ both hold. This consequence is clearly a contradiction against the difference between $(a,b)$ and $(a',b')$.

(2) In what follows, we wish to prove that (a) $T \subseteq \bigcup_{e\in D} (P_{e}\times Q_{e})$ and (b) $\bigcup_{e\in D} (P_{e}\times Q_{e})\subseteq T$.

(a) Take any pair $(x,y)\in T$ with $x=x_1x_2x_3$ and $y=y_1y_2$, where $x_1\in\Sigma^{i}$, $x_2\in\Sigma^{n-i}$, $x_3\in\Sigma^{n}$, $y_1\in\Sigma^{3n-i-j}$, and $y_2\in\Sigma^{i+j-n}$. By the definition of $P_{a,b}$'s and $Q_{a,b}$'s, the pair $(x,y)$ obviously belongs to $P_{x_2,y_1}\times Q_{x_2,y_1}$, and therefore $(x,y)$ should be in $\bigcup_{e\in D} (P_{e}\times Q_{e})$.

(b) Fixing $(a,b)\in D$ arbitrarily, we plan to show that $P_{a,b}\times Q_{a,b}\subseteq T$. For this purpose, take an arbitrary pair $(x,y)$ in $P_{a,b}\times Q_{a,b}$. Since $x\in P_{a,b}$, there are three strings $x_1,x_3,y'_2$ such that $x=x_1ax_3$ and $(x_1ax_3,by'_2)\in T$.
The definition of $T$ implies both $x_1b^Rx_3\in A$ and $a(y'_2)^R\in B$. Similarly, since $y\in Q_{a,b}$, we obtain
 $y=by_2$ and $(x'_1ax'_3,by_2)\in T$ for certain strings $x'_1,x'_3,y_2$, and therefore $x'_1b^Rx'_3\in A$ and $ay_2^R\in B$ hold.
{}From  $x_1b^Rx_3\in A$ and $ay_2^R\in B$, $T$ should contain  $(x_1ax_3,by_2)$; therefore, $(x,y)\in T$ holds.
\end{proof}

Finally, we shall estimate the discrepancy $\ell = Disc_{f}(T)$. Claim \ref{T-equal-P-times-Q} helps us obtain
\begin{eqnarray*}
\ell &=&  2^{-4n}\left| \sum_{(x,y)\in T}f(x,y)  \right|
\;\;=\;\; 2^{-4n}\left| \sum_{e\in D}\sum_{(x,y)\in P_e\times Q_e}f(x,y)  \right| \\
&\leq& 2^{-4n}\sum_{e\in D}\left| \sum_{(x,y)\in P_{e}\times Q_{e}}f(x,y)  \right|
\;\;=\;\; \sum_{e\in D}Disc_{f}(P_{e}\times Q_{e}),
\end{eqnarray*}
where $f(x,y)=(-1)^{x\odot y}$. Since $|P_{e}||Q_{e}|\leq 2^{2i+j}$ for any $e\in D$,
using Lemma \ref{discrepancy-bound}, $\ell$ is further upper-bounded as
\[
\ell  \leq 2^{-3n}\sum_{e\in D}\sqrt{|P_{e}||Q_{e}|}
\leq 2^{-3n}|D|\max_{e\in D}\left\{ \sqrt{|P_{e}||Q_{e}|} \right\}
\leq 2^{-3n}\cdot 2^{4n-2i-j}\cdot 2^{i+j/2} = 2^{n-i-j/2}.
\]
Since the assumption $2i+j\geq3n$ implies $i+j/2\geq 3n/2$, it follows that  $\ell\leq 2^{n-3n/2} = 2^{-n/2}$, as requested.

\begin{figure}[t]
\begin{center}
\includegraphics*[width=13.0cm]{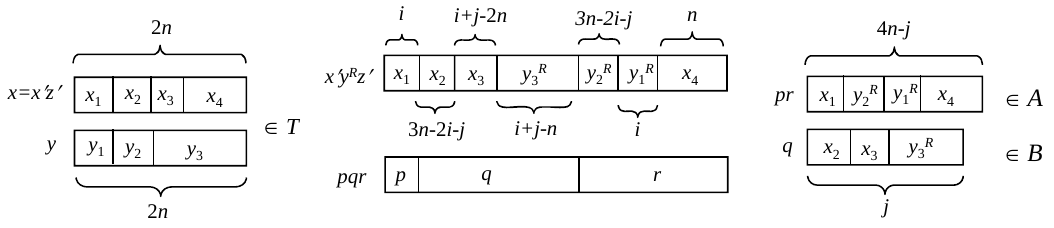}
\caption{A decomposition of $(x,y)\in T$ and its relationship to $(A,B)$
in Subcase 2 of Case 1, where  $x=x'z'$. Strings $p,q,r$ satisfy that $pqr = x'y^Rz'$ and $(pr,q)\in A\times B$.}\label{fig:case-1-subcase-2}
\end{center}
\end{figure}

%%%%
%%%%
\s
\n{\bf Subcase 2:} Next, we assume that $2i+j< 3n$. Note that, for any
element $(x,y)\in\Sigma^{2n}\times\Sigma^{2n}$ in $T$, $x$ and $y$ are always  decomposed as $x=x_1x_2x_3x_4$ and $y=y_1y_2y_3$ with $|x_1|=|y_1|=i$, $|x_2|=|y_2|=3n-2i-j$, $|x_3|=i+j-2n$, $|x_4|=n$, and $|y_3|= i+j-n$ to satisfy both $x_1y_2^Ry_1^Rx_4\in A$ and $x_2x_3y_3^R\in B$. In Figure \ref{fig:case-1-subcase-2}, we
illustrate this decomposition and also a relationship between $T$ and $A\times B$.

%%%%%
%%%%%

%%%%%
%%%%%

To estimate $\ell$, however, we cannot apply the same argument as in the previous subcase since $T$ is no longer expressed as a union of product sets.
In this subcase, therefore, we want to transform $T$ to another set by the following mapping $\mu$ in which we swap certain portions of two strings.
To be more precise, let $(x,y)$ be any element in $T$ having the form $x=x_1x_2x_3x_4$ and $y=y_1y_2y_3$, as described above using the fixed triplet $(i,j,n)$. We define $\mu(x,y) = (\hat{x},\hat{y})$, where $\hat{x}= x_1y_2x_3x_4$ and $\hat{y}=y_1x_2y_3$ are obtained from $x$ and $y$ by swapping $x_2$ and $y_2$.  Associated with $\mu$, we write  $T^{\mu}$ for the  range of $\mu$, namely, $T^{\mu} = \{\mu(x,y)\mid (x,y)\in T\}$.  An important role of $\mu$ is demonstrated in the next simple claim.

\begin{claim}\label{second-T-equal-P-times-Q}
The mapping $\mu$ from $T$ to $T^{\mu}$ is a bijection and satisfies the following condition: for any pair $(x,y)\in T$, if $\mu(x,y)=(\hat{x},\hat{y})$ then $f(x,y) = f(\hat{x},\hat{y})$.
\end{claim}

\begin{proof}
The bijective property of $\mu$ is obvious from its definition. For any  pair $(x,y)\in T$, let $\mu(x,y) = (\hat{x},\hat{y})$. The value $x\odot y$ is calculated as follows:
\begin{eqnarray*}
x\odot y &=& (x_1x_2x_3x_4)\odot (y_1y_2y_3)
\;\;=\;\; x_1\odot y_1 + x_2\odot y_2 + (x_3x_4)\odot y_3 \\
&=& (x_1y_2x_3x_4)\odot (y_1x_2y_3)
\;\;=\;\;  \hat{x}\odot \hat{y}\;(\mathrm{mod}\;2).
\end{eqnarray*}
Obviously, the above equalities yield $f(x,y) = f(\hat{x},\hat{y})$.
\end{proof}

Henceforth, we shall be focused on $T^{\mu}$ instead of $T$.
For convenience, we define an index set $D$ as  $D = \{(x_3,y_1)\mid x_3\in \Sigma^{i+j-2n},y_1\in\Sigma^{i}\}$. Clearly, $|D| = 2^{2i+j-2n}$ holds.
Given an arbitrary pair $(a,b)\in D$, let us introduce the following two sets $P_{a,b}$ and $Q_{a,b}$.

\begin{itemize}\vs{-1}
  \setlength{\topsep}{-2mm}%
  \setlength{\itemsep}{0mm}% original = 1mm
  \setlength{\parskip}{0cm}%

\item $P_{a,b}$ consists of all strings of the form $x_1y_2ax_4$ with $x_1\in\Sigma^i$,  $y_2\in\Sigma^{3n-2i-j}$, and $x_4\in \Sigma^n$ satisfying the following: there exist strings $x_2\in\Sigma^{3n-2i-j}$  and  $y_3\in\Sigma^{i+j-n}$ for which $(x_1y_2ax_4,bx_2y_3)\in T^{\mu}$.

\item $Q_{a,b}$ consists of all strings of the form $bx_2y_3$ with $x_2\in\Sigma^{3n-2i-j}$ and $y_3\in\Sigma^{i+j-n}$ satisfying the following: there exist strings  $x_1\in\Sigma^{i}$, $y_2\in\Sigma^{3n-2i-j}$ and $x_4\in\Sigma^{n}$ for which  $(x_1y_2ax_4,bx_2y_3)\in T^{\mu}$.
\end{itemize}\vs{-1}
It thus follows that
$|P_{a,b}||Q_{a,b}| \leq 2^{4n-i-j}\cdot 2^{2n-i} = 2^{6n-2i-j}$.
We then wish to prove that (i) all product sets in $\{P_{e}\times Q_{e}\}_{e\in D}$ are mutually disjoint and (ii) $T^{\mu}$ equals the union $\bigcup_{e\in D}(P_{e}\times Q_{e})$. Those two properties can be proven in a way similar to Claim \ref{T-equal-P-times-Q} and their proofs are therefore omitted.

The bijection $\mu$ together with the above properties (i)--(ii) helps us calculate the value $\ell$ as
\begin{eqnarray}
\ell &=&  2^{-4n}\left| \sum_{(x,y)\in T}f(x,y)  \right|
\;\;=\;\; 2^{-4n}\left| \sum_{(\hat{x},\hat{y})\in T^{\mu}}f(\hat{x},\hat{y})  \right|
{\nonumber} \\
&\leq& 2^{-4n} \sum_{e\in D} \left| \sum_{(\hat{x},\hat{y})\in P_{e}\times Q_{e}}f(\hat{x},\hat{y}) \right|
\;\;=\;\;  \sum_{e\in D}Disc_{f}(P_{e}\times Q_{e}),
\label{ell-Disc-equation}
\end{eqnarray}
where the second equality comes from Claim \ref{second-T-equal-P-times-Q}.
By Lemma \ref{discrepancy-bound}, we  can further upper-bound $\ell$ by
\begin{eqnarray*}
\ell \leq  2^{-3n}\sum_{e\in D} \sqrt{|P_e||Q_e|}
\leq 2^{-3n}\cdot 2^{2i+j-2n} \cdot 2^{3n-i-j/2} = 2^{i+j/2-2n}.
\end{eqnarray*}
Finally, from $i+j/2<3n/2$, we conclude that  $\ell \leq 2^{i+j/2-2n}\leq 2^{3n/2-2n} = 2^{-n/2}$.

%%%
%%%
\s
\n{\bf Case 2:}
In this second case, we assume that $0\leq i\leq n$  and $3n< i+j\leq 4n$.
Slightly different from Case 1, $T$ is composed of all pairs $(x,y)$ with $x=x_1x_2x_3x_4$ satisfying that $|x_1|=i$,  $|x_2|=n-i$, $|x_3| = i+j-3n$, $|x_4| = 4n-i-j$, $|y|=2n$, $x_1x_4\in A$, and $x_2y^Rx_3\in B$.

As in the previous case, we define an index set $D$ as $D = \{ (x_2,x_3)\mid |x_2|=n-i,|x_3|=i+j-3n\}$ of cardinality $2^{j-2n}$. Given each pair $(a,b)\in D$, two sets $P_{a,b}$ and $Q_{a,b}$ are naturally introduced from $T$ as given below.
\begin{itemize}\vs{-1}
\item $P_{a,b} = \{x_1abx_4\mid x_1\in \Sigma^i, x_4\in\Sigma^{4n-i-j}, \exists y\in\Sigma^{2n}\, [\, (x_1abx_4,y)\in T \,] \}$.
\vs{-2}
\item $Q_{a,b} = \{y\in\Sigma^{2n} \mid \exists x_1\in\Sigma^{i}\, \exists x_4\in\Sigma^{4n-i-j}\, [\, (x_1abx_4,y)\in T \,] \}$.
\end{itemize}\vs{-1}
It follows that $|P_{a,b}||Q_{a,b}|\leq 2^{6n-j}$ since $|P_{a,b}|\leq 2^{4n-j}$ and $|Q_{a,b}|\leq 2^{2n}$.

Similarly to Claim \ref{T-equal-P-times-Q} of Case 1, it is not difficult to show that (i) all product sets in $\{P_{e}\times Q_{e}\}_{e\in D}$ are mutually disjoint and (ii) $T$ coincides with the union $\bigcup_{e\in D}(P_{e}\times Q_{e})$.
 Using these two properties, we can estimate $\ell$  as
\begin{eqnarray*}
\ell =   \sum_{e\in D}Disc_{f}(P_{e}\times Q_{e})  \leq 2^{-3n}\sum_{e\in D}\sqrt{|P_{e}||Q_{e}|}
\leq 2^{-3n}\cdot 2^{j-2n} \cdot 2^{3n-j/2} = 2^{j/2-2n}.
\end{eqnarray*}
Since $j\leq 3n$, $\ell$ clearly satisfies that
$\ell \leq 2^{j/2-2n} \leq 2^{3n/2-2n} = 2^{-n/2}$.

%%%%%
%%%%%
\s
\n{\bf Case 3:}
Assume that  $n<i\leq 2n$ and  $2n < i+j \leq 3n$.  Recall that $j\in[n,4n]_{\integer}$.
Any element $(x,y)\in\Sigma^{2n}\times\Sigma^{2n}$ in $T$ satisfies that  $x=x_1x_2$, $y=y_1y_2y_3$, $|x_1|=|x_2|=n$, $|y_1|=3n-i-j$, $|y_2| = j$, $|y_3|=i-n$, $x_1y_3^Ry_1^Rx_2\in A$, and $y_2^R\in B$.

Our index set $D$ is now  set to be $\{ (y_1,y_3) \mid |y_1|=3n-i-j,|y_3|=i-n\}$, yielding $|D| = 2^{2n-j}$. For each pair $(a,b)\in D$, we  define two sets $P_{a,b}$ and $Q_{a,b}$ as follows.
\begin{itemize}\vs{-1}
\item $P_{a,b}=\{ x_1x_2 \mid x_1,x_2\in\Sigma^n, \exists y_2\in\Sigma^{j}\,  [\, (x_1x_2,ay_2b)\in T \,] \}$.
\vs{-2}
\item $Q_{a,b}=\{ ay_2b \mid y_2\in\Sigma^{j}, \exists x_1,x_2\in\Sigma^n \, [\,
(x_1x_2,ay_2b)\in T \,] \}$.
\end{itemize}\vs{-1}
Since $|a|=3n-i-j$ and $|b|=i-n$, we obtain $|P_{a,b}|\leq 2^{2n}$ and  $|Q_{a,b}|\leq 2^{j}$, from which $|P_{a,b}||Q_{a,b}|\leq 2^{2n+j}$ follows instantly.

The series $\{P_{e}\times Q_{e}\}_{e\in D}$ satisfies  that
(i) all product sets in the series are mutually disjoint and (ii) $T = \bigcup_{e\in D}(P_{e}\times Q_{e})$.
{}From these properties and the inequality $j\geq n$, we deduce  that
\[
\ell = \sum_{e\in D}Disc_{f}(P_{e}\times Q_{e})
\leq 2^{-3n}\sum_{e\in D}\sqrt{|P_{e}||Q_{e}|}
\leq 2^{-3n}\cdot 2^{2n-j}\cdot 2^{n+j/2} =
2^{-j/2} \leq 2^{-n/2}.
\]

%%%%%
%%%%%
%\s
\n{\bf Case 4:}
In this final case, we assume that $n<i\leq 3n$ and  $3n < i+ j \leq 4n$.
In essence, this case is symmetric to Case 1. We shall discuss two subcases, depending on the value of $2i+j$.

\s
\n{\bf Subcase 1:}
Let us consider the case where $2i+j\leq 5n$. Note that $T$ is composed of all pairs $(x,y)$ with $x=x_1x_2x_3$, $y=y_1y_2$, $|x_1|=n$, $|x_2|=i+j-3n$, $|x_3|=4n-i-j$, $|y_1|=3n-i$, and $|y_2|=i-n$ satisfying both $x_1x_3\in A$ and $y_2^Ry_1^Rx_2\in B$.

Take a set $D = \{(x_2,y_2)\mid x_2\in\Sigma^{i+j-3n},y_2\in\Sigma^{i-n}\}$ as our index set with   $|D|=2^{2i+j-4n}$.
Given a pair $(a,b)$ in $D$, two sets $P_{a,b}$ and $Q_{a,b}$ are defined in the following way.
\begin{itemize}\vs{-1}
\item $P_{a,b}=\{ x_1ax_3 \mid x_1\in\Sigma^n, x_3\in\Sigma^{4n-i-j},  \exists y_1\in\Sigma^{3n-i}\,  [\, (x_1ax_3,y_1b)\in T \,] \}$.
\vs{-2}
\item $Q_{a,b}=\{ y_1b \mid y_1\in\Sigma^{3n-i}, \exists x_1\in\Sigma^{n} \exists x_3\in\Sigma^{4n-i-j} \, [\,
(x_1ax_3,y_1b)\in T \,] \}$.
\end{itemize}\vs{-1}
We then obtain $|P_{a,b}||Q_{a,b}|\leq 2^{8n-2i-j}$ from $|P_{a,b}|\leq 2^{5n-i-j}$ and $|Q_{a,b}|\leq 2^{3n-i}$.

Following an argument similar to the one given in Subcase 1 of Case 1, we draw a conclusion that
\[
\ell\leq 2^{-3n}\cdot 2^{2i+j-4n}\cdot 2^{4n-i-j/2} = 2^{i+j/2-3n}.
\]
The assumption $i+j/2\leq 5n/2$ further implies that
$\ell \leq 2^{i+j/2-3n} \leq 2^{5n/2-3n} = 2^{-n/2}$.

\s
\n{\bf Subcase 2:} Under the assumption $2i+j>5n$, $T$ consists of all pairs $(x,y)$ with $x=x_1x_2x_3x_4$, $y=y_1y_2y_3y_4$, $|x_1|=|y_1|=n$, $|x_2|=|y_2|=i+j-3n$, $|x_3|=|y_3|=5n-2i-j$, and $|x_4|=|y_4|=i-n$ for which  $x_1y_4^Rx_3x_4\in A$ and $y_3^Ry_2^Ry_1^Rx_2\in B$ hold.

A bijection $\mu$ from $T$ to $T^{\mu}=\{\mu(x,y)\mid (x,y)\in T\}$ is defined as follows.  Let $(x,y)$ be in $T$ with $x=x_1x_2x_3x_4$ and $y=y_1y_2y_3y_4$ defined above. For this pair $(x,y)$, we set $\mu(x,y) = (\hat{x},\hat{y})$, where $\hat{x}=x_1x_2y_3x_4$ and $\hat{y}=y_1y_2x_3y_4$. Notice that, similarly to Claim \ref{second-T-equal-P-times-Q},   $\mu(x,y)=(\hat{x},\hat{y})$ implies $f(x,y)=f(\hat{x},\hat{y})$.

Here, we set an index set $D$ as $D=\{(a,b)\mid a\in\Sigma^{i+j-3n},b\in\Sigma^{i-n}\}$ of cardinality $2^{2i+j-4n}$. Letting  $(a,b)$ be any pair in $D$, we further define $P_{a,b}$ and $Q_{a,b}$ as follows.
\begin{itemize}\vs{-1}
\item $P_{a,b}$ consists of $x_1ay_3x_4$ with $x_1\in\Sigma^n$, $y_3\in\Sigma^{5n-2i-j}$, and $x_4\in\Sigma^{i-n}$ such that there are strings $y_1\in\Sigma^n$, $y_2\in\Sigma^{i+j-3n}$, and $x_3\in\Sigma^{5n-2i-j}$ satisfying $(x_1ay_3x_4,y_1y_2x_3b)\in T^{\mu}$.
\vs{-2}
\item  $Q_{a,b}$ consists of $y_1y_2x_3b$ with $y_1\in\Sigma^n$, $y_2\in\Sigma^{i+j-3n}$, and $x_3\in\Sigma^{5n-2i-j}$ such that there are strings $x_1\in\Sigma^n$, $y_3\in\Sigma^{5n-2i-j}$, and $x_4\in\Sigma^{i-n}$ satisfying $(x_1ay_3x_4,y_1y_2x_3b)\in T^{\mu}$.
\end{itemize}\vs{-1}
{}From $|P_{a,b}|\leq 2^{5n-i-j}$ and $|Q_{a,b}|\leq 2^{3n-i}$,   the inequality $|P_{a,b}||Q_{a,b}|\leq 2^{8n-2i-j}$ follows.  Note that
Eq.(\ref{ell-Disc-equation}) also holds in this case. Using this equation, $\ell$ is upper-bounded by
\[
\ell \leq 2^{-3n}\cdot 2^{2i+j-4n}\cdot 2^{4n-i-j/2} = 2^{i+j/2-3n}.
\]
From our assumption $i+j/2\leq 5n/2$, it easily follows that $\ell\leq 2^{i+j/2-3n} \leq 2^{5n/2-3n} = 2^{-n/2}$.

%%%
%%%

\ms

In all the possible cases, the desired inequality $\ell\leq 2^{-n/2}$ always holds. This conclusion finishes the proof of Lemma \ref{disc_M(T)-bound}. At last, the entire proof of Theorem \ref{generator-CFL} is completed.

%%%%%%%%%%%%%%%%%%%%%%%%%%%%%%%%%%%
\section{Summary and Future Work}

{\em Pseudorandom generators} have played an essential role in modern cryptography and also have impacted the development of computational complexity theory. Throughout this paper, we have discussed such generators in a slightly different framework of ``formal language and automata theory.'' The first discussion in this framework was made in  \cite{Yam11}, in which an almost 1-1 pseudorandom generator against $\reg/n$  is constructed in $\cflsvt$ but no pseudorandom generator against $\reg$ with stretch factor $n+1$ is shown to exist in $\oneflin$. In this paper, we have taken a further step toward a full understanding of pseudorandomness in this framework.
In particular, we have proven that an almost 1-1  pseudorandom generator against $\cfl/n$ actually exists in $\mathrm{FL}\cap\cflmvtwo/n$ (Theorem \ref{generator-CFL}) but  no almost 1-1 pseudorandom  generator against $\cfl$ stretching $n$-symbol seeds to $(n+1)$-symbol strings over a certain alphabet exists in $\cflmv$ (Theorem \ref{no-generator}).
Notably, a core of our proof of Theorem \ref{generator-CFL} is a demonstration of the $\cfl/n$-pseudorandomness of $IP_{3}$ (and thus $IP^{+}$).

Beyond the above-mentioned results, there still remain numerous questions concerning the pseudorandomness of languages and the efficiency of pseudorandom generators. For instance, we can raise the following basic questions.

\begin{itemize}\vs{-1}
  \setlength{\topsep}{-2mm}%
  \setlength{\itemsep}{0mm}% original = 1mm
  \setlength{\parskip}{0cm}%

\item[1.] Our $\cfl/n$-pseudorandom language $IP_3$ belongs to $\dl\cap \cfl(2)/n$ (Proposition \ref{IP-dtime}). Is there any $\cfl/n$-pseudorandom language in $\cfl(2)$, instead of $\cfl(2)/n$?  An affirmative answer to this question exemplifies a seemingly larger gap between $\cfl/n$ (and thus $\cfl$) and $\cfl(2)$.

\item[2.] As discussed in Section \ref{sec:random-generator}, the  generator $G(x)=x\cdot \chi_{IP_3}(x)$ is one-to-one and also pseudorandom against $\cfl/n$; however, it is unlikely to belong to  $\cflmvtwo/n$. Does a one-to-one pseudorandom generator against $\cfl/n$ actually exist in $\cflmvtwo/n$?

\item[3.]  Find much more efficient pseudorandom generators against $\cfl/n$, which is, for example, computable in $\cflsvtwo/n$ or even $\cflsvtwo$ (or a much lower complexity class). To achieve such efficiency, we may need to seek generators that are not even almost 1-1.
\end{itemize}

Besides the language family $\cfl(2)$, we can consider a more general language family $\cfl(k)$ for any number $k\geq2$. Note that $\cfl(k)$ ($k$-conjunctive closure)
is a collection of all languages, each of which is made by  the intersection of $k$ context-free languages (see, \eg \cite{Yam11,Yam14a}).
Its advised version, $\cfl(k)/n$, contains all languages $L$ of the form
$\{x\mid \track{x}{h(|x|)}\in S\}$ for
certain languages $S\in\cfl(k)$ and certain length-preserving advice functions  $h$. In Corollary \ref{CFL(2)-vs-CFL/n}, we have shown that $\cfl(2)\nsubseteq \cfl/n$.

\begin{itemize}\vs{-1}
  \setlength{\topsep}{-2mm}%
  \setlength{\itemsep}{0mm}% original = 1mm
  \setlength{\parskip}{0cm}%

\item[4.] For each given index $k\geq2$, is there any efficient  pseudorandom generator  against $\cfl(k)/n$ (with or without the almost one-to-oneness)?

\item[5.] For every index $k\geq2$, is there any $\cfl(k)/n$-pseudorandom language in $\cfl(k+1)$? An affirmative answer also settles an open question of whether $\cfl(k+1)\nsubseteq \cfl(k)/n$ for all numbers  $k\geq2$.
\end{itemize}

Structural properties of functions that are computed by simple-structured  ``one-way''  machines (such as npda's) with write-only output tapes are largely unexplored in formal language and automata theory. In a polynomial-time setting, it is well known that  the behaviors of functions are quite different in nature from those of languages (see, \eg \cite{BLS84,Sel94}).
Naturally, we expect that a similar difference is present in ``low-complexity'' counterparts.  We strongly believe that it is possible to develop an exciting theory of functions in various low-complexity function classes, including $\cflmv$,  $\cflsv$,  $\cflsvt$, and moreover $\mathrm{CFL}(k)\mathrm{SV_t}$, where $\mathrm{CFL}(k)\mathrm{SV_t}$ is a functional version of $\cfl(k)$, and their advised analogues.
For other interesting function classes, refer to \cite{Yam14a,Yam14b}.

\begin{itemize}
\item[6.] Find interesting properties and useful applications of multi-valued partial functions that are computed by simple-structured one-way machines with write-only output tapes.
\end{itemize}

%%%%%%%%%%%%%%%%%%%%%%%%%%
%%%%%%%%%%%%%%%%%%%%%%%%%%
%%%%%%%%%%%%%%%%%%%%%%%%%%
\bibliographystyle{alpha}

%%%%%%%%%%%%%%%%%%
\end{document}
%%%%%%%%%%%%%%%%%%%%%%%%%%%%%%
%%%%%%%%%%%%%%%%%%%%%%%%%%%%
%%%%%%%%%%%%%%%%%%%%%%%%%%%%

%%%%%%%%%%%%%%%%%%%%%%%%%%%
%%%%%%%%%%%%%%%%%%%%%%%%%%%
\end{document}